%% file: main.tex
\documentclass[twocolumn,onecolappendix]{aastex63}

\usepackage{makecell}
\usepackage{xcolor}

\usepackage{amsmath,amsthm,amssymb,amsfonts}
\usepackage{graphics}
\usepackage{listings}
\usepackage{enumitem}
\usepackage{url}
\usepackage{appendix}
\usepackage{hyperref}
\usepackage{mathtools}

\usepackage[super]{nth}

\newcommand{\eps}{e$^-$ s$^{-1}$}
\newcommand{\zem}{\texttt{Z14}}

\newcommand{\sbunit}{nW m$^{-2}$ sr$^{-1}$}

\newcommand{\bootes}{Bo{\"o}tes}

\begin{document}
\correspondingauthor{Richard M. Feder}
\email{rmfeder@berkeley.edu}

\author[0000-0002-9330-8738]{Richard M. Feder}
\affiliation{California Institute of Technology, 1200 E California Blvd, 91125, USA}
\affiliation{Berkeley Center for Cosmological Physics, University of California, Berkeley, CA 94720,
USA}
\affiliation{Lawrence Berkeley National Laboratory, Berkeley, California 94720, USA}

\author[0000-0002-5710-5212]{James J. Bock}
\affiliation{California Institute of Technology, 1200 E California Blvd, 91125, USA}

\author[0000-0002-5437-0504]{Yun-Ting Cheng}
\affiliation{California Institute of Technology, 1200 E California Blvd, 91125, USA}

\author[0000-0002-3892-0190]{Asantha Cooray}
\affiliation{Department of Physics \& Astronomy, University of California
Irvine, Irvine CA 92697, USA}

\author{Phillip M. Korngut}
\affiliation{California Institute of Technology, 1200 E California Blvd, 91125, USA}

\author[0000-0002-5698-9634]{Shuji Matsuura}
\affiliation{School of Science and Technology, Kwansei Gakuin University, Sanda, Hyogo 669-1337, Japan}

\author[0000-0002-8802-5581]{Jordan Mirocha}
\affiliation{Jet Propulsion Laboratory, California Institute of Technology, 4800 Oak Grove Drive, Pasadena, CA 91109, USA}
\affiliation{California Institute of Technology, 1200 E California Blvd, 91125, USA}

\author[0000-0001-9368-3186]{Chi H. Nguyen}
\affiliation{California Institute of Technology, 1200 E California Blvd, 91125, USA}

\author[0000-0002-8405-9549]{Kohji Takimoto}
\affiliation{Department of Solar System Sciences, Institute of Space and Astronautical Science, Japan Aerospace Exploration Agency, 3-1-1 Yoshinodai, Chuo-ku, Sagamihara, Kanagawa 252-5210, Japan}

\author[0000-0001-7143-6520]{Kohji Tsumura}
\affiliation{Department of Natural Sciences, Faculty of Science and Engineering,
Tokyo City University, 158-8557 Tokyo, Japan}

\author[0009-0005-2265-2506]{Ryan Wills}
\affiliation{Center for Detectors, School of Physics and Astronomy, Rochester Institute of Technology, 1 Lomb Memorial Drive, Rochester, New York 14623, USA}

\author[0000-0001-8253-1451]{Michael Zemcov}
\affiliation{Center for Detectors, School of Physics and Astronomy, Rochester Institute of Technology, 1 Lomb Memorial Drive, Rochester, New York 14623, USA}
\affiliation{Jet Propulsion Laboratory, California Institute of Technology, 4800 Oak Grove Drive, Pasadena, CA 91109, USA}

\author{CIBER collaboration}
\title{\emph{CIBER} \nth{4} flight fluctuation analysis: Measurements of near-IR auto- and cross-power spectra on arcminute to sub-degree scales}
\shorttitle{CIBER \nth{4} Flight Fluctuation Analysis}
\begin{abstract}
    We present new anisotropy measurements in the near-infrared (NIR) for angular multipoles $300<\ell<10^5$ using imaging data at 1.1 $\mu$m and 1.8 $\mu$m from the fourth flight of the Cosmic Infrared Background ExpeRiment (\emph{CIBER}). Using improved analysis methods and higher quality fourth flight data, we detect surface brightness fluctuations on scales $\ell<2000$ with \emph{CIBER} auto-power spectra at $\sim14\sigma$ and 18$\sigma$ for 1.1 and 1.8 $\mu$m, respectively, and at $\sim10\sigma$ in cross-power spectra. 
    The \emph{CIBER} measurements pass internal consistency tests and represent a $5-10\times$ improvement in power spectrum sensitivity on several-arcminute scales relative to that of existing studies. Through cross-correlations with tracers of diffuse galactic light (DGL), we determine that scattered DGL contributes $< 10\%$ to the observed fluctuation power at high confidence. On scales $\theta > 5\arcmin$, the \emph{CIBER} auto- and cross-power spectra exceed predictions for integrated galactic light (IGL) and integrated stellar light (ISL) by over an order of magnitude, and are inconsistent with our baseline IGL+ISL+DGL model at high significance. We cross-correlate two of the \emph{CIBER} fields with 3.6 $\mu$m and 4.5 $\mu$m mosaics from the \emph{Spitzer} Deep Wide-Field Survey and find similar evidence for departures from Poisson noise in \emph{Spitzer}-internal power spectra and \emph{CIBER} $\times$ \emph{Spitzer} cross-power spectra. A multi-wavelength analysis indicates that the auto-power of the fluctuations at low-$\ell$ is bluer than the Poisson noise from IGL and ISL; however, for $1\arcmin <\theta < 10\arcmin$, the cross-correlation coefficient $r_{\ell}$ of nearly all band combinations decreases with increasing $\theta$, disfavoring astrophysical explanations that invoke a single correlated sky component.   
\end{abstract}



\tableofcontents

\input{sections/introduction}
\input{sections/ciber}

\input{sections/data_preprocess}

\input{sections/mask}
\input{sections/noise_model}
\input{sections/ps_estimation}
\input{sections/ciber_results}

\input{sections/dgl}

\input{sections/overview_spec}
\input{sections/modl_interp}

\input{sections/conclusion}
\input{sections/acknowledgements}
\newpage
\appendix
\input{sections/field_diff}
\input{sections/halfexp_crosscorr}

\input{sections/image_filter}
\input{sections/dgl_zl_mono_bpcorr}
\input{sections/spitzer_mopex_selfcal}

\bibliography{references}{}
\bibliographystyle{aasjournal}

\end{document}

%% file: sections/introduction.tex
\section{Introduction}
\label{S:introduction}
Characterizing the NIR extragalactic background light (EBL) through its spatial and spectral distribution offers a window into the astrophysical processes that drive cosmic light production \citep{hauser_dwek_review, cooray_ebl}. Over the course of several decades, experiments using instruments with precisely-determined absolute calibration have made progress toward measuring the EBL intensity monopole and its frequency spectrum as a means of interpreting EBL components. However, these measurements are plagued by issues when disentangling true extragalactic components from local foregrounds. In particular, errors in zodiacal light (ZL) and diffuse galactic light (DGL) subtraction can lead to order-unity errors on the EBL at optical and NIR wavelengths, as these foregrounds are much brighter than the EBL. Measurements at large heliocentric distance benefit from reduced ZL brightness, but have not converged and still require great care in the presence of other foregrounds \citep{new_horizons_22, zemcov_nh, symons_nh, postman_24}.

Fluctuation-based measurements largely bypass these challenges due to the fact that ZL is measured to be smooth on few-arcminute angular scales by \emph{Spitzer}, \emph{ISO} and \emph{AKARI} spanning 2.4-25 $\mu$m \citep{arendt16, abraham97, pyo_2012} and in the near-infrared on degree scales by \emph{DIRBE} \citep{kelsall}. Fluctuations measurements contain information beyond the intensity monopole related to the scale-dependent clustering and abundance of galaxies. Fluctuations on small scales are driven by the Poisson noise of discrete sources and can in principle be used to constrain the galaxy luminosity function \citep[e.g.,][]{helgason}. On arcminute to degree scales, EBL fluctuations encode the linear (``two-halo") clustering of galaxies, which traces large-scale structure and non-linear galaxy clustering (``one-halo"), which traces the formation of galaxies within a dark matter halo \citep{cooray_sheth}. Through cross-correlations, fluctuation measurements are immune to noise biases and enable a spatial-spectral decomposition that can test existing models and potentially facilitate astrophysical component separation. 

NIR fluctuation measurements have been pursued with deep imaging data from \emph{Spitzer} \citep{kashlinsky_2005, cooray12, kashlinsky_12}, the \emph{AKARI} satellite \citep{matsumoto_akari, seo_akari}, \emph{HST} \citep{thompson_2007, mitchellwynne, matsumoto_hst}, \emph{2MASS} \citep{kashlinsky_2MASS, odenwald_03} and the Cosmic Infrared Background ExpeRiment \citep[\emph{CIBER};][]{zemcov14}. While EBL fluctuation studies can carry their own sets of systematics, a growing body of measurements has pointed to departures in large-angle fluctuation power from the Poisson fluctuations of integrated galaxy light (IGL) alone. These studies typically probe fluctuations on few-arcminute scales and finer, due to challenges of mitigating instrumental systematics and recovering modes much larger than the instantaneous FOV. The existing auto-power spectrum measurements from different instruments imply a blue wavelength dependence \citep[e.g.,][]{seo_akari}; however, measurements of the coherence of fluctuations across NIR wavelengths, as accessed through cross-correlation, have not been pursued within a uniform analysis, precluding further interpretation of its origin(s).

Many astrophysical scenarios have been proposed to explain the measured NIR intensity fluctuations. One possibility is that some known astrophysical populations (e.g., galaxies, dwarf galaxies) are not fully accounted for by current models \citep{helgason14}. Another explanation is that diffuse intra-halo light (IHL) in the outskirts of galaxies is abundant and contributes to EBL fluctuations on linear and non-linear scales. Fluctuation analyses \citep{zemcov14, cooray12} and stacking analyses \citep{chengihl} have estimated IHL fractions ranging from a few per cent to as much as 50\% (relative to the total IGL+IHL intensity), however it should be noted that these estimates rely on assumptions about the mass and redshift dependence of the IHL signal, the boundary radius beyond which extended galaxy light is considered ``IHL", as well as assumptions about the contributions from other components such as non-linear IGL clustering \citep{cheng22} and DGL. At higher redshifts, a signal from the Lyman-$\alpha$ break associated with epoch of reionization (EoR) galaxies is expected \citep{cooray12_model}, however its amplitude is expected to be several orders of magnitude smaller than existing \emph{CIBER} fluctuation measurements \citep{zemcov14}. Some studies have posited emission from direct collapse black holes (DCBHs) at high redshift \citep{yue13, dcbh_14, cappelluti_13, cappelluti_17}, motivated by non-zero cross-correlations between \emph{Spitzer} infrared and \emph{Chandra} X-ray maps. Single species (monoenergetic) eV-scale dark matter (DM) models with a coupling to light through decay and/or annihilation \citep{gong_axion, cyril_18, kalashev19, caputo21} have also been invoked as an explanation for observed fluctuations, as the range of masses and axion-photon coupling strengths is currently unconstrained \citep{bernal_21, bernal_22}. 


In order to distinguish between the plethora of astrophysical components -- both ``local" and extragalactic -- that could contribute to diffuse intensity fluctuations, more sensitive measurements spanning a broad range of angular scales and wavelengths are needed. In this work we apply an improved methodology for measuring EBL anisotropies on sub-degree angular scales (detailed in Feder et al. 2025a, hereafter Paper I) to 1.1 $\mu$m and 1.8 $\mu$m imager data from the fourth and final flight of \emph{CIBER}-1. In Paper I, we developed a power spectrum formalism to quantify and correct for flat field (FF) errors using in situ flight data. We also made improvements in achieving deep source masking, to better separate large-angle diffuse fluctuations from the Poisson fluctuations of discrete sources. We also re-analyze IRAC 3.6 $\mu$m and 4.5 $\mu$m mosaics from the \emph{Spitzer} Deep Wide-Field Survey \citep[SDWFS,][]{sdwfs} that overlap with two of the five \emph{CIBER} fields, enabling a full measurement of auto- and cross-power spectra across $1-5$ $\mu$m.

The paper is organized as follows. We introduce \emph{CIBER} in \S \ref{S:ciber} and describe the observations and map pre-processing steps in \S \ref{S:data_preprocess}. In \S \ref{Sec:mask} we detail the construction of instrument and astronomical masks, validating our source masking with stacked aperture photometry. In \S \ref{Sec:noise_model} we present the read and photon noise models, which we validate using flight differences. We summarize the pseudo-power spectrum formalism applied in this work (and developed in Paper I) in \S \ref{Sec:ps_estimation}. We then present the \emph{CIBER} 1.1 $\mu$m and 1.8 $\mu$m auto-power spectrum measurements in \S \ref{sec:ciber_results}, assessing their internal consistency and sensitivity to various instrumental and analysis systematic effects. In \S \ref{sec:cross_spectra} we present \emph{CIBER} 1.1 $\times$ 1.8 $\mu$m cross-power spectra and cross-correlations with \emph{Spitzer} 3.6 $\mu$m and 4.5 $\mu$m mosaics that overlap with the two \emph{CIBER} \bootes\ fields. We detail \emph{CIBER} and \emph{Spitzer} cross-correlations with tracers of DGL in \S \ref{sec:ciber_iris_xcorr}. We then adopt a single astronomical source mask and characterize the color/correlation structure of the \emph{CIBER} and \emph{Spitzer} fluctuations using the full set of auto- and cross-power spectra in \S \ref{sec:super_cl}. Lastly, we describe the astrophysical predictions presented in this work and discuss several interpretations of the multi-wavelength mesaurements in \S \ref{sec:modl_interp}. 

Throughout this work we quote all fluxes in the Vega magnitude system unless otherwise specified.

%% file: sections/ciber.tex
\section{Cosmic Infrared Background Experiment (CIBER)}
\label{S:ciber}

\emph{CIBER}\footnote{\url{https://ciberrocket.github.io/}} is a rocket-borne instrument designed to characterize the NIR EBL through measurements of its spatial fluctuations and electromagnetic spectrum \citep{zemcov13}. Four instruments were flown as part of \emph{CIBER}: two wide-field imagers \citep[][the focus of this work]{bock13}, a narrow-band spectrometer \citep{ciber_nbs, korngut_zl}, and a low-resolution spectrometer \citep{ciber_lrs, matsuura_ciber}. Each imager consisted of a wide-field refracting telescope with an 11 cm aperture, band-defining filters, a cold shutter, and a 1024$\times$1024 HgCdTe Hawaii-1 detector. The instantaneous field of view for each imager is $2^{\circ}\times 2^{\circ}$, which corresponds to a pixel size of 7\arcsec\ $\times$ 7\arcsec.

\begin{figure*}
\centering
    \includegraphics[width=0.85\linewidth]{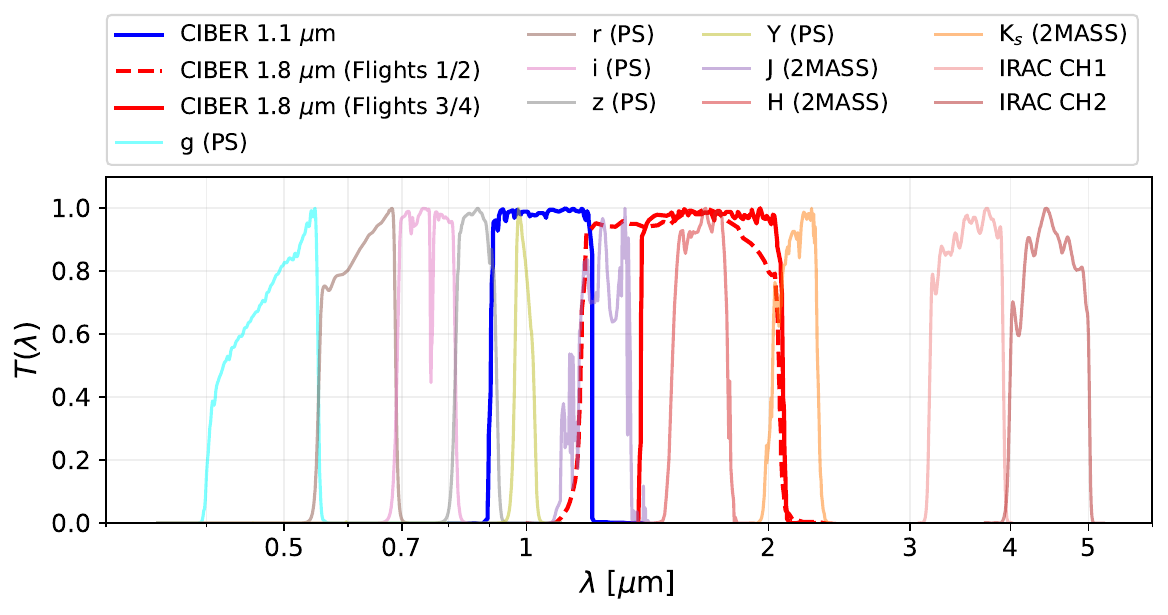}
    \caption{Peak-normalized filter transmission curves for the datasets used in this work. The \emph{CIBER} imagers (\textbf{bold}) are centered at 1.05 $\mu$m and 1.79 $\mu$m (third and fourth flights). We also plot the filters from PanSTARRS \citep[PS;][]{panstarrs} in the optical and 2MASS in the NIR. In \S \ref{sec:spitzer_noise} we cross-correlate against IRAC mosaics from the \emph{Spitzer} Deep Wide-Field Survey \citep[SDWFS;][]{sdwfs} at 3.6 $\mu$m and 4.5 $\mu$m.}
    \label{fig:bandpass_filters}
\end{figure*}

\emph{CIBER}-1 was flown four times. In the first flight, scattered thermal emission from the rocket skin contaminated imager measurements \citep[described in][]{tsumura_10}, leading to improvements in telescope baffling for subsequent flights \citep{bock13}. Between the second and third flights, the long wavelength filter was swapped, shifting the central wavelength from 1.6 $\mu$m to 1.79 $\mu$m. In Figure \ref{fig:bandpass_filters} we show the CIBER filters alongside those of other datasets used in this work. We use imager data from the fourth \emph{CIBER} flight that launched at 3:05 UTC 2013 June 6 from Wallops Flight Facility on a four-stage Black Brant XII rocket. Unlike in the first three flights, the payload during the fourth flight achieved an altitude of 550 km compared to $\sim 330$ km, resulting in a longer total exposure time and lower levels of airglow contamination, though the payload itself was not recovered. 

\emph{CIBER}-1 observed eight consecutive fields during the fourth flight. The first and second fields are not used in this work due to airglow contamination (most prominently near 1.6 $\mu$m). The third field observed was Lockman Hole, which we excluded after identifying evidence of sunlight contamination due to lower solar elongation angle $(\epsilon = 69.33^{\circ})$ than the other fields. After these cuts, we retain the last five fields for our analysis. Table \ref{tab:ciber_fields} contains information on the coordinates and integration times of the fields, which reside at high Galactic latitude $(b > 44^{\circ})$ and span a range of ecliptic latitudes $(11.5^{\circ} \leq \beta \leq 72.5^{\circ})$. 

\begin{table*}
    \centering
    \begin{tabular}{c|c|c|c|c|c|c|c}
        Field & (R.A., Decl.) & ($l, b$) & ($\lambda, \beta$) & $t_{exp}$ & N$_{\textrm{frame}}$ & ZL at 1.1 $\mu$m & ZL at 1.8 $\mu$m \\
        & (deg.) & (deg.) & (deg.) & (s) & & (\sbunit) & (\sbunit) \\
        \hline
         elat10 & (190.5, 8.0) & (295.8, 70.7) & (186.5,11.5) & 42.7 & 24 & 660 & 350 \\
         elat30 & (193.1, 28.0) & (109.2, 89.1) & (179.8,30.7) & 17.8 & 10 & 470 & 260 \\
         \bootes\ A & (218.4, 34.8) & (58.6, 66.9) & (200.5,46.6) & 51.6 & 29 & 350 & 190 \\
         \bootes\ B & (217.2, 33.2) & (55.0, 68.2)& (200.3,44.7) & 49.8 & 28 & 345 & 185  \\
         SWIRE & (242.8, 54.8) & (84.6, 44.7) & (208.9,72.5) & 44.5 & 25 & 285 & 150 
    \end{tabular}
    \caption{Table of science fields used in analysis. We include the celestial, galactic and ecliptic coordinates of our fields, the exposure times $t_{exp}$, and the mean Zodiacal light levels as determined by the \cite{kelsall} model that determine the relative photon noise levels across fields. Coordinates are in reference to the central pointing of the 1.1 $\mu$m imager.}
    \label{tab:ciber_fields}
\end{table*}

%% file: sections/data_preprocess.tex
\section{Data Pre-processing}
\label{S:data_preprocess}

\subsection{Time stream filtering and slope fits}

We employ time stream filtering to remove pickup noise in readout electronics, which is typically most severe in dark exposures taken with an external power supply. The Hawaii-1 detectors are read out in four-channel mode, where each channel corresponds to one detector quadrant. Inspecting the time-ordered data from each separate quadrant and computing its 1D power spectrum, we identify noisy frequencies for the TM1 (1.1 $\mu$m) and TM2 (1.8 $\mu$m) detectors at $9.503$ Hz and 9.538 Hz respectively, corresponding to an angular scale of $\ell \sim 6200$ \citep{chi_thesis}. We use a notch filter centered on the noisy frequencies with a bandwidth of 0.1 Hz, which reduces the associated noise power by a factor of $\gtrsim 10^2$. The transfer function to a slope for these narrow-band filters is near unity. 

We then perform slope fits starting two frames after each global reset. Due to pointing instability during the first half of the elat30 field exposure, we only use the last ten frames of the integration. For a very small fraction of pixels with high photocurrent, we correct the slope fits for ADU register wraps.

\subsection{Dark current subtraction}
We correct for the non-zero response of the detectors in the absence of photons, known as dark current, which is sourced by thermally produced charge carriers and multiplexer glow. To correct for the dark current from each imager we average the set of dark exposures taken on the rail before flight and subtract the mean template from each photocurrent map. The mean dark current level is $\sim 0.2$ \eps\ for both detectors. The raw dark current templates have a fluctuation amplitude that is $<10\%$ of that from the observed CIBER signal on all scales, so residual errors in the templates have a negligible impact.

\subsection{Gain calibration}

The observed surface brightness $\lambda I_{\lambda}$ is related to the digitized detector output current $i$ through
\begin{equation}
    \lambda I_{\lambda} \, [\textrm{nW m}^{-2} \textrm{sr}^{-1}] = g_a g_1 g_2 i \, [\textrm{ADU fr}^{-1}],
\end{equation}
where $g_a$ is the amplification gain from ADU to Volts from the electronics design \citep{zemcov13}. The second factor, $g_1$, converts from units of Volts frame$^{-1}$ to photocurrent (\eps), while $g_2$ denotes the conversion from photocurrent to surface brightness. In contrast to \cite{zemcov14}, which used factory values for $g_1$, we use the CIBER data to directly estimate both $g_1$ and $g_2$, in order to accurately model the contribution of photon noise.

\subsubsection{Estimation of $g_1$}
\label{sec:g1}
Following App.B in \cite{chi_thesis} we estimate the gain factor $g_1$ using the noise statistics of the flight data. In the photon-noise dominated limit, the per-pixel root-mean-squared (RMS) in digital units, $\sigma_{pix}^{dig}$ (ADU frame$^{-1}$) is linearly proportional to the photocurrent noise RMS by $g_1$, such that measuring the $\sigma_{pix}^{dig}$ in exposures with varying mean levels allows us to fit directly for $g_1$.

We estimate $g_1$ using 22 readout frames from each of the science fields. For each field, we compute a difference image from 11-frame masked half-exposures. We restrict the set of pixels from each exposure to those with relative gain in the range $0.95 \leq FF \leq 1.05$ and also make a cut on pixels with correlated double sample (CDS) noise more than 3$\sigma$ away from the median CDS. Once the noise variance in each field is calculated we perform jackknife resampling across the fields to estimate $g_1$ and its uncertainty to be $-2.67\pm 0.02$ at 1.1 $\mu$m and $-3.04\pm 0.02$ at 1.8 $\mu$m. We find similar results using pre-flight laboratory test data ($g_1 = -2.50\pm 0.05$ and $-2.80 \pm 0.10$ for 1.1 $\mu$m and 1.8 $\mu$m, respectively), however we use $g_1$ estimates from the flight data as they capture the detector condition more faithfully. In \S \ref{sec:ps_vs_g1} we show the impact of different assumed $g_1$ factors on the \emph{CIBER} auto power spectra.

\subsubsection{Absolute gain calibration}
\label{sec:abscal}
We perform point source flux calibration for each imager to estimate the total calibration factor (i.e., $g_ag_1g_2$) going from ADU frame$^{-1}$ to \sbunit. We begin by identifying all 2MASS sources with magnitudes $11 < J < 14.5$. The bright end is set to avoid non-linear detector response above an integrated charge of $7.5\times10^4$ e-, while the faint end ensures well-determined fluxes for calibrator sources. We then perform aperture photometry in 13$\times$13 pixel sub-regions centered on the sources. The sub-region size is chosen to be large enough to obtain reliable background estimates; however, to mitigate biases from nearby bright sources we remove any calibration sources with bright ($J < 16$) neighbors in the same region. After cutting sources with neighbors and postage stamps with masking fractions $>5\%$, we are left with between 500 and 800 sources per field. 

We estimate the absolute gain by comparing predicted \emph{CIBER} measurements with 2MASS photometry, which are related by

\begin{equation}
    \lambda F_{\lambda} [\textrm{nW m}^{-2}] = \int \lambda I_{\lambda}(\theta, \phi) d\Omega.
\end{equation}
Choosing a suitable region of integration around the source, we perform a change of variables to detector coordinates,

\begin{equation}
    \lambda F_{\lambda} = \int \int \lambda I_{\lambda}(x,y)\Omega_{pix} dx dy.
\end{equation}
While we have access to fluxes from the 2MASS catalog, the \emph{CIBER} maps are in digital units. The intensity map is then $\lambda I_{\lambda}(x,y) = g_2(\lambda)i_{phot}^{CIBER}(x, y)$, meaning we can solve the gain from each 2MASS source $i$ with

\begin{equation}
    g_ag_1g_2(\lambda) = \frac{\hat{\lambda F_{\lambda}}^{2MASS}}{\Omega_{pix}\int\int i_{phot}^{CIBER}(x,y)dxdy}.
\end{equation}
The \emph{CIBER} PSF is sum normalized, i.e., $\int_{d\Omega}P(\theta, \phi) \equiv 1$.

Accurate surface brightness calibration is complicated both because the 2MASS fluxes are observed over a different integrated bandpass than the \emph{CIBER} filters, and because the diffuse sky component may in general have a different spectrum, and thus different effective wavelength, compared to the point sources used for calibration. For each calibration source, we use cross-matched photometry from 2MASS, PanSTARRS and WISE (described further in \S 6 of Paper I) to convert to monochromatic flux densities, fitting a smooth spline to model the source spectrum. We use the model SEDs to interpolate to the \emph{CIBER} filter central wavelengths and to calculate the bandpass correction needed to predict the flux response $R$,
\begin{equation}
    R = \frac{\int \lambda I(\lambda)F(\lambda)d\lambda}{I(\lambda_{eff})\Delta \lambda}.
\end{equation}
The interpolation to \emph{CIBER} fluxes affects the calibration at the 10-20\% level and depends on the spectral type of the calibrator. 

There is an additional correction related to field distortions across each detector which modify the per-pixel etendue $(A\Omega)_{pix}$. Using the WCS solutions for each field, we find $d\theta/dx = 6.99\arcsec-7.03\arcsec$, with smooth variation as a function of detector position. This corresponds to a variation in $(A\Omega)_{pix}$ of $<1\%$, which is negligible compared to our total calibration uncertainty.  

\begin{figure}
    \label{fig:abscal_srcs}
    \centering
    \includegraphics[width=0.95\linewidth]{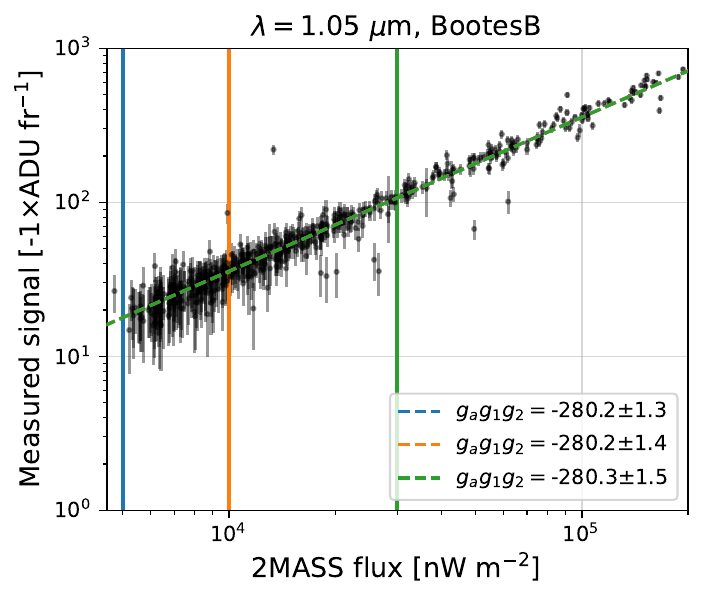}
    \caption{Comparison of \emph{CIBER} measured source fluxes and color-corrected 2MASS fluxes, before outlier rejection is applied. The colored vertical lines indicate the minimum flux thresholds used to obtain the gain estimates in the legend. These indicate the stability of our gain solutions to choice of minimum 2MASS flux. The quantity $g_ag_1g_2$ has units of (nW m$^{-2}$ sr$^{-1}$)/(ADU frame$^{-1}$).}
    \label{fig:pred_vs_measured_flux_sbcal}
\end{figure}

We compare the set of measured \emph{CIBER} signal against color-corrected fluxes in Fig. \ref{fig:pred_vs_measured_flux_sbcal}. We fit a slope to each set of sources and perform iterative sigma clipping to reduce the impact of flux outliers on each gain solution. We obtain consistent gain estimates across science fields and confirm that the gain is insensitive to the chosen minimum flux cut. We then combine the per-field estimates with a weighted mean to obtain the final gain estimates.

To validate our absolute calibration, we convert the \emph{CIBER} maps to surface brightness units, mask bright sources, calculate the mean surface brightness in each map, and correlate these with ZL estimates predicted by a modified Kelsall model that accounts for the solar spectrum and ZL reddening to predict intensity over each \emph{CIBER} bandpass \citep{Crill2020}. The ZL intensity varies by a factor of $\sim2$ across the five \emph{CIBER} fields, which enables this comparison. We compute an additional bandpass correction given a fiducial ZL spectrum; this correction is at the few per cent level. In Table \ref{tab:cal_result_byquad} we report the best-fit slopes from this comparison, using jackknife resampling to estimate uncertainties. For both bands we find scalings consistent with the ZL model to within $\sim 10\%$. We then perform the same exercise using separate fits from the four quadrants of each detector, for which the calibration of each quadrant is estimated independently. The distribution of slopes from separate quadrants similarly suggest consistency of our gain solutions at the $\sim10\%$ level.


\begin{table}
    \centering
    \begin{tabular}{c|c|c|c}
    \emph{CIBER} band & Detector subset & N$_{src}$ & Best fit slope\\
     & & & \\
    \hline
    TM1 (1.05 $\mu$m) & Quadrant A & 735 & 1.04 $\pm$ 0.03\\
        & Quadrant B & 715 & 1.09 $\pm$ 0.03 \\
        & Quadrant C & 668 & 1.13 $\pm$ 0.02\\
        & Quadrant D & 710 & 1.17 $\pm$ 0.03\\
        & \textbf{Full array} & 2828 & \textbf{1.10 $\pm$ 0.02}\\

        \hline 
        TM2 (1.79 $\mu$m) & Quadrant A & 818 & 0.89 $\pm$ 0.05\\
        & Quadrant B & 638 & 0.95 $\pm$ 0.06\\
        & Quadrant C & 730 & 0.90 $\pm$ 0.05\\
        & Quadrant D & 588 & 0.99 $\pm$ 0.06\\
        & \textbf{Full array} & 2774 & \textbf{0.91 $\pm$ 0.05}\\
    \end{tabular}
    \caption{\emph{CIBER} - ZL model surface brightness comparison, estimated in separate quadrants from the full arrays. Slopes consistent with unity indicate correlation between the measured and predicted surface brightness. $N_{src}$ denotes the total number of sources across the five fields in each subset.}
    \label{tab:cal_result_byquad}
\end{table}



\subsection{Flat field correction and image filtering}
\label{sec:ffgrad}
Two more corrections are needed to estimate the \emph{CIBER} power spectrum. The first is correction for relative gain variations across each imager, otherwise known as the FF responsivity. As discussed in Paper I, our ability to reliably measure the power spectra of individual fields comes from our revised treatment of the FF using in situ flight data. We directly estimate and correct for the FF with a stacking estimator, in which each field's mean background (which is dominated by ZL) acts as an approximately uniform source of illumination. This estimator has errors related to instrument noise and sample variance between sky fields that scale inversely with the total number of exposures. In Paper I, we modeled the effects of FF errors on the pseudo-power spectrum and showed that the underlying fluctuation power can be recovered through a noise bias and mode mixing correction. 

The second correction is removal of large scale ZL variations and other foregrounds with an image filtering step. Following \S 7.4 of Paper I, we opt to use a second-order Fourier component model to regress out low-$\ell$ power, which we find effective at removing large-scale contaminants and mitigating sample variance leakage through mode coupling. We adopt this filtering at the cost of the lowest two bandpowers, leading to an effective $\ell_{min}=300$. 

While inspecting the slope fits from both laboratory and flight data, we identified a quadrant-specific electrical effect in the images. The slope fits from separate quadrants exhibit a ``flickering" effect that coherently shifts the ADU values across frames, and is most prominent in one quadrant of the 1.8 $\mu$m detector. This resembles a form of two-state noise, potentially in the common portion of the readout. The observed effect is coherent across pixels in each quadrant and leads to different DC offsets in signal. To avoid spurious anisotropic power, we correct for the effect by adding per-quadrant offsets into the image filtering step through a simultaneous fit. The offsets are in general well constrained by the data. We show in App. \ref{sec:filterchoice} that the per-quadrant treatment improves the fidelity of our read noise model, as we found the effect present in dark integrations as well.

\begin{figure*}
        \centering
\includegraphics[width=0.95\linewidth]{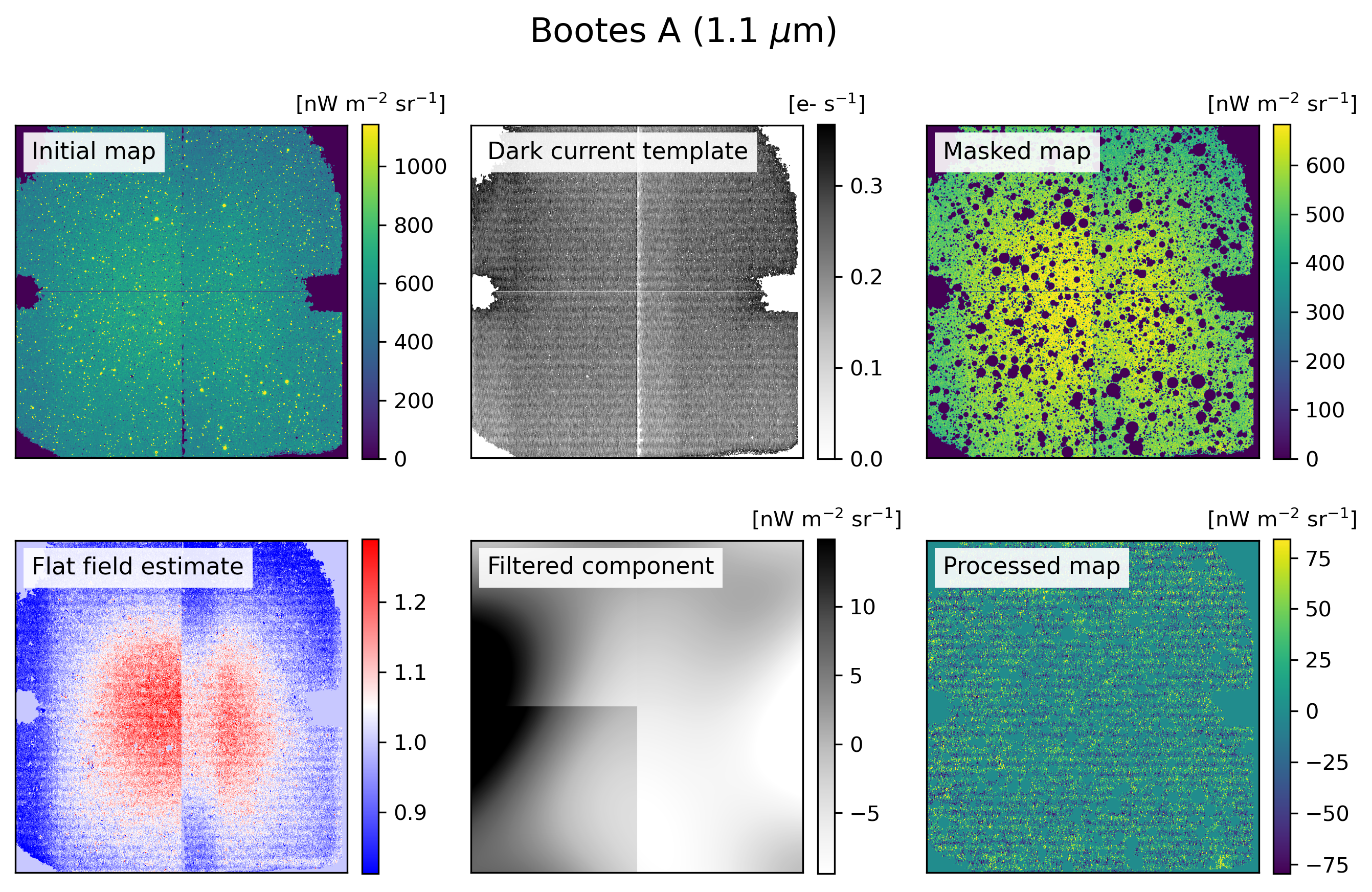} 
    \caption{Example of data processing for \emph{CIBER} 1.1 $\mu$m observations of the \bootes\ B field. The top left panel shows the calibrated image after applying the instrument mask, which removes errant pixels and regions affected by multiplexer glow. We then subtract the dark current template (top middle). The top right panel shows the same field after DC subtraction and application of the astronomical mask. The bottom left panel shows the FF estimated from the other four fields, while the bottom middle panel shows the best-fit filtered component. Finally, the bottom right panel shows the FF-corrected, filtered, mean-subtracted map.}
    \label{fig:ff_estimate_proc}
\end{figure*}

In Figure \ref{fig:ff_estimate_proc} we show an example of our pre-processing for the \bootes\ B field at 1.1 $\mu$m.

%% file: sections/mask.tex
\section{Mask construction}
\label{Sec:mask}

\subsection{Instrument and FF mask}
\label{sec:instmask}
The ``instrument mask" flags pixels that are unresponsive, have unusually high photocurrent, or reside near detector edge effects (e.g., multiplexer glow). Some pixels are identified by computing outliers from dark exposure differences. The instrument mask comprises $\sim 10\%$ and $7\%$ of pixels in the 1.1 $\mu$m and 1.8 $\mu$m imagers, respectively.

An additional mask, unique to each field, is required for pixels with an undefined FF estimate (i.e., all ``off-field" pixels that would contribute to the FF are already masked). This increases the total masking fraction by $\lesssim 0.5\%$, depending on the field. To avoid non-linear effects from large FF errors, we also mask pixels with FF estimates that deviate by $>3\sigma$ relative to the mean local FF value, comprising an $1-2\%$ of pixels in each field.

\subsection{Astronomical mask}
\label{sec:astromask}
Each set of astronomical source masks is defined down to fixed depth in $J$- and $H$-band (with additional cuts on IRAC 3.6 $\mu$m in \S \ref{sec:ciber_spitzer} and \S \ref{sec:super_cl}). We refer the reader to \S 6.4 of Paper I for details of our source masking procedure. In brief, we use a combination of direct NIR photometry from 2MASS along with ancillary PanSTARRS/unWISE photometry to predict NIR magnitudes fainter than 2MASS through random forest regression. Through validation tests and comparisons with directly measured/model-based number counts in Paper I, we determine that we are able to reliably mask over two magnitudes deeper than \zem, which relied on 2MASS photometry alone. The source masking radius is defined as a piecewise function of magnitude and is optimized in Paper I using point source mocks to minimize contributions from the extended PSF.

\subsubsection{Stacking validation on CIBER images}
We perform a final validation of our faint-end source masks by comparing the fluxes predicted by our masking catalogs to direct aperture photometric fluxes from the \emph{CIBER} maps. We consider sources with $J\in[16, 18.5]$ and $H\in[15.5, 18.0]$. While individual sources in this range have low SNR ($\leq 3-5$, with variation across fields due to ZL and read noise), we stack sources in magnitude bins of width $\Delta m = 0.5$ to obtain estimates of the mean flux within each bin. For each field, we omit sources within 50 pixels of the detector edge and any source with a neighbor brighter than $J=20$ or $H=19$. After these cuts we are left in each field with 100-1000 sources for each magnitude bin.

\begin{figure}
    \centering
    \includegraphics[width=0.9\linewidth]{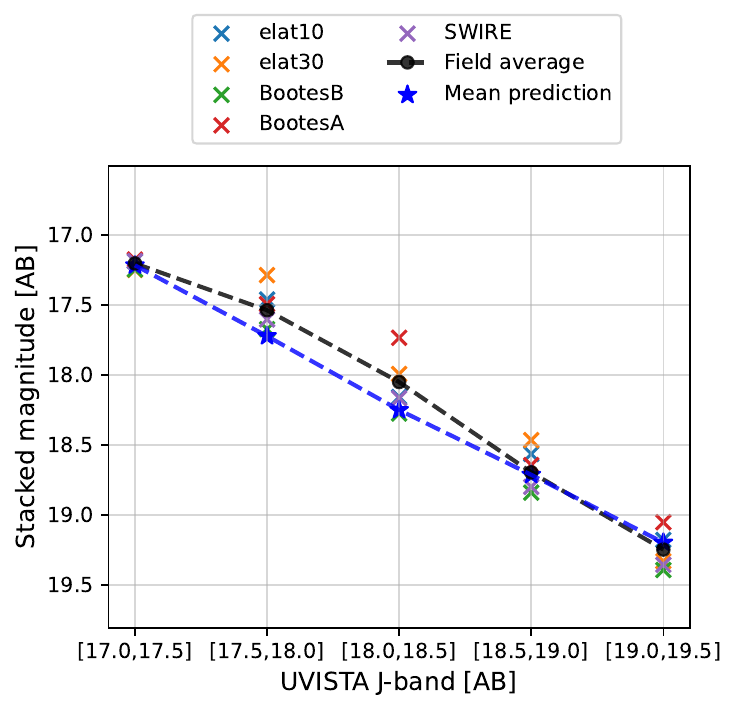}
    \includegraphics[width=0.9\linewidth]{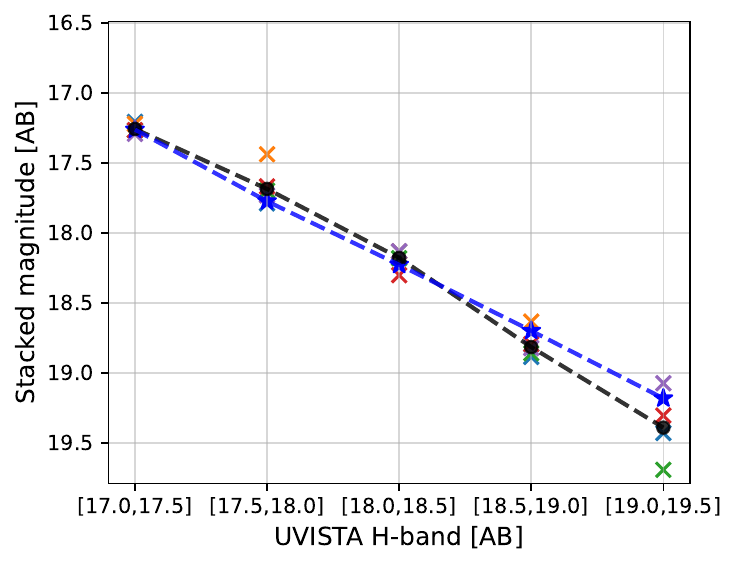}
    \caption{Comparison of mean stacked aperture fluxes (colored crosses) for individual fields in magnitude bins of width $\Delta m=0.5$, field-averaged magnitudes (black) and masking catalog predictions (blue). We find close agreement between measurements and predictions across the range $m_{AB}=17-19.5$, with the exception of $J=17.5-18.5$ in which we measure slightly higher average fluxes.}
    \label{fig:ciber_stack_flux_mask}
\end{figure}

In Figure \ref{fig:ciber_stack_flux_mask} we show the results of this test for both imagers. We perform an additional color correction to relate the observed \emph{CIBER} fluxes at 1.05 $\mu$m and 1.79 $\mu$m to the predictions, which are reported in UVISTA magnitudes (1.25 $\mu$m and 1.65 $\mu$m for $J$ and $H$ band, respectively). The mean fluxes from individual fields are consistent with one another, though elat30 has much noisier estimates due to its shorter integration time. For both bands we find close agreement between measured and predicted fluxes, with the exception of bins $J\in[17.5, 18.0]$ and $J\in[18.0, 18.5]$ where our measured fluxes are $\sim 0.2$ mag brighter.

\subsubsection{Bright stars}
Bright ($J<7$) stars within or near the field of view can increase the light distribution due to scattering in the \emph{CIBER} optical chain. We perform a search for such stars in or near each field of view (FOV). The flux of particularly bright stars is spread over many pixels due to the extended PSF. Within each FOV, we find an average of $\sim 3$ $J$-band sources brighter than \nth{7} magnitude and only one field (elat30) with two sources brighter than \nth{5} magnitude. We confirm through visual inspection that these stars are sufficiently masked, with no clear signs of extended PSF halos surrounding each source mask.

\subsubsection{Additional sources}
Beyond the catalogs described above, we mask galaxy clusters identified in \cite{wen12_clusters} using SDSS photometric data. The number of clusters in each \emph{CIBER} field is small, ranging from 32 to 52. By comparing results with and without the mask on these clusters, we find that they have a negligible contribution to the observed \emph{CIBER} fluctuation power. Lastly, in the observed data there are a handful of low-redshift, extended sources that we identify and mask.

\subsubsection{Final science masks}
After combining all mask components, we find that the science masks of the first four fields (excluding SWIRE) have mean unmasked pixel fractions of 72\%/65\% for cuts of $J<16$ and $H<16$, respectively, which reach 64\%/58\% for $J<18.5$/$H<18.0$ (the most aggressive masking depths probed in this work). As a result of higher source density, the SWIRE masks leave 60\%/53\% and 50\%/43\% of pixels unmasked, for the same masking cuts as above. As discussed in Paper I, the FF error formalism applied in this work relieves the need to perform field differences. As a result we are able to mask bright sources (which typically dominate the astronomical mask fraction) more aggressively and push to deeper masking depths, while avoiding large errors in the mode coupling correction due to prohibitively high masking fractions.  



%% file: sections/noise_model.tex
\section{Noise Model}
\label{Sec:noise_model}

Our noise model is constructed from two components. The first is read noise from the detector and readout electronics, which we estimate from dark exposures just prior to flight. The second is photon noise. We use the derived models to simulate noise in our mock observations (see Paper I) and to estimate noise biases and uncertainties on observed auto- and cross-power spectra.

\subsection{Noise model construction}
\label{sec:rdnoise_modl}

To construct our read noise model, we use a series of twelve exposures taken while the rocket was on the launcher shortly before flight and follow a similar procedure to that employed in \texttt{Z14}. These exposures most closely match the electrical environment during flight, and were taken with the shutter closed, so that read noise is the dominant contribution. To obtain a read noise model for each science field, we perform time-stream filtering and slope fits on the exposures matching the number of frames in each science field integration. We then compute the difference between dark exposure pairs and compute the two-dimensional power spectrum. In these differences, any ambient non-zero signals should cancel out, while the noise power should double.

\begin{figure*}
    \centering
    \includegraphics[width=0.9\linewidth]{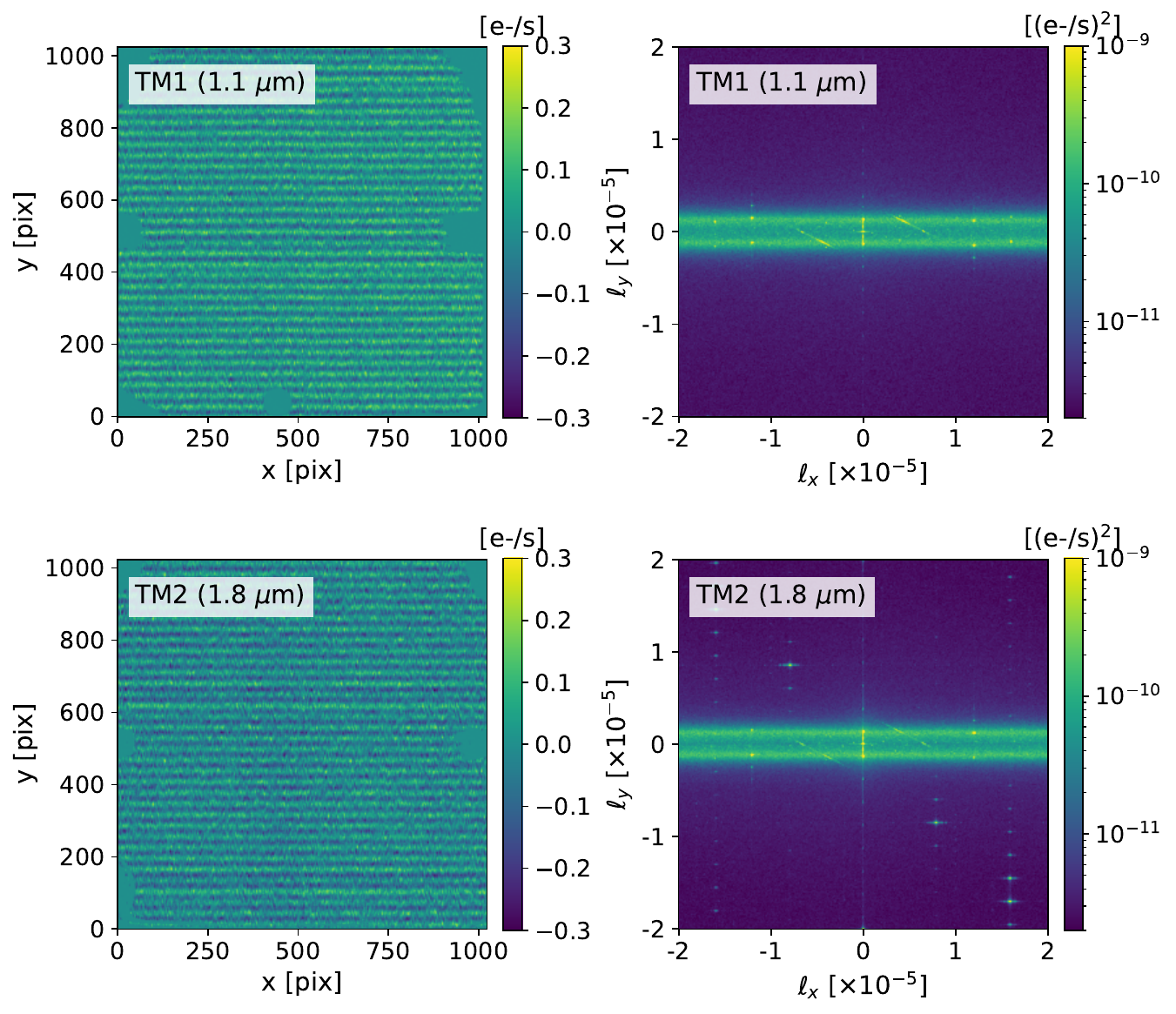}
    \caption{Example dark differences (left column) and two-dimensional power spectra averaged over exposure pairs (right column). These match the integration time of science field elat10. Read noise is highly anisotropic in Fourier space, which we take advantage of by applying inverse-variance weighting of the measured two-dimensional power spectrum before computing azimuthally averaged bandpowers.}
    \label{fig:read_noise_model_2d}
\end{figure*}

In Figure \ref{fig:read_noise_model_2d} we show pairwise dark exposure differences and the read noise models for the elat10 field for both \emph{CIBER} imagers. Due to the orientation of the readout amplifiers, there is anisotropic power in the two-dimensional power spectrum along $\ell_y$, peaking at specific ($\ell_x,\ell_y$) modes. This motivates Fourier weighting of individual $(\ell_x,\ell_y)$ modes, calculated after masking and before computing azimuthally-averaged bandpowers.

We estimate the photon noise component of each \emph{CIBER} field by converting the images to units of photocurrent, computing the mean in unmasked pixels and relating this to the photon noise RMS, which is derived in \cite{garnett_forrest},
\begin{equation}
    \sigma_{\gamma}^2 = \frac{6}{5}\frac{i_{phot}}{T_{int}}\left(\frac{N^2+1}{N^2-1}\right),
\end{equation}
where $i_{phot}$ is the photocurrent, $T_{int}$ is the integration time and $N$ is the number of frames in the integration.

\subsection{Validation tests}
\label{sec:noisemodl_validation}
\subsubsection{Read noise model consistency with dark data}
For each field integration, we generate a set of 500 Gaussian noise realizations drawn from the read noise model and calculate their weighted 1D power spectra. We then compare these against the power spectra of the initial dark exposure differences. We show the results of this test in Fig. \ref{fig:darkdiff_read_TM1_diff_filter} of App. \ref{sec:filterchoice}. Our model employing per-quadrant offset fitting and full-array gradient filtering closely reproduces the power spectra of the dark data, including sharp, anisotropic features such as the peak near $\ell \sim 6200$. While the paucity of dark exposures taken on battery power limits our ability to precisely compare the variance of the read noise power spectra on large scales, the simulated read noise realizations yield a dispersion consistent with the data to within sample variance. 




\subsubsection{Consistency with flight data}

Due to the non-destructive readout of the Hawaii detector electronics, we are able to construct flight half-exposures from subsets of the readout frames. For each field, we compute the angular power spectrum of the flight half-exposure difference and compare this with differences generated from a noise model defined at the half-exposure integration time. For both the flight data and the noise model realizations, we apply the full masks (which are detailed in \S \ref{Sec:mask}) to remove Poisson fluctuations from point sources in the observed data. 

We use the publicly available tool \texttt{astrometry.net} \citep{lang10} to compute separate astrometry for each set of flight half-exposures and find differences as large as 10\arcsec\ between pointing solutions. We use our set of IGL+ISL mocks to estimate the level of leakage power due to drift and then subtract this contribution from the power spectrum of each flight difference. We estimate the magnitude of correction from drift to be at most 10\% on scales $\ell > 10^4$ after masking.

\begin{figure*}
    \centering
    \includegraphics[width=0.8\linewidth]{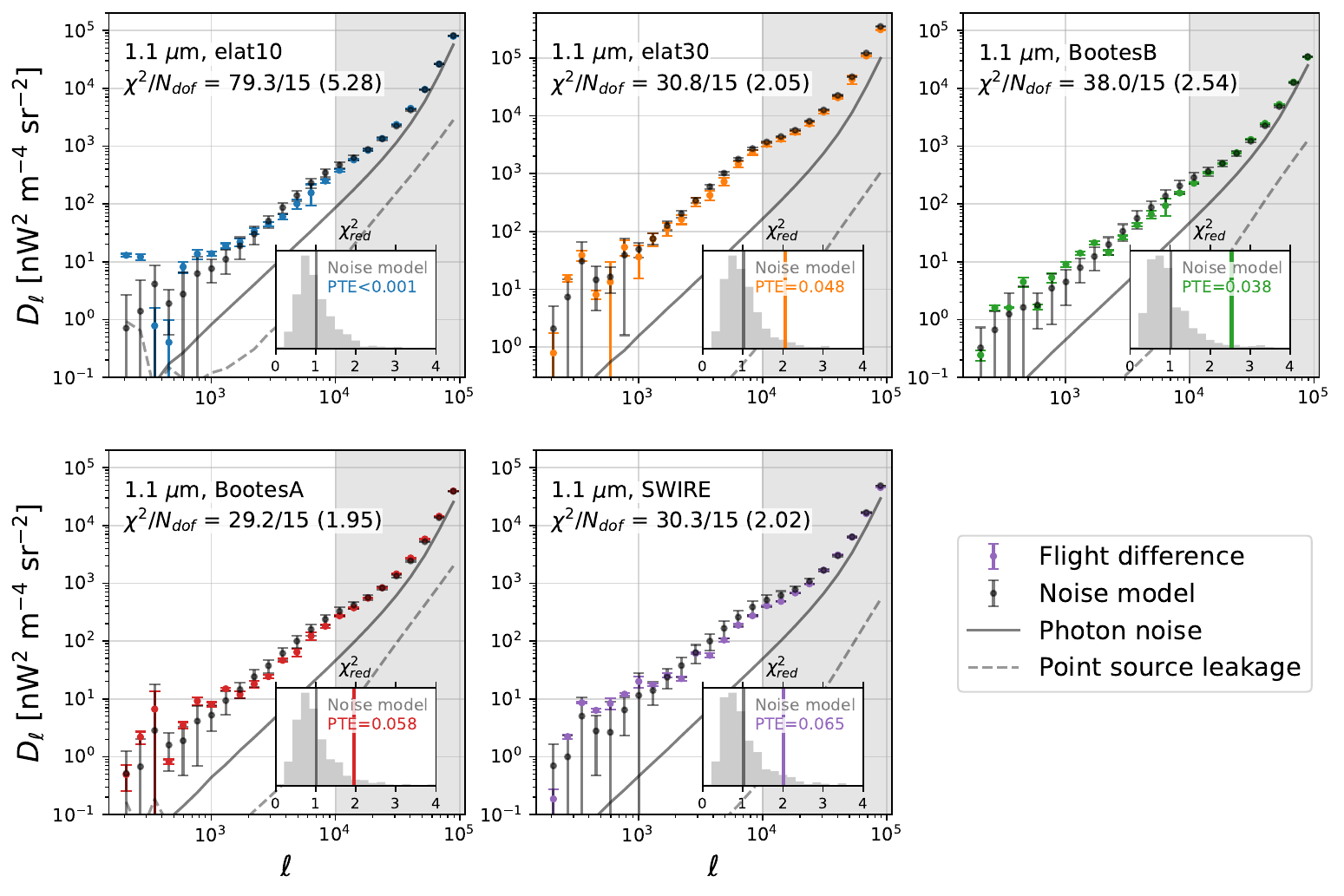}
    \includegraphics[width=0.8\linewidth]{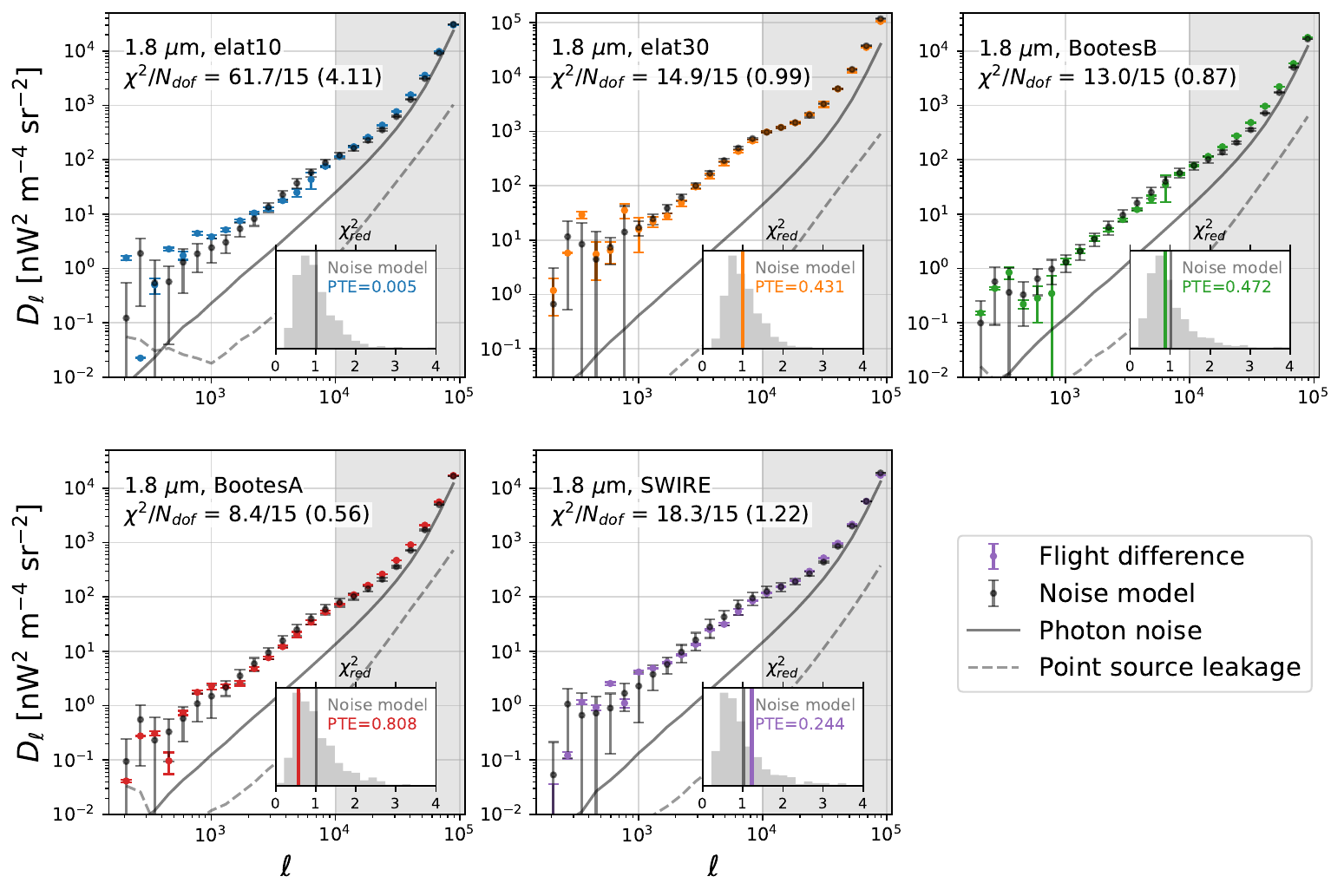} 
    \caption{Noise model validation for 1.1 $\mu$m (top) and 1.8 $\mu$m using flight half-exposure differences. We show the power spectra of the flight differences (colored points) and those from 1000 simulated noise differences (black points). The grey solid lines indicate the photon noise contribution to the difference spectra, while the dashed lines show the point source leakage due to pointing drift. All power spectra have been corrected for mode coupling from masking and filtering. The shaded region indicates scales above multipole $\ell > 10000$ that we do not include in the $\chi^2$ calculation. In each inset we plot the reduced $\chi^2$ distribution of noise realizations relative to the mean of the noise model, along with that of the observed data.}
    \label{fig:nv_halfexp_tm1}

\end{figure*}

We show the results of this consistency test in Figure \ref{fig:nv_halfexp_tm1}, comparing the mean and dispersion of simulated difference spectra (black) to those from flight (colored points). We apply mode-mixing and beam corrections to each set of power spectra to place the results in sky units, and find the following:
\begin{itemize}
\item On small ($\ell > 20000$) scales, the difference spectra (aside from elat30) are photon noise-dominated and are consistent with our noise model to within 10-15\% for 1.1 $\mu$m and 20\% for 1.8 $\mu$m. 
\item On intermediate scales ($2000 < \ell < 20000$), the difference spectra are dominated by read noise. For both bands, our noise model over-predicts power relative to the flight differences, by up to 25\% in some bandpowers. We find such differences to be larger for the 1.1 $\mu$m array than for 1.8 $\mu$m. One possibility is that pickup noise during flight was lower than during ground measurements. Another potential explanation is that non-stationarity in the read noise leads to discrepancies when assuming a stationary noise model. 

\item 
On large scales ($\ell < 2000$), almost all fields have flight differences that are consistent with our simulated noise differences. The one exception is elat10, in which the lowest two bandpowers of the flight differences have higher power than those from other fields and from the elat10 noise models.    
\end{itemize}



We quantify noise model consistency for the five fields by computing a $\chi^2$ statistic between our noise model and the flight difference data on scales $\ell<10^4$. To account for bandpower correlations we estimate a covariance matrix for each field derived from the ensemble of 1000 weighted noise realizations. Using these covariance matrices we calculate 1) $\chi^2$ of the flight difference spectra relative to their noise models; and 2) the distribution of $\chi^2$ values for each of the 1000 noise realizations for each field. These are both shown in the insets of Fig. \ref{fig:nv_halfexp_tm1}. We then compute probability-to-exceed (PTE) values for each field using the empirical CDF of the flight $\chi^2$ relative to the distribution from noise model realizations. With the exception of the elat10 field, the PTE values obtained for 1.8 $\mu$m support the null hypothesis that the flight data have noise statistics consistent with our noise model. Our 1.1 $\mu$m results have lower PTEs, due largely to overestimation of the noise model on read noise-dominated scales. Since this test quantifies noise model errors at half-exposure integration length, residual errors are expected to be smaller for the full exposure integrations, as read noise power decreases $\propto 1/T^3$ and photon noise $\propto 1/T$ \citep{bock13}. This behavior is observed in the flight data, for example by comparing the half-exposure difference spectra of elat30 with those of the other four fields.

\subsubsection{Cross-correlation of half-exposure differences}
\begin{figure}
    \centering
    \includegraphics[width=0.9\linewidth]{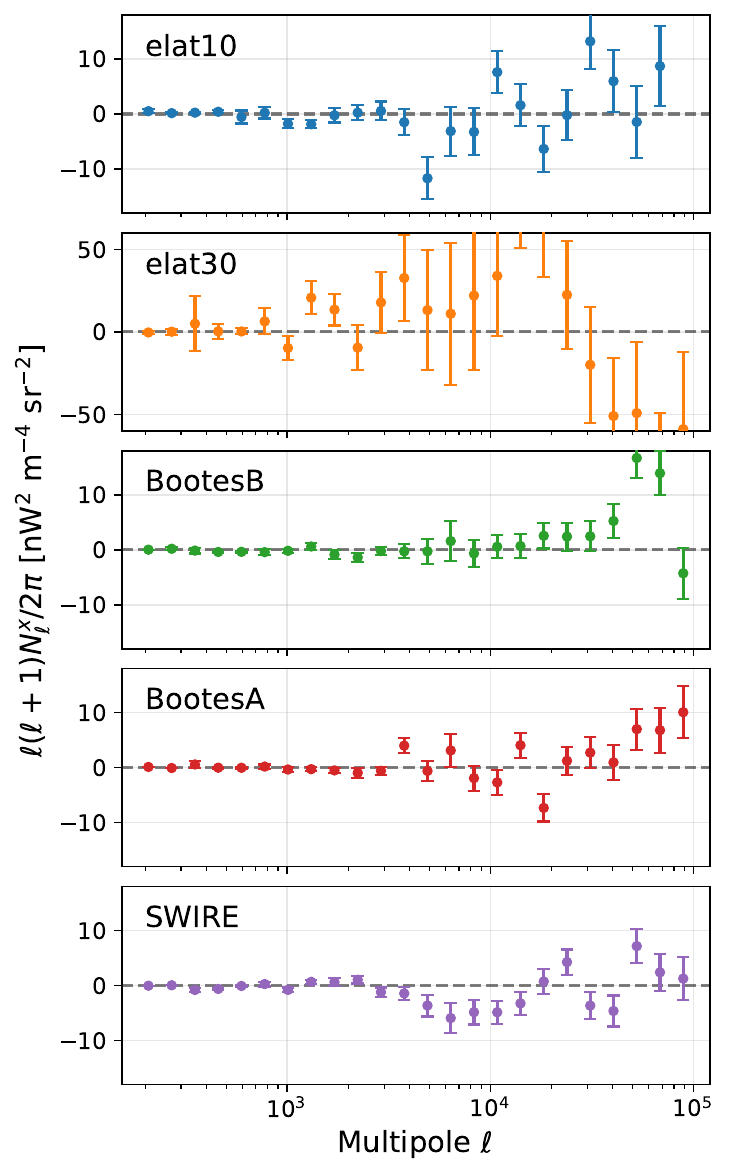}
    \caption{Cross-power spectra of \emph{CIBER} 1.1 $\mu$m and 1.8 $\mu$m exposure half-differences. These constrain the presence of any time-variable signals (over the course of one exposure) seen in common across imagers. We find that all cross-power spectra are consistent with zero.}
    \label{fig:cross_nl}
\end{figure}

Time-variable signals during the \emph{CIBER} observations that are correlated across wavelengths may leave a signal in the cross-power spectrum of half-exposure differences. We check this for 1.1 $\mu$m and 1.8 $\mu$m and show the resulting cross-power spectra in Fig. \ref{fig:cross_nl}. For all fields the cross-spectra are consistent with zero. This further supports the conclusion that time-variable phenomena, such as fast-decay airglow, are at a negligible level for the \emph{CIBER} science fields.

%% file: sections/ps_estimation.tex
\section{Power spectrum estimation}
\label{Sec:ps_estimation}

\subsection{Pseudo-$C_{\ell}$ formalism}
We relate an underlying sky power spectrum to bandpower estimates of the observed pseudo power spectrum. (see \S 4 and 5 from Paper I). The sky signal with power spectrum $C_{\ell}^{sky}$ is related to the observed pseudo power spectrum $C_{\ell}^{obs}$ as
\begin{equation}
    C_{\ell}^{obs} = \sum_{\ell'}M_{\ell\ell^{\prime}}(B_{\ell'}^2 C_{\ell'}^{sky} + N_{\ell'}).
\end{equation}
Here, $M_{\ell\ell'}$ denotes the mode mixing matrix, $B_{\ell}$ is the beam transfer function and $N_{\ell}$ is the noise bias. One can determine $C_{\ell}^{sky}$ by inverting the equation above, i.e.,
\begin{equation}
    \hat{C}_{\ell}^{sky} = B_{\ell}^{-2}\sum_{\ell'}M_{\ell\ell'}^{-1}(C_{\ell'}^{obs} - N_{\ell'}).
\end{equation}

\begin{figure}
    \centering
    \includegraphics[width=\linewidth]{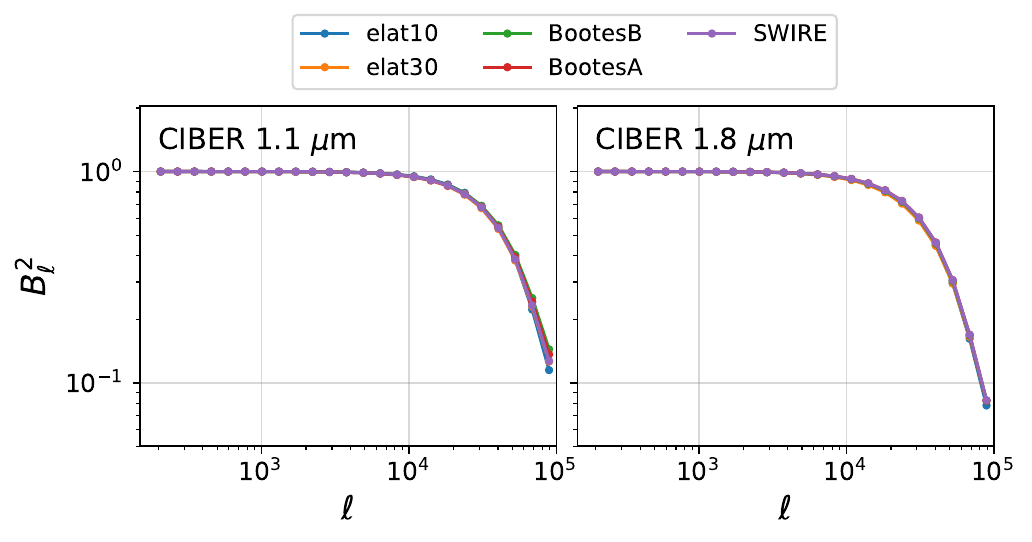}
    \caption{\emph{CIBER} pixel-convolved beam functions for both imagers. These are derived from the best-fit PSF models obtained from stacking in \cite{chengihl}.}
    \label{fig:b_ell_estimates}
\end{figure}

The following steps are required to estimate the underlying sky power spectrum for each field:
\begin{enumerate}
    \item Estimate the noise bias (and Fourier noise weights) using an ensemble of read+photon noise realizations. We include realizations of the FF noise bias, which comes from instrument noise in the stacked FF estimates coupled to the mean sky brightnesses of the target fields. The noise bias takes the form
    \begin{equation}
    N_{\ell} =\sum_{\ell'}M_{\ell\ell'}(N_{\ell'}^{read}+N_{\ell'}^{\gamma} + N_{\ell'}^{\delta FF}).
    \end{equation}
In practice, we calculate the mean and variance of masked 2D noise spectra from a set of 500 Monte Carlo realizations. We use a modification of Welford's online algorithm \citep{welford62}, which alleviates the memory requirements associated with storing 2D power spectra from a large number of realizations. We then calculate the inverse variance of the 2D power spectra that defines our Fourier weights $w(\ell_x, \ell_y)$.

    \item Compute the masked 2D pseudo-power spectrum of the data and apply the derived Fourier noise weights to obtain bandpower estimates,
    \begin{equation}
    C_{\ell}^{obs} = \frac{\sum_{(\ell_x, \ell_y)} w(\ell_x, \ell_y)M(\ell_x, \ell_y) }{\sum_{(\ell_x, \ell_y)} w(\ell_x, \ell_y)},
\end{equation}
where $M(\ell_x,\ell_y)$ denotes the two-dimensional observed power spectrum.
\item Apply noise de-biasing to $C_{\ell}^{obs}$ using the mean, Fourier-weighted noise power spectrum $N_{\ell}$.

\item Compute mixing matrices $M_{\ell\ell^{\prime}}$ and apply $M_{\ell\ell'}^{-1}$ to the noise de-biased power spectrum. As described in Paper I, $M_{\ell\ell^{\prime}}$ accounts for mode coupling induced by the masks, image filtering and FF errors. We use 500 Monte Carlo realizations per bandpower to estimate each set of mixing matrices (see Appendix A.2.3 of Paper I for more details).

\item Correct for the beam transfer function. For each field, we estimate $B_{\ell}$ by taking the best-fit PSF, generating a set of 100 sub-pixel shifted PSFs and computing the ensemble-averaged power spectrum, which is normalized to unity on large scales. We show these $B_{\ell}$ estimates in Fig. \ref{fig:b_ell_estimates}.
\end{enumerate}

\subsection{Mock power spectrum recovery}
\label{sec:mock_ps_recov}
In Paper I we detail the mock data constructed for this analysis. The mocks include random realizations of IGL, ISL and ZL unique to each \emph{CIBER} field\footnote{For the mocks used to produce the results in this work, we re-scale each Kelsall ZL realization to match the observed mean surface brightness in order to simulate representative photon noise levels (see \S \ref{sec:abscal})} and EBL fluctuations that match the observed power spectrum from \zem. The synthetic observations include simulated read noise, photon noise and a realistic flat field measured using laboratory data  that modulates the per-pixel responsivity across each array. We validate our ability to recover unbiased estimates of sky fluctuation power by testing our pseudo-$C_{\ell}$ pipeline on one thousand sets of mocks, down to masking depths of $J=18.5$ and $H=18.0$, and find minimal biases at the level of the \emph{CIBER} \nth{4} flight power spectrum sensitivity. We use the dispersion of mock recovered power spectra to define per-bandpower weights that we use to combine estimates from the five \emph{CIBER} fields and to quantify statistical measurement uncertainties for the observed power spectra. The latter application is particularly important at low-$\ell$, where estimating uncertainties from the observed set of modes in each bandpower suffers from large sample variance. Lastly, we use the mocks to estimate power spectrum covariances that we apply in \S \ref{sec:auto_field_consistency} to assess internal consistency across the five \emph{CIBER} fields.

\subsection{Cross-power spectra}

Cross-power spectra isolate the common signal between imagers while mitigating the effect of uncorrelated noise biases. For maps $A$ and $B$, the cross-power spectrum variance is given by
\begin{equation}
    (\delta C_{\ell}^{A\times B})^2 = \frac{1}{n_{\ell}}\left[(C_{\ell}^A+N_{\ell}^A)(C_{\ell}^B+N_{\ell}^B)+(C_{\ell}^{A\times B})^2\right],
    \label{eq:cross_knox}
\end{equation}
where $n_{\ell}$ is the effective number of modes in the field of view. For the cross-power spectra, we estimate the noise contributions ($N_{\ell}^A(C_{\ell}^B+N_{\ell}^B)$ and $N_{\ell}^B(C_{\ell}^A+N_{\ell}^A)$) by generating noise realizations from one noise model, computing the cross power spectra against the other map, and then calculating the dispersion of noise spectra.

%% file: sections/ciber_results.tex
\section{CIBER auto-power spectrum results}
\label{sec:ciber_results}

In this section we present the \emph{CIBER} auto-power spectrum measurements and compare against predictions from IGL, ISL and DGL. The DGL predictions are derived in \S \ref{sec:ciber_iris_xcorr}, while the IGL and ISL predictions are detailed in \S \ref{sec:astro_modl}. 

\subsection{Power spectrum measurements}

\begin{figure*}
    \centering
    \includegraphics[width=\linewidth]{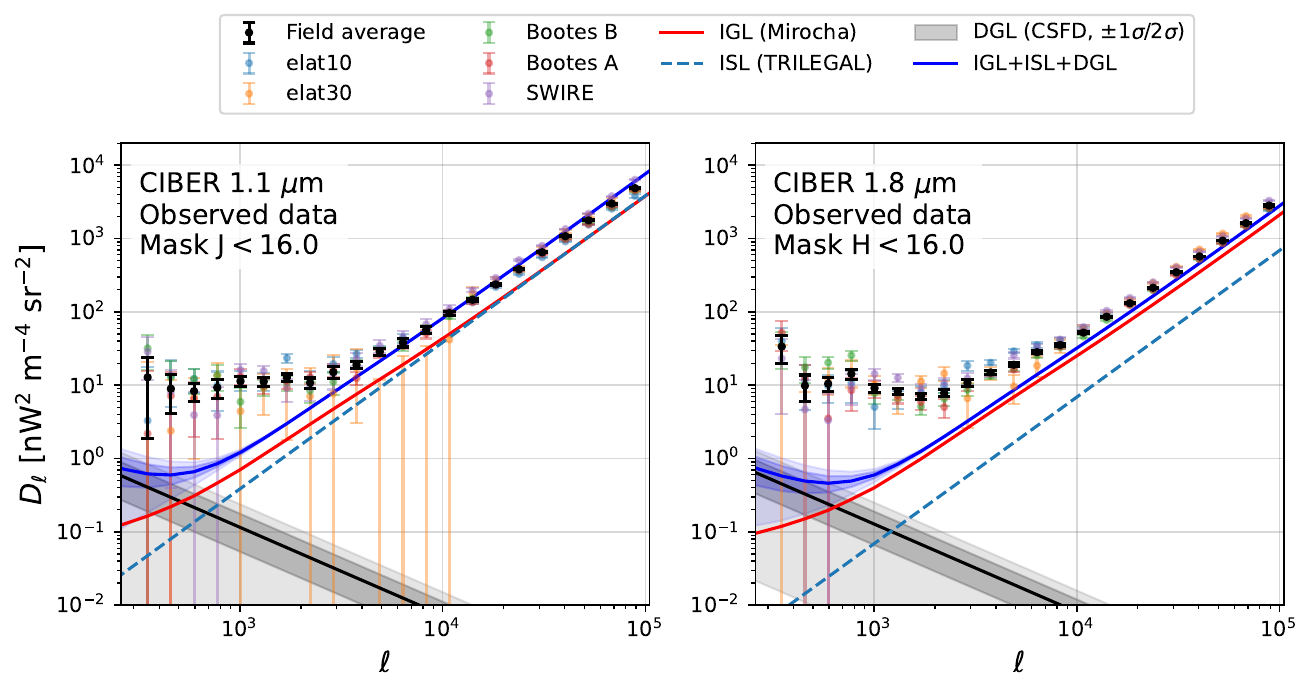}
    \caption{\emph{CIBER} auto-power spectrum measurements for 1.1 $\mu$m (left) and 1.8 $\mu$m (right), with $D_{\ell}=\ell(\ell+1)C_{\ell}/2\pi$. The per-field measurements are shown with colored points while the weighted field averages are plotted in black. Also shown is the estimated DGL contribution (black solid lines) with shaded regions spanning 1$\sigma$ and 2$\sigma$ uncertainties (see \S \ref{sec:ciber_iris_xcorr}). The dashed blue and solid red curves show the predicted ISL and IGL angular power spectra, described in \S \ref{sec:astro_modl}. The measurements show large-angle fluctuation power exceeding that from our IGL+ISL+DGL model (dark blue curves). We do not include model-based uncertainties for IGL or ISL, however these are discussed in \S \ref{sec:modl_interp}. Angular multipoles $100<\ell<300$ are technically accessible within the \emph{CIBER} FOV, but suppressed by aggressive image filtering.}
    \label{fig:ciber_auto}
\end{figure*}

Figure \ref{fig:ciber_auto} shows the per-field \emph{CIBER} auto-power spectra at 1.1 $\mu$m and 1.8 $\mu$m, along with each weighted field average. Notably, we find a clear detection of fluctuation power in both \emph{CIBER} bands on scales $\ell < 5000$ that exceeds the predicted Poisson fluctuations from IGL and ISL. For scales $\theta > 5'$, we measure non-zero fluctuation power at $14.2\sigma$ and $18.1\sigma$ for 1.1 $\mu$m and 1.8 $\mu$m, respectively. Our DGL estimates come from cross-correlations of the \emph{CIBER} maps with the ``corrected SFD" extinction maps. The \emph{CIBER} auto-power spectra show increasing departures at large angular scales, with observed fluctuations on scales $\ell < 1000$ that are an order of magnitude higher than our combined model predictions. After folding in DGL uncertainties, the 1.1 $\mu$m and 1.8 $\mu$m auto-power spectra are inconsistent with the baseline IGL+ISL+DGL model at 11.9$\sigma$ and 16.3$\sigma$, respectively. This neglects uncertainties associated with the IGL and ISL models, however these are discussed in \S \ref{sec:modl_interp}.

We find consistent power spectrum estimates across the five \emph{CIBER} fields, which we assess statistically in \S \ref{sec:auto_field_consistency}. The elat30 field (orange) has larger errors due to the shorter integration used (10 frames, compared to 24-28 frames for the other four fields). For 1.1 $\mu$m, we find that the SWIRE field (purple) has a slightly higher Poisson level than the other fields, which we attribute to the field's higher stellar density and corresponding ISL fluctuations. We find further evidence for this in the \emph{CIBER} cross-power spectra (see \S \ref{sec:cross_spectra}).


\subsection{Data consistency tests}

\subsubsection{Field-field consistency}
\label{sec:auto_field_consistency}
\begin{figure*}
    \centering
    \includegraphics[width=0.8\linewidth]{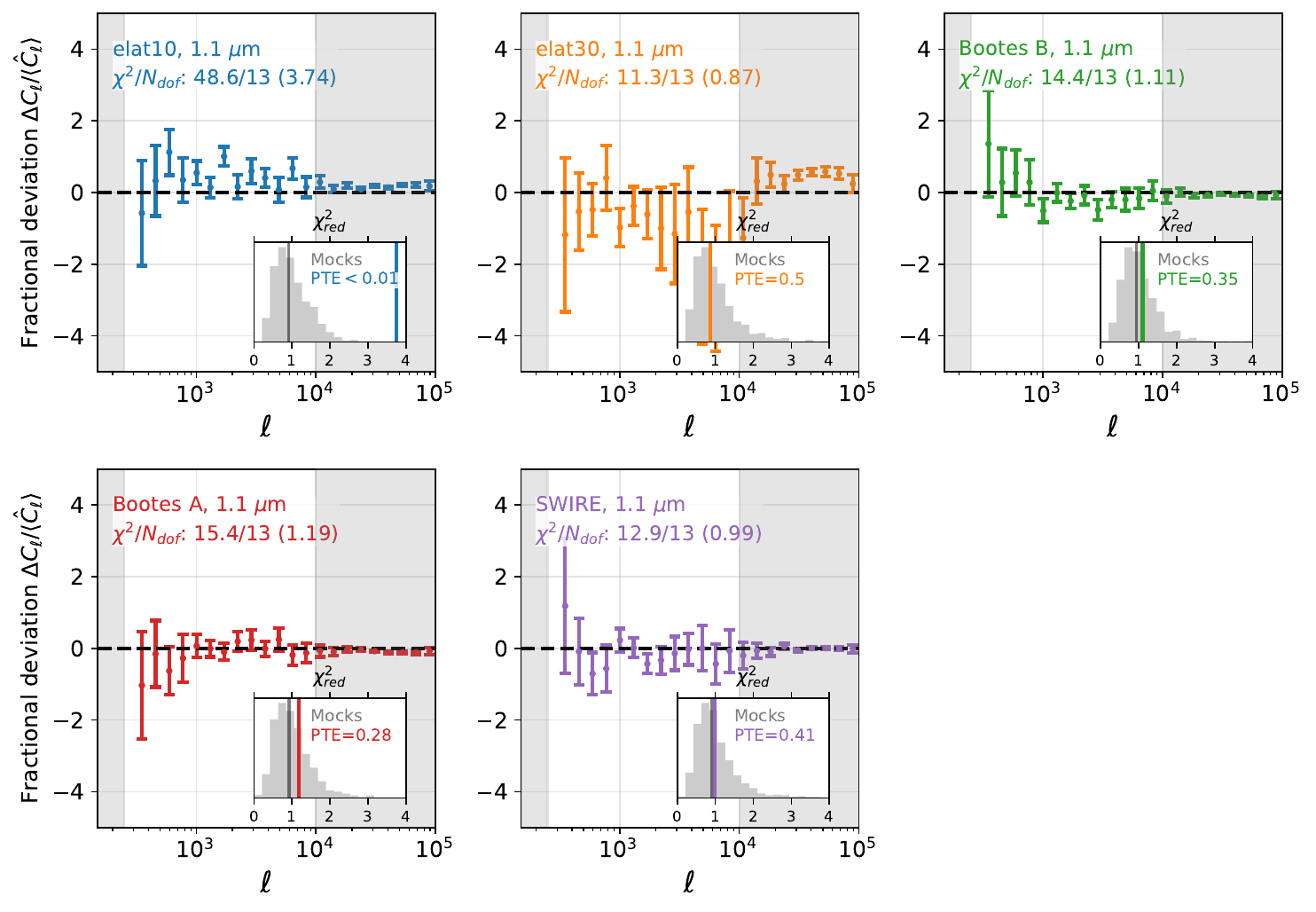}
    \caption{Fractional deviation of the individual field \emph{CIBER} 1.1 $\mu$m power spectra relative to the weighted field average. We test consistency of the per-field auto-power spectra on scales $\ell < 10000$ by computing a $\chi^2$ statistic between each field and the average, using bandpower covariances derived from an ensemble of 1000 mock recovered power spectra. In the insets, we plot the distribution of $\chi^2$ values from the mock ensemble recovered power spectra and compare against the $\chi^2$ of the observed data (colored vertical lines) to determine a probability-to-exceed (PTE) value for each field.}
    \label{fig:ciber_auto_field_consistency_TM1}
\end{figure*}

\begin{figure*}
    \centering
    \includegraphics[width=0.8\linewidth]{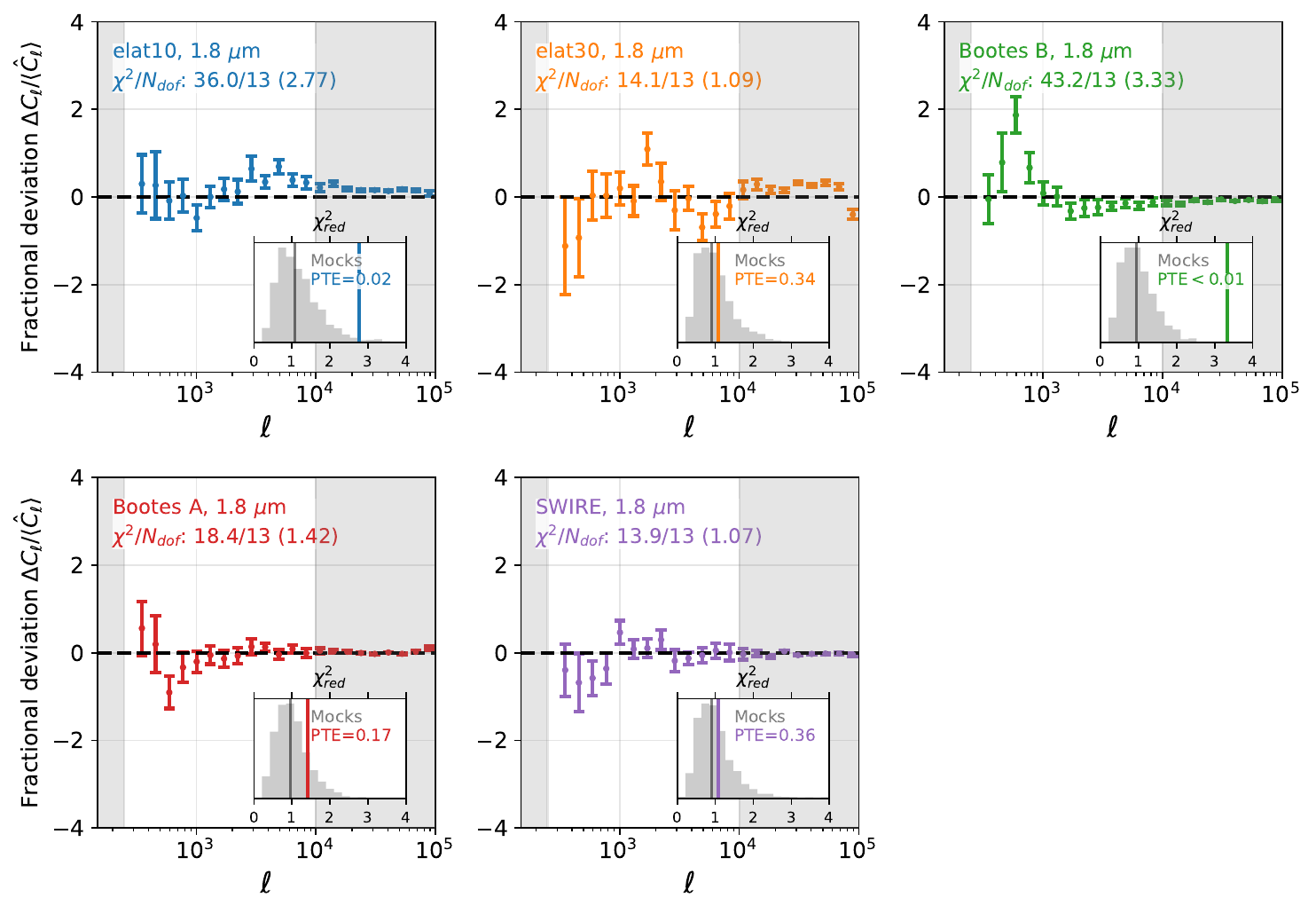}
    \caption{Same as Fig. \ref{fig:ciber_auto_field_consistency_TM1} but for 1.8 $\mu$m.}
    \label{fig:ciber_auto_field_consistency_TM2}
\end{figure*}

In Figures \ref{fig:ciber_auto_field_consistency_TM1} and \ref{fig:ciber_auto_field_consistency_TM2} we plot the deviations of the per-field 1.1 $\mu$m and 1.8 $\mu$m auto-power spectra from their weighted field averages. We evaluate field-field consistency using slightly deeper masks ($J<17.5$ and $H<17.0$) in order to minimize differences in ISL that are seen at our fiducial masking depths. On small scales ($\ell > 10000$), the dispersion across fields is roughly $\pm 15\%$ and $\pm 10\%$ for 1.1 $\mu$m and 1.8 $\mu$m respectively, with the exception of elat30 which has slightly larger departures. The noise bias for elat30 is $\sim 10\times$ the amplitude of the sky signal on intermediate to small scales due to its shorter integration time, meaning the recovered power spectra for this field are most sensitive to noise de-biasing errors. The observed field consistency confirms that variations in Galactic stellar foregrounds are small at this masking depth.

For bandpowers $\ell < 10000$, we calculate a $\chi^2$ statistic for each field's recovered power spectrum relative to the weighted field average and assuming Gaussian uncertainties. For field $i$,
\begin{equation}
    \chi^2_i = (C_{\ell,i}^{obs} - C_{\ell, av}^{obs})^T\hat{\mathcal{C}}_{i}^{-1}(C_{\ell,i}^{obs} - C_{\ell, av}^{obs}).
\end{equation}
The covariance $\hat{\mathcal{C}}_{i}$ is estimated using the mock recovered power spectra described in \S \ref{sec:mock_ps_recov}, for which we compute the same statistic across realizations $j$,
\begin{equation}
    \hat{\mathcal{C}}_{i} = \langle (C_{\ell, i}^j - C_{\ell, av}^j)^2\rangle_j.
\end{equation}
For the low-$\ell$ bandpowers that are dominated by sample variance ($\ell < 1000$), we rescale the covariance to reduce errors from  mismatch between the \zem\ power spectrum amplitude assumed in the mocks and that of the observed data. For 1.1 $\mu$m, the reduced $\chi^2$ of our fields range between 0.9 (elat30) and 3.7 (elat10), while for 1.8 $\mu$m they span 1.1 (SWIRE) to 3.3 (\bootes\ B). 

We then obtain a probability-to-exceed (PTE) statistic for each field by performing the same $\chi^2$ calculation for all sets of mocks using them to compute the rank statistic of the observed data. We find that the mean $\chi^2_{red}$ of the mocks is $\sim 0.9$ (shown in each panel's inset by the grey vertical lines). This is explained by our use of the block-diagonal components $\mathcal{C}_{i}$ to compute per-field PTEs, which neglects correlations across fields that are introduced through our FF stacking estimator (see \S 7.3 of Paper I). Our use of the rank statistic mitigates the impact of the cross-field covariance on our results, as the same $\chi^2$ calculation is applied to the mock realizations as the observed data.

In both bands, we recover higher fluctuation power in elat10 than the other fields, most prominently for angular multipoles $1000<\ell<10000$. This field has both the highest ZL intensity and, as a result, the largest FF errors, which may explain the modest disagreement. The large $\chi^2_{red}$ and low PTE of the \bootes\ B field at 1.8 $\mu$m is driven by an outlier in the third lowest bandpower ($516 < \ell < 672$). With these exceptions, the CIBER fields have reasonably uniform PTEs and a dispersion consistent with that expected from realizations of a single underlying sky power spectrum.

\subsubsection{Variation with source masking depth}

\begin{figure*}
\centering
\includegraphics[width=0.49\linewidth]{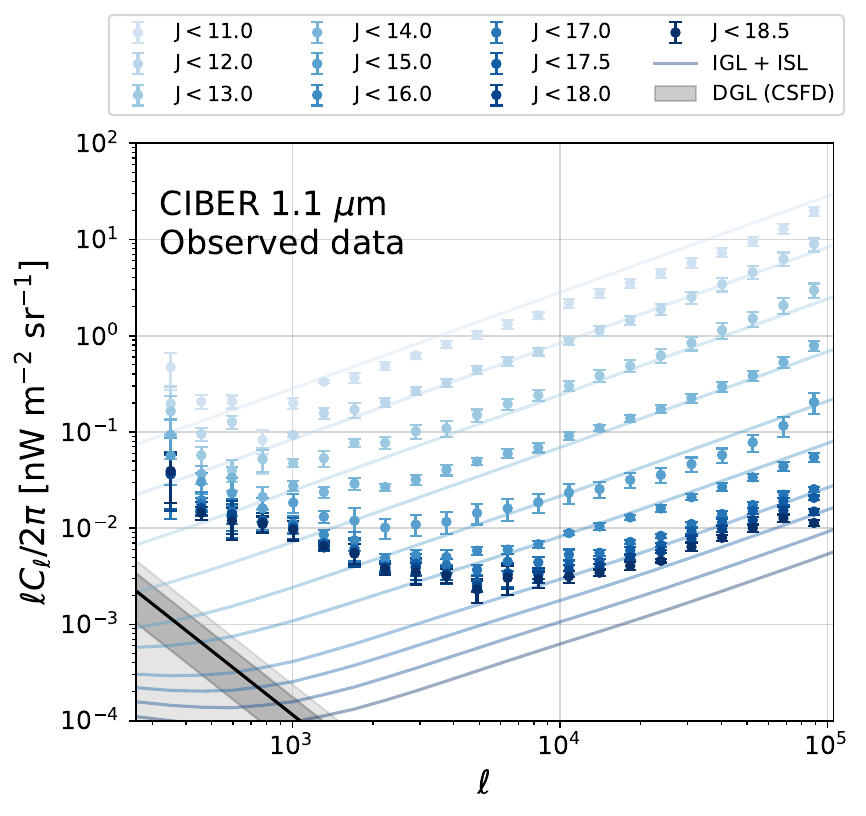}
\includegraphics[width=0.49\linewidth]{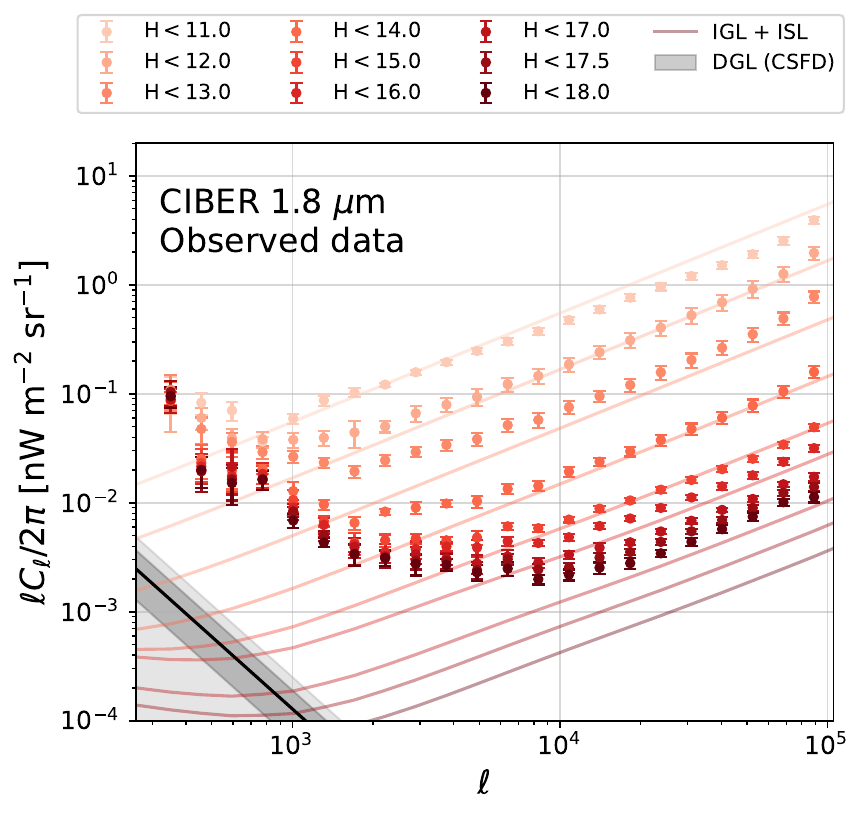}
\caption{\emph{CIBER} 1.1 $\mu$m (left) and 1.8 $\mu$m (right) power spectra at varying source masking depths, along with IGL+ISL model predictions (colored curves, see \S \ref{sec:astro_modl}). DGL constraints are shown in black with $\pm1\sigma/2\sigma$ uncertainties (see \S \ref{sec:ciber_iris_xcorr}). Here we plot $\ell C_{\ell}/2\pi$ rather than $D_{\ell}$ in order to reduce the dynamic range between cases. On large angular scales ($\ell < 2000$), the fluctuation power varies for the shallowest masking depths but is insensitive to source masking beyond $J=16$ and $H=16$.} 
\label{fig:auto_ps_vs_mask_depth}
\end{figure*}

We compute auto-power spectra at a range of masking depths and compare against IGL+ISL predictions in Fig. \ref{fig:auto_ps_vs_mask_depth}. For the brightest sources, non-linearity in the detector response suppresses the measured flux of bright pixels. Non-linearity suppresses the flux by 10\% for an integrated charge of 7.5$\times 10^4$ electrons, which corresponds to sources with Vega magnitudes $J,H=10.5-11.5$ depending on the exact sub-pixel source position. The power spectra in both bands decrease monotonically as a function of masking depth, however we find that $\ell < 2000$ fluctuations remain unaffected beyond $J=16$ and $H=15$. The observed small-scale power matches Poisson predictions down to these depths as well, however for deeper masking thresholds the observed spectra show progressively larger fractional departures from predictions. 

In \S 6.3 of Paper I, we validated our masking procedure down to $J=18.5$ and $H=18.0$ using the COSMOS 2015 catalog as a test set and determined that masking errors have at most a $\sim 15\%$ effect on the recovered Poisson signal relative to ground truth. We also demonstrated that the number counts of our masking catalogs match those predicted from mocks and deeper UKIDSS catalogs (where available). In Figure \ref{fig:ciber_stack_flux_mask} of this work, we find that the predicted catalog fluxes agree closely with direct stacked aperture photometry on the calibrated \emph{CIBER} maps at the locations of masking catalog sources. Nonetheless, the recovered small-scale power is considerably higher than predictions at our deepest masking cuts. 

In Appendix \ref{sec:halfexp_cross} we perform half-exposure cross-correlations from each field integration as an alternate estimator of the intensity auto-power spectrum. We find that these estimates are consistent with our fiducial results, ruling out explanations of the small-scale power that invoke transient phenomena (e.g., cosmic rays) that are uncorrelated in time. Furthermore, by measuring the fractional response of the cross-power spectrum to integer pixel shifts between the first and second half-exposures, we find the observed de-correlation response is consistent with that estimated from mocks containing Poisson fluctuations from stars and galaxies. These results provide strong evidence for the observed small-scale power originating from static sources, and imply that there is a population of sources missed by our masking catalogs. 

\subsection{Instrumental and analysis systematic uncertainties}
We consider several systematic effects and re-run our analysis in different configurations in order to assess the sensitivity of the presented measurements. These include errors in the instrument noise model (\S \ref{sec:noisemodl_errors}), uncertainties in the gain solution (\S \ref{sec:ps_vs_g1}) and optics (\S \ref{sec:bl_errors}-\ref{sec:persistence}), along with analysis choices and power spectrum modeling assumptions (\S \ref{sec:auto_vary_mask}). 

\subsubsection{Noise model errors}
\label{sec:noisemodl_errors}
Noise model errors can manifest in the form of non-optimal mode weighting, mis-estimated FF errors, and errors in the noise bias correction. As the \nth{4} flight science integrations are limited to $\sim50$ seconds per field, it is important to model read noise accurately. By representing the read noise model as a Gaussian random field in Fourier space, we assume stationarity of the noise. However, this may not necessarily be the case for the CIBER exposures, which could have different read noise properties than the laboratory data taken on the rail. This is most difficult to characterize on scales comparable to the array, and it is possible that the dispersion of read noise power across fields is larger than predicted.

We perform two tests to check the robustness of the results to noise model mis-specification and residual noise biases. In Appendix \ref{sec:field_diff} we compute auto-power spectra from \bootes\ field differences as in \zem\, which bypass the FF correction and corresponding FF noise bias. For both 1.1 $\mu$m and 1.8 $\mu$m we find strong agreement between difference spectra and our fiducial results. In Appendix \ref{sec:halfexp_cross} we present half-exposure cross-correlations for all fields (excluding elat30), for which noise bias only enters through the FF correction. Here we also find strong agreement between methods, the exception being a handful of bandpowers near $\ell \sim 5000$ in 1.8 $\mu$m that are read noise-dominated at half-exposure integration time. 

\subsubsection{Gain errors}
\label{sec:ps_vs_g1}
\begin{figure}
    \centering 
    \includegraphics[width=0.95\linewidth]{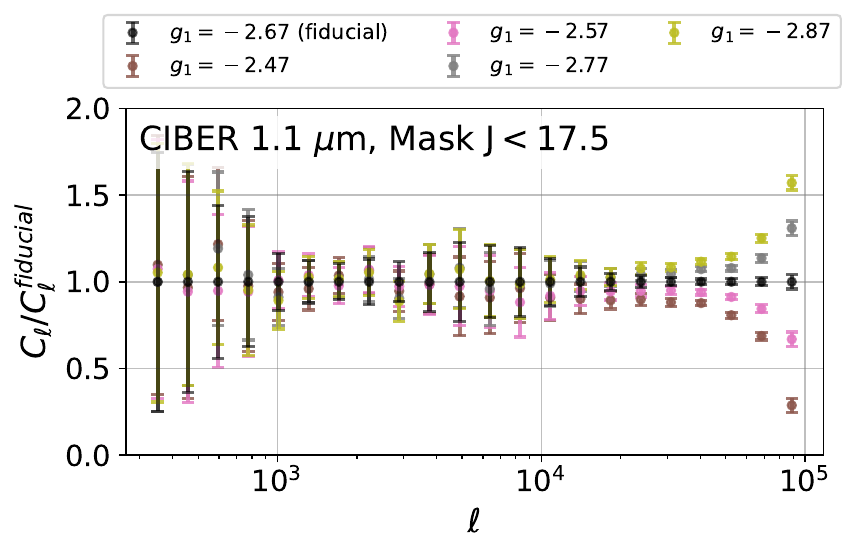}
    \includegraphics[width=0.95\linewidth]{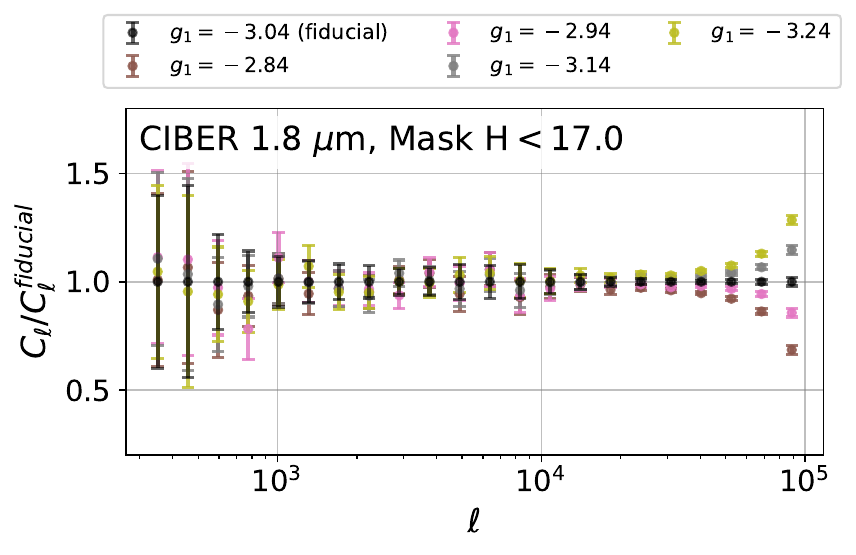}
    \caption{Sensitivity of  \emph{CIBER} auto-power spectra to different assumed gain factors $g_1$, for 1.1 $\mu$m (top) and 1.8 $\mu$m (bottom). This gain factor determines the ratio between read noise and photon noise. The recovered power spectra on scales $\ell<30000$ are largely insensitive to changes in $g_1$.}
    \label{fig:cl_vary_g1}
\end{figure}

We test the sensitivity of the power spectrum results to the assumed gain calibration by varying $g_1$, which converts the maps from units of ADU frame$^{-1}$ to electrons per second. For a fixed absolute gain (ADU frame$^{-1}$ to \sbunit), modifying $g_1$ has the effect of changing the relative contributions of photon noise and read noise, which affects the estimated noise bias. Figure \ref{fig:cl_vary_g1} shows the \emph{CIBER} observed power spectra for different assumed $g_1$ factors within $\pm 0.2$ of  our fiducial values ($g_1=-2.67$ for 1.1 $\mu$m, $g_1=-3.04$ for 1.8 $\mu$m). Varying $g_1$ has little impact on scales $\ell<30000$. The largest impact is on small scales, which are most sensitive to photon noise. As these departures follow the beam function, errors in $g_1$ much larger than our uncertainties would lead to a strongly non-Poissonian small-scale power spectrum. This is not the case for our fiducial results (c.f. Fig. \ref{fig:ciber_auto}), indicating the assumed $g_1$ is near the true value.  

There is an additional overall uncertainty from our determination of the absolute calibration, $g_1 g_2$. Because the power scales as the square of the intensity, its amplitude varies quadratically with the derived calibration factor; however, this does not impact the shape of the power spectrum.

\subsubsection{Errors in beam correction}
\label{sec:bl_errors}
The \emph{CIBER} power spectra are sensitive to beam errors that scale as $\delta C_{\ell} \propto 1/(B_{\ell}^2+\delta B_{\ell}^2)$ and are largest at high-$\ell$.  There is some dispersion in the estimated small-scale 1.1 $\mu$m beam transfer function across the five fields. The fact that the same is not seen in the 1.8 $\mu$m beam correction suggests it is not due to pointing jitter, which would be coherent across imagers. However, based on the consistency of the observed small-scale power spectra we determine that field-dependent beam errors are not significant at our sensitivity. As in \cite{chengihl} we adopt a single beam model for each field, i.e., we do not attempt to quantify or correct for PSF variations across each detector.



\subsubsection{Scattered and reflected stray light}
\label{sec:scattered_light}
An important systematic for any analysis of diffuse light measurements is the presence of light reflected and/or scattered off the telescope optical elements and focal plane. As described in \cite{bock13}, baffling was added for the two \emph{CIBER} imagers following the identification of stray light paths in the first flight data. The telescope off-axis gain function $g(\theta)$ with baffling is measured to be small ($g(\theta=5^{\circ})/g(0) \sim 4 \times 10^{-3}$ and $g(\theta\geq 15^{\circ})/g(0) < 10^{-4}$). However, contributions from off-axis specular and diffuse reflections of bright sources that land within the focal plane are more difficult to quantify, as their structure depends on the geometry of focal plane elements and their reflectance, along with the astronomical scene of each pointing. We do not attempt to simulate this directly, however such asymmetric contributions are expected to de-correlate between imagers.



\subsubsection{Image persistence}
\label{sec:persistence}
Image persistence describes the effect of remnant charge in a detector following one or several exposures, and is typically explained by traps in the depletion regions of photodiodes within each pixel \citep{smith08}. The amplitude of persistence decays exponentially with time and is determined to be at the sub-per cent level for HgCdTe detectors. In place of constructing a full persistence model for the Hawaii-1 detectors, we check for effects of persistence in the power spectrum measurements by taking the brightest sources ($J<12$) from each field and masking them in the subsequent exposures. We calculate power spectra at our deepest masking cuts ($J<18.5$ and $H<18.0$) using the additional persistence masks and find consistent results to within $<5$\% on small scales, suggesting the effect is negligible at \emph{CIBER} sensitivity.

\subsubsection{Sensitivity to astronomical mask}
\label{sec:auto_vary_mask}
\begin{figure}
    \centering 
    \includegraphics[width=0.95\linewidth]{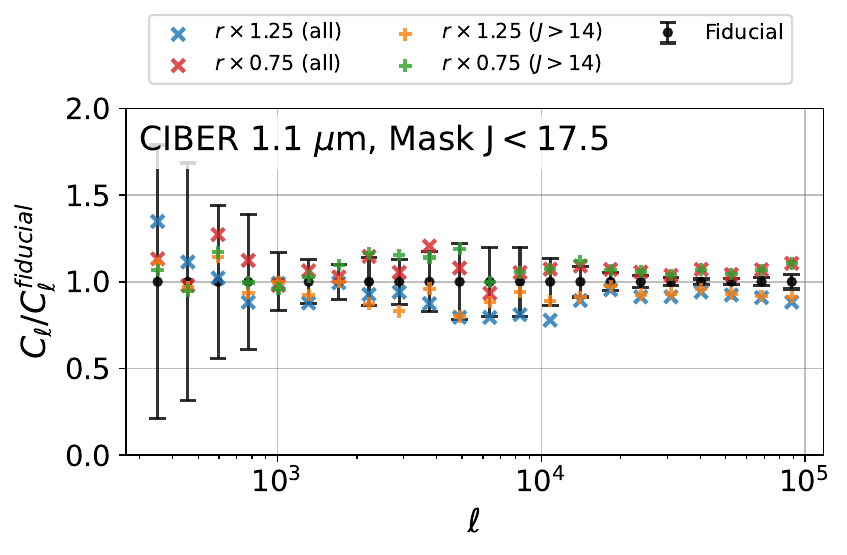}
    \includegraphics[width=0.95\linewidth]{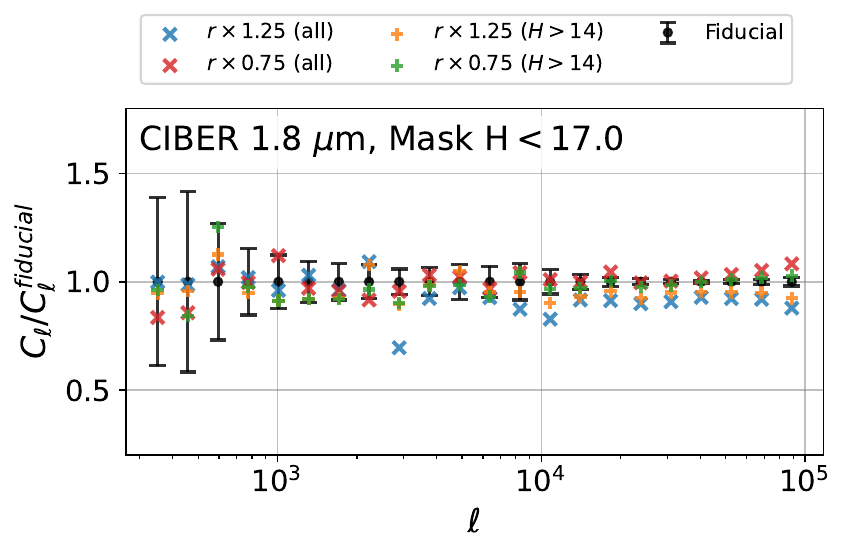}
    \caption{Sensitivity of \emph{CIBER} auto-power spectra to source masking radius, for 1.1 $\mu$m (top) and 1.8 $\mu$m (bottom). The first set of results come from perturbing all source radii, while the second come from perturbing only the fainter ($J>14$ and $H>14$) source radii. In both cases, the power spectra are consistent with the fiducial results to within uncertainties on large scales and within $\pm 15\%$ on scales $\ell > 10000$.}
    \label{fig:cl_vary_mask}
\end{figure}

If the observed signal is uncorrelated with the astronomical mask, the de-convolved power spectra for masks with varying radius around each source should be consistent with each other. In order to probe the correlation between the observed signals and masks, we scale the masking radii by $\pm25\%$. For each case we recompute mode mixing matrices. Figure \ref{fig:cl_vary_mask} shows the fractional power spectrum deviation relative to our fiducial results (denoted $C_{\ell}^{fiducial}$). For both bands, modifying the source masking radii has the effect of altering the small-scale power by 5-10\%. On larger scales, the variations are slightly larger in some bandpowers but largely within the reported uncertainties of the fiducial results. These results indicate there may be some leakage of sources outside of the masked area, although much smaller than the discrepancy with the modelled Poisson signal in Fig. \ref{fig:auto_ps_vs_mask_depth}.

To probe the variation with source mask further, we perform the same exercise but only modify the mask surrounding fainter sources ($J>14$ for 1.1 $\mu$m, $H > 14$ for 1.8 $\mu$m). The resulting power spectra agree well with those from perturbing the mask around all sources, suggesting that the extended PSF of bright sources does not impact our results significantly.

\subsection{Comparison with \texttt{Z14}}

Our closest point of comparison is the \emph{CIBER} auto-power spectrum results from \cite{zemcov14} (\zem), which used imaging data from \emph{CIBER}-1's second and third flights. There are several choices that differ between \zem\ and this analysis which may impact the measured clustering. These include, but are not limited to: the use of field differences in \zem\ vs. our use of individual flight exposures; treatment of the FF correction and its errors and modeling of read noise variance used for noise bias subtraction. There is an additional color correction which we do not quantify due to the fact that the \emph{CIBER} $H$-band filters differed slightly (1.6 $\mu$m to 1.8 $\mu$m) between the second and third flights. To place the results of this work in context with \zem, we rescale the \zem\ power spectra to the new gain derived in this analysis. To test the impact of analysis variations, we reprocess the power spectra in two configurations:
\begin{itemize}
    \item Restricting our masking catalog to detected sources 2MASS, as in \zem. 2MASS has an integrated completeness of 75\% down to $J=17.5$ and $H=17.0$, however the completeness falls off rapidly beyond $J=16$ and $H=15.0$.
    \item Applying the 2MASS-only catalog and using the masking radius function from \zem\ with the following parametrization:
    \begin{equation}
        r(m) = \alpha_m m + \beta_m,
    \end{equation}
    where $\alpha_m=-6.25$ and $\beta_m=110\arcsec$, which is less aggressive for bright sources than used in this work (see \S \ref{sec:astromask}). 
\end{itemize}

\begin{figure*}
    \centering
    \includegraphics[width=0.95\linewidth]{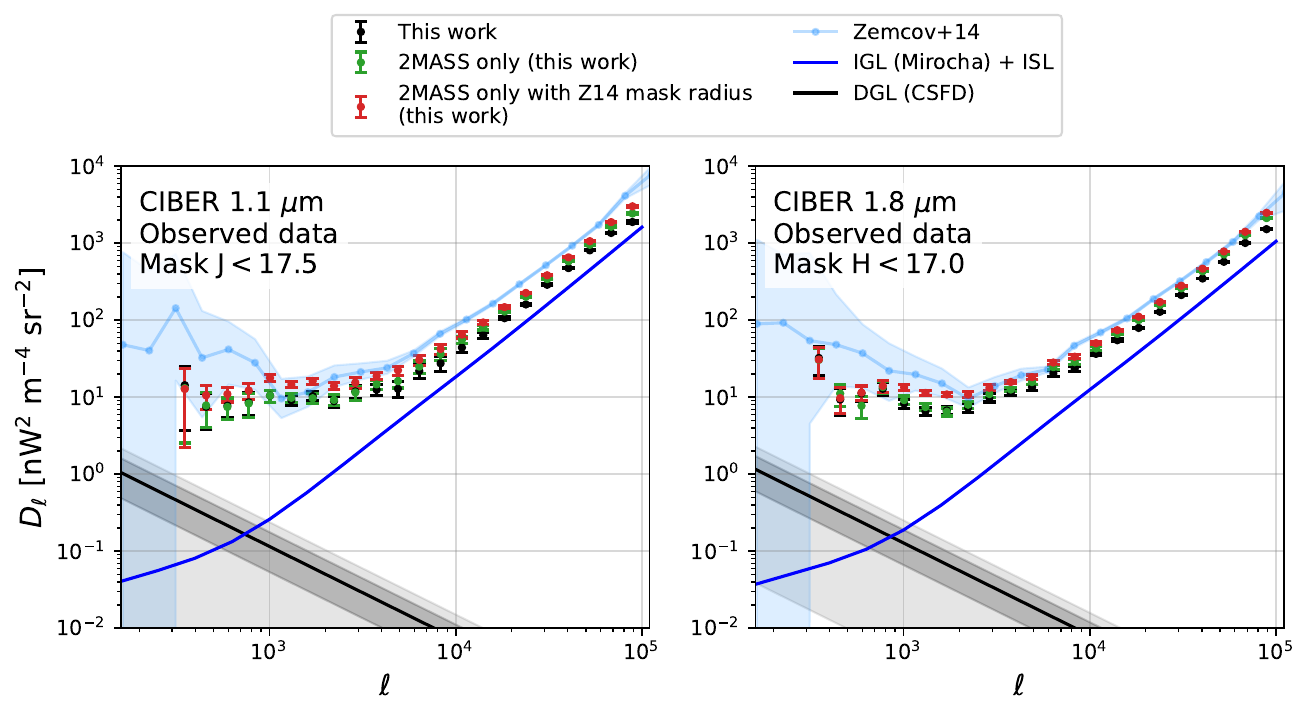}
    \caption{Comparison of \nth{4} flight \emph{CIBER} auto-power spectra (this work) with results from \zem\ (light blue). We show our fiducial results in black alongside those using two analysis variations: 1) restricting our masking catalog to 2MASS only (green) and 2) additionally reverting to the masking radius prescription from \zem\ (red). These results agree with \zem\ with caveats noted in the text. DGL constraints (black shaded region) come from CIBER $\times$ CSFD cross-correlations in \S \ref{sec:ciber_iris_xcorr}.}
    \label{fig:ps_result_2MASS_only}
\end{figure*}

We show the results of this comparison in Fig. \ref{fig:ps_result_2MASS_only}. Our fiducial results are broadly consistent with \zem, however we do find differences. In particular our measurements are consistently lower by $30-50\%$ in both bands for scales $\ell > 1000$. Both measurements show evidence for fluctuation power on scales $\ell < 1000$. Our estimates are a factor of two to three lower than \zem\ at $\ell<1000$, but within the large uncertainties of the \zem\ measurement, which are order unity and greater on these scales.

We find that retaining our masking radius prescription but restricting the masking catalog to 2MASS sources increases the small-scale fluctuation power by 20\% and 30\% for 1.1 and 1.8 $\mu$m, respectively, but does not strongly affect scales $\ell < 5000$. When we additionally revert to the \zem\ masking prescription, the power on scales $1000 < \ell < 5000$ is 50\% higher on average than our fiducial measurements. This analysis variation also increases the small-scale power by $\sim 10\%$ for 1.1 $\mu$m. These results suggest some leakage of extended PSF from the less aggressive \zem\ mask. There remains some discrepancy between measurements for $5000 < \ell < 20000$, scales for which both analyses are sensitive to read noise.

\section{Consistency across wavelengths through cross spectra}
\label{sec:cross_spectra}

\subsection{Correlation between CIBER bands}
\label{sec:ciber_ciber_cross}
To compute \emph{CIBER} 1.1 $\mu$m $\times$ 1.8 $\mu$m cross power spectra, we re-project the \emph{CIBER} 1.8 $\mu$m maps and masks to 1.1 $\mu$m detector coordinates using the WCS solutions and a bilinear interpolation step. We perform the reprojection after the FF correction and image filtering in native 1.8 $\mu$m detector coordinates. This step imposes a transfer function on the 1.8 $\mu$m maps that primarily impacts the small-scale cross-spectrum. We estimate the transfer function using 100 point source mocks, for which we perform the same procedure and calculate the ratio between initial and reprojected power spectra,
\begin{equation}
    \hat{T}_{\ell}^{reproj} = C_{\ell}^{post}/C_{\ell}^{pre}. 
\end{equation}
As we only apply the reprojection to 1.8 $\mu$m, the transfer function for the cross spectrum is $\sqrt{T_{\ell}^{reproj}}$. We assume any aliasing effects due to the undersampled \emph{CIBER} PSFs are captured by $T_{\ell}^{reproj}$. There are potential errors associated with the re-projection due to relative astrometric errors between arrays. We do not attempt to correct for such errors or quantify them in detail, as they generally persist at the sub-pixel level or smaller. 

We mask sources using both $J$- and $H$-band magnitudes in order to probe the sub-threshold fluctuations common to both bands. The relative boresight between imagers is small, however there is a modest increase in the masking fraction due to the combination of instrument masks. While there is no noise de-biasing applied to the cross-power spectra, we do perform a correction for multiplicative FF bias, which arises from the common sky signal across \emph{CIBER} imagers entering both sets of FF estimates (explained further in Paper I). We estimate mode coupling matrices with a similar prescription to that of the auto-power spectra, using the union masks and 1.1/1.8 $\mu$m field sky brightnesses.

\begin{figure}
    \centering
    \includegraphics[width=\linewidth]{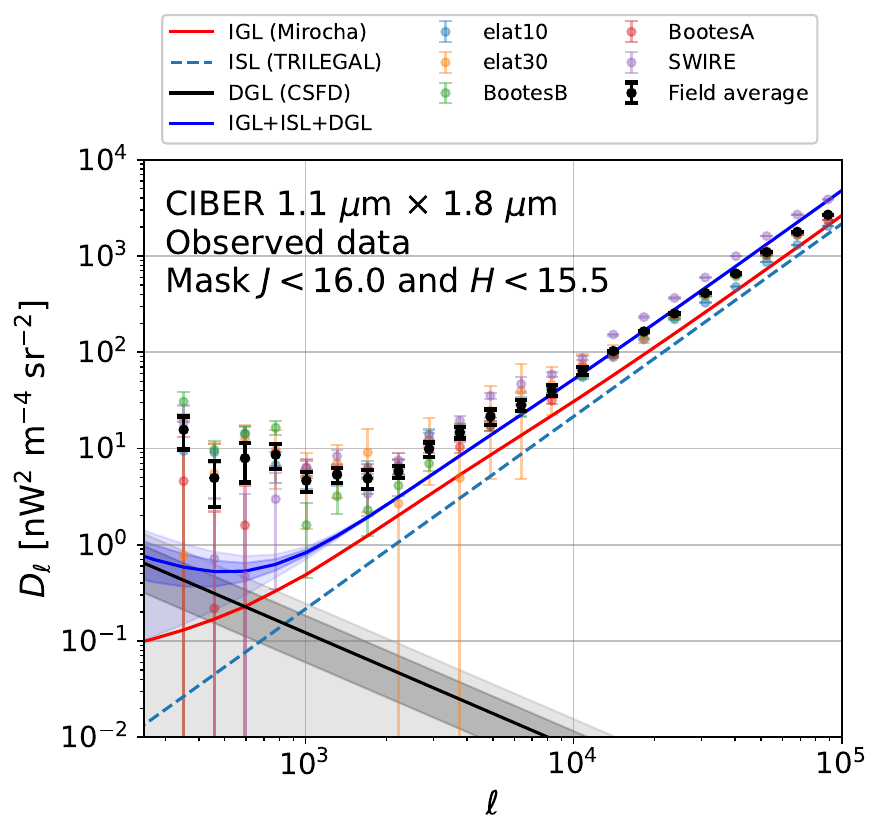}
    \caption{1.1 $\times$ 1.8 $\mu$m cross-power spectra, for individual fields (colored points) and weighted field average (black). The blue solid curve shows the IGL+ISL cross-power spectrum predictions described in \S \ref{sec:astro_modl}. The estimated DGL contribution is taken as the geometric mean of the 1.1 $\mu$m and 1.8 $\mu$m DGL auto-power spectrum predictions in \S \ref{sec:ciber_iris_xcorr}. The SWIRE field (purple) has $\sim50\%$ higher stellar density than the other fields, which explains the slightly higher Poisson level.}
    \label{fig:ciber_cross}
\end{figure}

In Figure \ref{fig:ciber_cross} we show the cross-power spectrum measurements of the five \emph{CIBER} fields along with their weighted average. The cross-spectra are similar in structure to the auto-spectra, exhibiting departures from model predictions that increase with angular scale. Seen more clearly in the cross-power spectrum than in the auto-power spectra, the SWIRE field has more small-scale power due to higher levels of ISL. We also find similar departures in the \bootes\ B field on large scales to those found in the auto-power spectra (c.f., Figs. \ref{fig:ciber_auto}, \ref{fig:ciber_auto_field_consistency_TM1}, \ref{fig:ciber_auto_field_consistency_TM2}). With these caveats, we find strong internal consistency across fields. On scales $\ell < 2000$, the cross-power spectra have a combined SNR of 9.9$\sigma$ and are in excess of the fiducial IGL+ISL+DGL model with a significance of $7.9\sigma$.

\begin{figure}
    \centering
    \includegraphics[width=\linewidth]{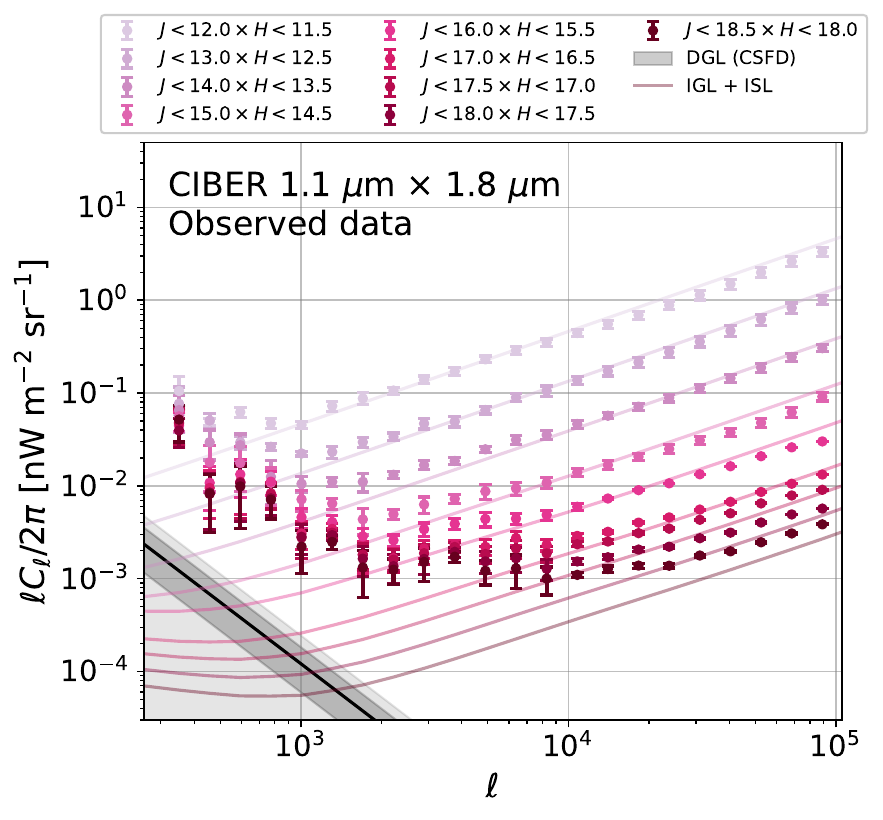}
    \caption{1.1 $\mu$m $\times$ 1.8 $\mu$m cross-power spectra for a range of source masking cuts indicated in the legend, alongside IGL+ISL (solid curves) and DGL (black) predictions. The large-angle fluctuation power evolves more mildly than that on small scales, with little to no variation beyond $J=17$. Note the different normalization of the power spectra $D_{\ell}/\ell$ as in Fig. \ref{fig:auto_ps_vs_mask_depth}.}
    \label{fig:ciber_cross_vs_mag}
\end{figure}

Figure \ref{fig:ciber_cross_vs_mag} shows the field-averaged 1.1 $\mu$m $\times$ 1.8 $\mu$m cross-power spectra for a set of progressively deeper masking cuts ranging from $(J, H)<(12, 11.5)$ to $(J,H)<(18.5, 18.0)$.  Unlike the \emph{CIBER} auto-spectra, the small-scale cross-spectra are much closer to IGL+ISL Poisson noise predictions for all cuts down to our deepest masking depth ($J=18.5$ and $H=18.0$). Similar to Fig. \ref{fig:auto_ps_vs_mask_depth}, the small-scale cross-spectrum varies rapidly with masking depth, while the large-angle fluctuations appear to be largely independent of masking depth beyond $J=17$.

To quantify the coherence of surface brightness fluctuations across imagers, we calculate the power spectrum cross-correlation coefficient as a function of multipole. For maps $A$ and $B$, we define the correlation coefficient $r_{\ell}$ (sometimes called the ``coherence") as
\begin{equation}
    r_{\ell} = \frac{C_{\ell}^{A\times B}}{\sqrt{C_{\ell}^AC_{\ell}^B}},
\end{equation}
which has variance
\begin{multline}
    \delta r_{\ell}^2 = \frac{1}{C_{\ell}^AC_{\ell}^B} \times \left[(\delta C_{\ell}^{A\times B})^2 \right. \\ \left. + \left(\frac{C_{\ell}^{A\times B}\delta C_{\ell}^A}{2C_{\ell}^A}\right)^2 + \left(\frac{C_{\ell}^{A\times B}\delta C_{\ell}^B}{2C_{\ell}^B}\right)^2 \right].
\label{eq:delta_corrcoef}
\end{multline}

\begin{figure}
    \centering
    \includegraphics[width=\linewidth]{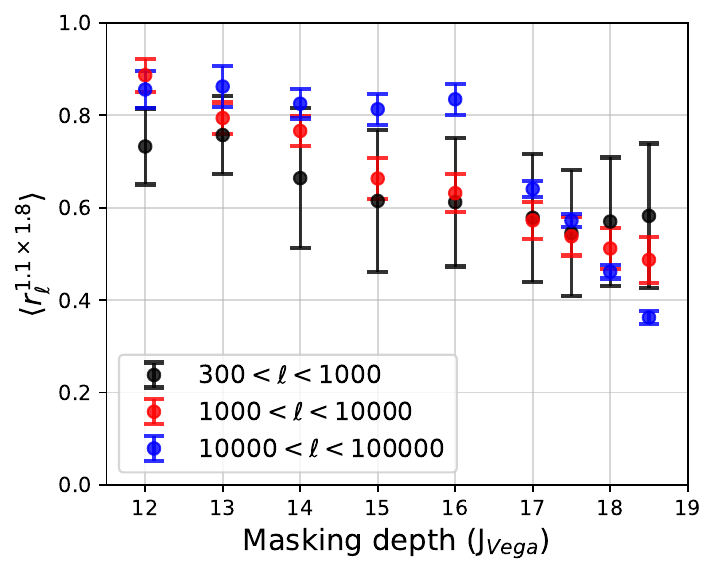}
    \caption{1.1 $\mu$m $\times$ 1.8 $\mu$m cross-correlation coefficient in broad bandpowers as a function of $J$-band masking depth. We use the masking cuts from Fig. \ref{fig:ciber_cross_vs_mag} to compute the auto- and cross-power spectra for these results. While we find strong de-correlation on small scales as the masking depth increases, the correlation coefficient on large scales ($\ell<1000$) remains well within uncertainties beyond $J=15$.}
    \label{fig:ciber_crosscoeff_vs_mag}
\end{figure}

We compute $r_{\ell}$ from auto- and cross-power spectra using consistent masks enumerated in Fig. \ref{fig:ciber_cross_vs_mag}. To highlight the scale dependence of $r_{\ell}$, we calculate estimates in three coarse bandpowers ($300<\ell<10^3$, $10^3<\ell<10^4$ and $10^4<\ell<10^5$). We plot these estimates as a function of masking magnitude in Fig. \ref{fig:ciber_crosscoeff_vs_mag}. For $300 < \ell < 1000$, $r_{\ell}$ decreases with masking depth until $J=16$/$H=15.5$, for which $\langle r_{\ell}^{1.1\times 1.8}\rangle = 0.61\pm 0.14$. This is slightly lower but consistent with \zem, which reported $r_{\ell} = 0.76 \pm 0.10$. The \nth{2} flight filter had a slightly lower central wavelength (1.6 $\mu$m) and spectral coverage extending down to 1.2 $\mu$m (compared to 1.5 $\mu$m for the \nth{4} flight filter), which may contribute to the reduced cross-correlation amplitude. On small scales ($\ell > 10000$), $\langle r_{\ell}^{1.1\times 1.8}\rangle \sim 0.85 \pm 0.05$ for point source-dominated maps but decreases rapidly between $J=16.0$ and $J=18.5$ down to $\langle r_{\ell}^{1.1\times 1.8}\rangle = 0.36$. This is not seen in model predictions and reflects the behavior in the Poisson amplitude in auto-power spectra for deeper masks. On intermediate scales ($1000<\ell<10000$), $r_{\ell}$ decreases beyond $J=16$, albeit less rapidly than on small scales. 


\subsection{CIBER - Spitzer correlation}
\label{sec:ciber_spitzer}

For a subset of fields (\bootes\ A and B) we have access to IRAC infrared imaging. The \emph{Spitzer} Deep, Wide-Field Survey \citep[SDWFS;][]{sdwfs} covers the majority of the \bootes\ field footprint observed by \emph{CIBER}. We use two versions of the SDWFS mosaics in order to assess consistency of the power spectrum estimates. The first are the official SDWFS mosaics which are constructed using a combination of \texttt{IRACProc} \citep[an augmentation of the MOPEX algorithm][]{schuster06} for 1 deg$^2$ sub-mosaics and \texttt{Montage} \citep{montage} for the final 10 deg$^2$ mosaic. The second version, which we use for our fiducial results (also used in \cite{cooray12} and \texttt{Z14}), were generated using the self-calibration algorithm of \cite{fixsen_selfcal}. The co-addition algorithms are similar in spirit; however the self-calibration mosaics use dithered exposures to solve for the background and relative gains simultaneously. We utilize mosaics from four epochs of imaging at 3.6 $\mu$m (IRAC Channel 1) and 4.5 $\mu$m (Channel 2).



\subsubsection{IRAC data reduction}

We first reproject the SDWFS mosaics at each epoch onto \emph{CIBER} detector coordinates, interpolating to \emph{CIBER} angular resolution and converting from MJy sr$^{-1}$ to \sbunit. The initial SDWFS mosaics have a pixel resolution of 0.86$\arcsec$ and PSF FWHM of 1.9$\arcsec$. To place the mosaics in the same surface brightness units as our \emph{CIBER} results we apply a multiplicative normalization factor $N_{eff}=11.2$ equal to the PSF pixel effective area for a Gaussian beam. We validate the calibration of our interpolated \emph{Spitzer} maps by comparing aperture photometry around sources with $13 < L < 16$ with the SDWFS catalog, finding close agreement in recovered flux. To reduce scatter in our cross-correlation measurements due to bright \emph{Spitzer} sources, we mask any source with $L<15.0$ ($m_{AB}^{3.6}=17.7$) in addition to our cuts on $J<16.0$ and $H<15.5$. After combining these with the SDWFS instrument and coverage masks, we re-compute mode coupling matrices for each mask. The beam transfer function for the cross-spectrum is taken to be the geometric mean of the individual beam functions, i.e.,
\begin{equation}
    B_{\ell}^{\times} = \sqrt{B_{\ell}^{CIBER}B_{\ell}^{IRAC}}
\end{equation}
where we use the beam correction $B_{\ell}^{IRAC}$ determined in \cite{cooray12}. We also correct for reprojection effects using $T_{\ell}^{reproj}$ as in \S \ref{sec:ciber_ciber_cross}.

\subsubsection{Spitzer noise model}
\label{sec:spitzer_noise}

The \emph{Spitzer} map making process does not produce a noise model, which limits our ability to check internal consistency, particularly on large scales. In the absence of a noise model, we follow the prescription from \zem\, which used the average 2D power spectrum of masked cross-epoch map differences. This approach assumes that any static sky signal cancels out while noise power adds linearly. This approach has the benefit of capturing the direct noise statistics of the data, however any residual systematics or non-cancellations will cause us to misestimate the errors.

We rectify the 2D Fourier modes within each bandpower using the diagonal elements of the $M_{\ell\ell^{\prime}}$ matrix calculated for each field. Assuming the SDWFS maps are photon noise-limited, we then scale the 2D noise power spectra to match the sensitivity of both the epoch pair averages used in the \emph{Spitzer} auto spectrum and the four-epoch averaged maps that we cross-correlate with \emph{CIBER}.

\subsubsection{Spitzer auto- and cross-power spectra}

\begin{figure*}
    \centering 
        \includegraphics[width=0.9\linewidth]{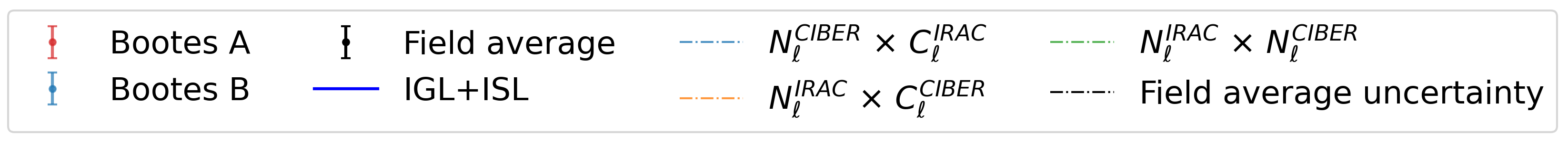}

    \includegraphics[width=0.48\linewidth]{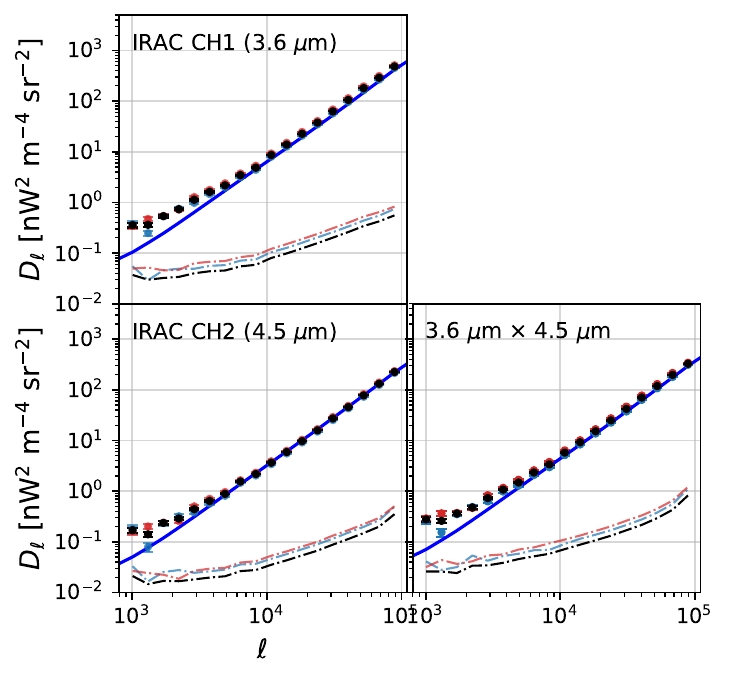}
    \includegraphics[width=0.48\linewidth]{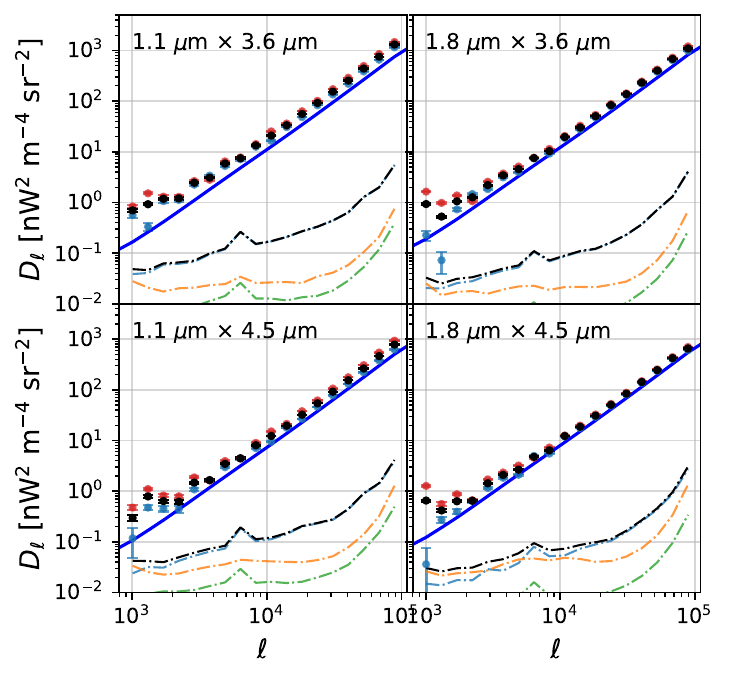}
    \caption{\emph{Spitzer} auto-power spectra (left) and \emph{CIBER} $\times$ \emph{Spitzer} cross-power spectra (right). We plot the individual \bootes\ field measurements in blue and red, while the field averages are plotted in black. In each panel, we plot the noise components that contribute to the derived uncertainties (dash-dotted curves) and the predicted IGL+ISL contribution (blue curves). The astronomical mask removes sources with $J<16$, $H<15.5$ or $L<15$.}
    \label{fig:ciber_spitzer_perfield}
\end{figure*}

We compute the \emph{Spitzer} auto-power spectrum from the four epoch mosaics $\lbrace{A,B,C,D\rbrace}$ as
\begin{equation}
    \hat{C}_{\ell}^{IRAC} = 0.5(A+B) \times 0.5(C+D)
\end{equation}
In other words, we average pairs of epochs and cross-correlate these against each other. This estimator is favorable because noise across the per-epoch maps is uncorrelated, along with any residual ZL in the maps after filtering. We estimate $\delta C_{\ell}^{IRAC}$ following Eq. \eqref{eq:cross_knox} by generating Gaussian random field realizations that match the statistics of the SDWFS differences and cross-correlating the realizations against each other and the pair-averaged maps. 

We calculate \emph{CIBER} $\times$ \emph{Spitzer} cross-power spectra using the four-epoch coadded mosaics and estimate uncertainties in a similar fashion, instead using using the \emph{CIBER} and IRAC (coadded depth) noise models. The \emph{CIBER} noise model realizations include read noise, photon noise, and Monte Carlo draws of the FF error specific to the \bootes\ fields. 

Lastly, we compute \emph{Spitzer} 3.6 $\mu$m $\times$ 4.5 $\mu$m cross spectra within the \bootes\ fields, using the coadded mosaics and derived noise models.

A significant source of systematic uncertainty in this measurement is the mosaicing used to combine individual IRAC exposures, which have an instantaneous field of view of $5\arcmin\times5\arcmin$. We compare the fiducial results with those using the official SDWFS mosaics (which use \texttt{IRACProc} and \texttt{Montage}) as a check of consistency. We plot the relative auto- and cross-power spectra in Fig. \ref{fig:sdwfs_mosaic_compare_ps} of App. \ref{sec:sdwfs_mosaic_app}. For both \emph{Spitzer} internal power spectra and \emph{CIBER} $\times$ \emph{Spitzer} cross-power spectra, we find significant differences between mosaicing algorithms on large angular scales, leading us to discard the lowest six bandpowers in our results. For $\ell>875$, both mosaicing algorithms recover consistent power spectra to within $5\%$.

In Figure \ref{fig:ciber_spitzer_perfield} we plot the measured \emph{Spitzer} and \emph{CIBER} auto- and cross-power spectra for \bootes\ A and B. The \emph{Spitzer}-internal auto- and cross-spectra show strong internal consistency between the fields, with the exception of the second lowest bandpower for which there is modest disagreement, and agree with IGL+ISL Poisson predictions on small scales. The \emph{CIBER} $\times$ \emph{Spitzer} cross-spectra are generally consistent for multipoles $\ell > 10000$, though the 1.1 $\mu$m cross-spectra show larger variation between fields and have slightly higher Poisson noise than predictions. On larger scales we find variations larger than the estimated uncertainties in different band combinations. In some cases, we find coherent variations across bandpowers common to both fields (e.g., 1.1 $\times$ 4.5 $\mu$m).

In each panel, we show the individual noise components that enter our uncertainty estimates for each band combination. For $875<\ell<3000$, the estimated cross-spectrum uncertainty sourced from \emph{CIBER} and \emph{Spitzer} noise is comparable, with \emph{CIBER} noise dominating the uncertainties on smaller scales and for the 1.1 $\mu$m maps. We caution, however, that our \emph{Spitzer} noise model is likely an incomplete estimate of the true systematic power spectrum uncertainty, and it is unclear how the \emph{Spitzer} mosaicing transfer function interacts with sky signals and noise in cross-correlation. Nevertheless, we do find evidence for departures from Poisson noise in the \emph{Spitzer} auto- and cross-power spectra, indicating that some portion of the \emph{CIBER} large-angle fluctuations are common across wavelengths and instruments on scales $\ell<5000$.

%% file: sections/dgl.tex
\section{Cross-correlation with tracers of Diffuse Galactic Light}
\label{sec:ciber_iris_xcorr}
We assess the contribution of DGL in our measurments by performing cross-correlations with external DGL tracer maps. This comes with the caveat that no single map is a perfect tracer of DGL. The DGL angular power spectrum falls steeply with multipole, with several studies finding $\gamma \sim -3$ \citep{dgl_powerlaw, dglps}. A joint analysis of HST and \emph{CIBER} clustering data at 1.6 $\mu$m yielded a best-fit spatial index $\gamma=-3.05\pm 0.07$ \citep{mitchellwynne}; however assuming DGL comprises all of the large-angle fluctuations. 

We perform cross-correlations against two versions of the well-used SFD 100 $\mu$m dust extinction maps, the original maps from \cite{sfd98} and the recent ``CIB-cleaned" SFD maps \citep[CSFD;][]{csfd}. Prior work has demonstrated the presence of extragalactic sources in SFD and other dust maps \citep{ykc_lss_in_dust}. Using cross-correlations with catalogs of known extragalactic sources, \cite{csfd} isolates and subtracts the two-halo contribution in existing SFD maps, which is validated by stacking the original and corrected maps on the positions of known CIB galaxies. We cross-correlate against both versions of SFD in order to assess the impact of CIB contamination on the DGL measurements, as seen by \emph{CIBER} and \emph{Spitzer}. 

We use the same cross-power spectrum formalism from the previous section. As the SFD maps have much coarser angular resolution than those from \emph{CIBER} and \emph{Spitzer}, the beam correction is well-approximated by $B^{\times}_{\ell} \approx \sqrt{B_{\ell}^{SFD}}$. We model $B_{\ell}^{SFD}$ assuming a Gaussian beam profile with FWHM $=6.1\arcmin$. We then estimate the noise associated with each cross-correlation following \eqref{eq:cross_knox}.

\begin{figure*}
    \centering
    \includegraphics[width=0.85\linewidth]{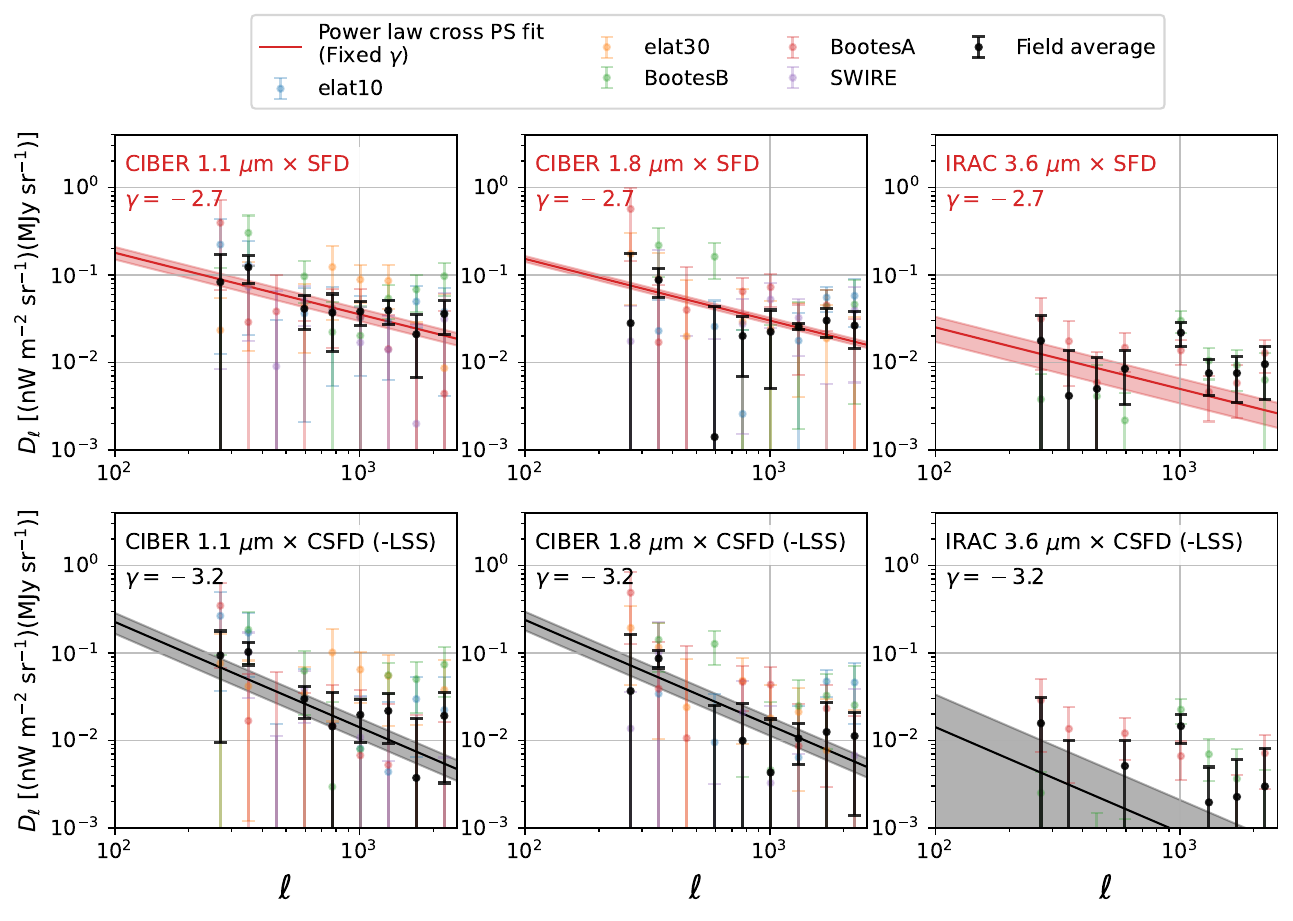}
    \caption{\emph{CIBER} cross-power spectra with SFD maps normalized to 100 $\mu$m (top row) and ``corrected" SFD maps that are cleaned of large-scale structure (CSFD, bottom row). Individual field measurements are shown by the colored points, while the weighted field averages are shown in black. The best-fit power spectra are calculated assuming a power-law form $C_{\ell}=A(\ell/\ell_{pivot})^{\gamma}$, for which the shaded regions span the $\pm 1\sigma$ uncertainties $\sigma(\hat{A})$. We fit the \emph{CIBER} $\times$ DGL cross-spectra with free amplitude $A$ but fixed $\gamma$. We find non-detections of cross-power between 4.5 $\mu$m and SFD/CSFD, and so we omit those results here.}
    \label{fig:ciber_dglmap_cross_compare}
\end{figure*}

In Figure \ref{fig:ciber_dglmap_cross_compare} we plot the measured cross-spectra from both sets of tracer maps. We obtain non-detections of cross-power between 4.5 $\mu$m and IRAS, and so we omit these results. While measurements for individual \emph{CIBER} fields are marginal, the field-averaged cross-spectra with SFD have total signal to noise ratios (SNRs) of 6.2, 11.8 and 5.4 for 1.1 $\mu$m, 1.8 $\mu$m and 3.6 $\mu$m, respectively. However, the bulk of the total SNR comes from small scales, which are susceptible to CIB contamination. In all cases, the cross-power spectra with CSFD are lower on small scales by a factor of two to three, with correspondingly lower SNR.

\begin{table}
    \centering
    \begin{tabular}{c|c|c|c}
        External map & $\lambda$ & Cross SNR & $\delta I_{\lambda}^{CIBER}/\delta I_{100\mu \textrm{m}}$ \\
        & ($\mu$m) & ($\ell < 2500$) & \\
        \hline 
        SFD & 1.1  & 6.2 & 9.1 $\pm$ 1.5\\
        ($\gamma=-2.7$) & 1.8 & 11.8 & 7.7 $\pm$ 0.6\\

         & 3.6 & 5.4 & 1.3 $\pm$ 0.4 \\
        \hline
        CSFD & 1.1 & 5.3 & \textbf{8.2 $\pm$ 2.2} \\
        ($\gamma=-3.2$) & 1.8 & 4.8 & \textbf{8.6 $\pm$ 2.1}\\
        & 3.6 & 3.9 & \textbf{0.5 $\pm$ 0.7}
        \end{tabular}
    \caption{Result of power-law fits to \emph{CIBER} $\times$ DGL and \emph{Spitzer} $\times$ DGL cross-power spectra. $\gamma$ indicates the best-fit spatial indices to SFD and CSFD angular fluctuations described in \S \ref{sec:ciber_iris_xcorr}. The conversion factors $\delta I_{\lambda}^{CIBER}/\delta I_{100\mu \textrm{m}}$ have units (nW m$^{-2}$ sr$^{-1}$)/(MJy sr$^{-1}$). We adopt the CSFD constraints for our fiducial results.}
    \label{tab:ciber_dgl_crossfit}
\end{table}

\begin{figure}
    \centering 
    \includegraphics[width=\linewidth]{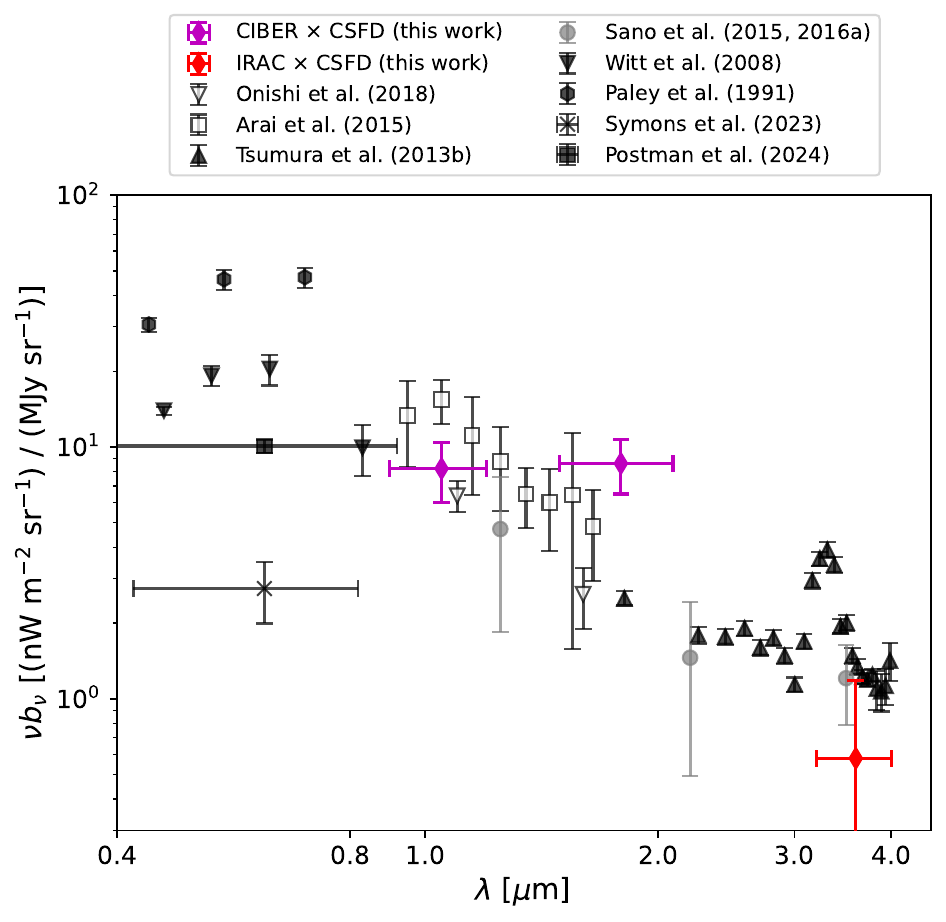}
    \caption{Comparison of measured DGL emissivity $\nu b_{\nu}$ with existing measurements in the literature. Unlike previous measurements that rely on correlations with the DGL mean intensity, the derived values from this work (magenta and red diamonds) come from the cross-power spectrum of \emph{CIBER} and CSFD.}
    \label{fig:dgl_agg}
\end{figure}

We model the \emph{CIBER} $\times$ SFD cross-spectra by fitting a power-law $C_{\ell} = A_{\times}\ell^{\gamma}$ for $\ell < 3000$. We measure separate spatial indices $\gamma$ for SFD and CSFD based on fits to $6^{\circ}\times 6^{\circ}$ maps centered on each of the \emph{CIBER} fields. Averaging over the five \emph{CIBER} fields we find $\gamma_{SFD} = -2.7$ and $\gamma_{CSFD}=-3.2$. This is consistent with CSFD containing less CIB contamination, which otherwise leads to a shallower power spectrum in the uncorrected SFD maps. In each fit we fix $\gamma$ to these values and float $A_{\times}$, the results of which are also shown in Fig. \ref{fig:ciber_dglmap_cross_compare} with 1$\sigma$ uncertainties overlaid. Table \ref{tab:ciber_dgl_crossfit} summarizes the cross-spectra and power-law fits for each map combination. The cross-spectra are clearly affected by sources in SFD, so we regard the SFD values in Table \ref{tab:ciber_dgl_crossfit} to over-estimate the DGL amplitude and to under-estimate the errors. Therefore we only show the DGL values from CSFD in Figs. \ref{fig:ciber_auto}, \ref{fig:auto_ps_vs_mask_depth}, \ref{fig:ps_result_2MASS_only}, \ref{fig:ciber_cross}, \ref{fig:ciber_cross_vs_mag} and \ref{fig:dgl_agg}.

In Figure \ref{fig:dgl_agg} we plot our DGL $\nu b_{\nu}$ measurements alongside existing measurements spanning $0.4-4$ $\mu$m \citep{arai_15, tsumura13b, sano15, sano16, witt_08, paley_91, symons_nh, postman_24}. Our fluctuation-based measurements differ qualitatively from these studies, which typically utilize correlations in the intensity monopole against SFD extinction maps re-scaled to 100 $\mu$m. Comparing both sets of measurements assumes that the color of the fluctuation component correlates with the color of the mean intensity, which may not be the case given variations in the starlight that produces DGL. Nonetheless, our \emph{CIBER} measurements are broadly consistent with \cite{onishi18}, in particular at 1.1 $\mu$m. While the \emph{IRAC} $\times$ SFD cross-correlations yield results in agreement with \cite{tsumura13b}, our CSFD-based measurements are lower at the 1$\sigma$ level.



With estimates of $\delta I_{\lambda}^{CIBER}/\delta I_{100\mu m}$ from the cross-power spectra and measured SFD/CSFD auto-power spectra, we make predictions for the contribution of DGL to observed \emph{CIBER} auto-power spectra. On large scales, the DGL auto- and cross-power spectra can be modeled as
\begin{align}
    C_{\ell}^{SFD} &= I_{100}^2 C_{\ell}^{DGL}, \\
    C_{\ell}^{CIBER \times SFD} &= I_{\lambda_{CIBER}}I_{100}C_{\ell}^{DGL}, \text{and} \\
    C_{\ell}^{\lambda_{CIBER}} &= \left(\frac{I_{\lambda_{CIBER}}}{I_{100}}\right)^2 C_{\ell}^{SFD}.
\end{align}
The scaled DGL predictions $C_{\ell}^{\lambda_{CIBER}}$ are what we show in the results of \S \ref{sec:ciber_results}. Our DGL predictions are roughly consistent with \cite{zemcov14}, which marginally detected cross-power between \emph{CIBER} and IRIS in one field difference. Our results establish with high confidence that scattered DGL does not explain the observed fluctuations. Indeed, our limits from CSFD are lower than the measured \emph{CIBER} fluctuations by over an order of magnitude, motivating the presence of a separate sky component to explain the power. In Appendix \ref{sec:dgl_zl_monopole_bpcorr} we further check that the large-scale bandpowers from the five \emph{CIBER} fields are uncorrelated with variations in DGL monopole intensity.

The predicted DGL contribution to the 3.6 $\mu$m auto-power spectrum is $\sim$2 orders of magnitude lower than the observed fluctuations, and are subdominant to IGL on the same scales and for \emph{CIBER} $\times$ \emph{Spitzer} cross-power spectra. For this reason we omit DGL in Fig. \ref{fig:ciber_spitzer_perfield}.


%% file: sections/overview_spec.tex
\section{Spectrum and coherence of diffuse light fluctuations from \emph{CIBER} and \emph{Spitzer}}
\label{sec:super_cl}

\begin{figure*}
    \centering
    \includegraphics[width=0.46\linewidth]{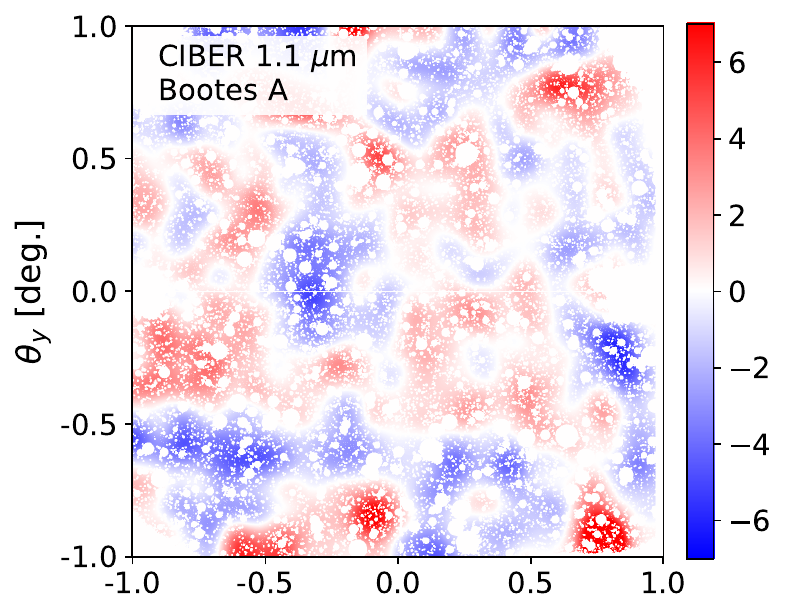}
    \includegraphics[width=0.47\linewidth]{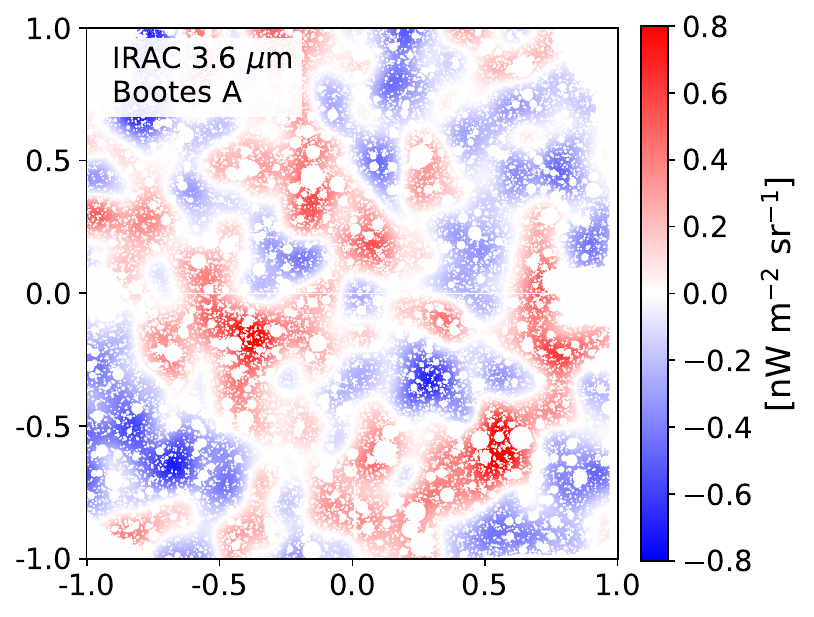}
    \includegraphics[width=0.46\linewidth]{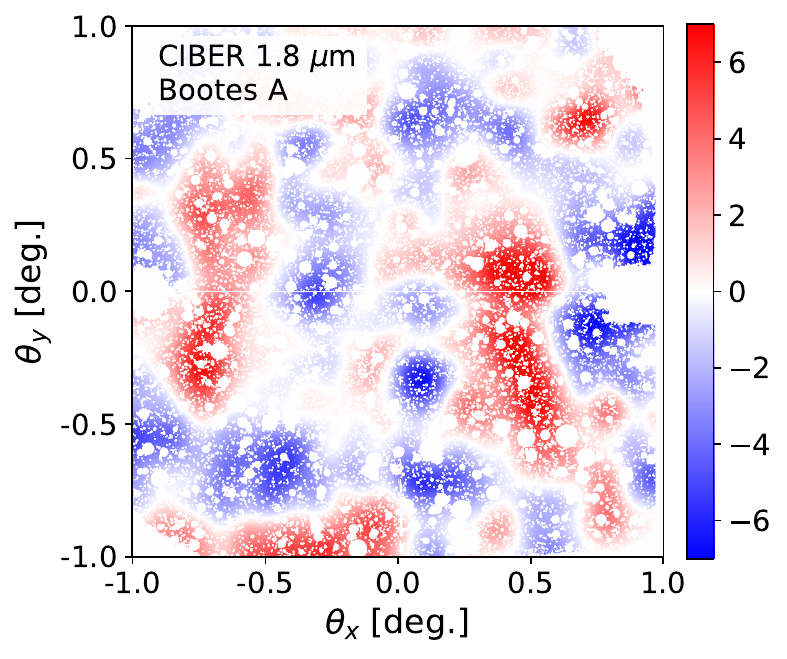}
    \includegraphics[width=0.47\linewidth]{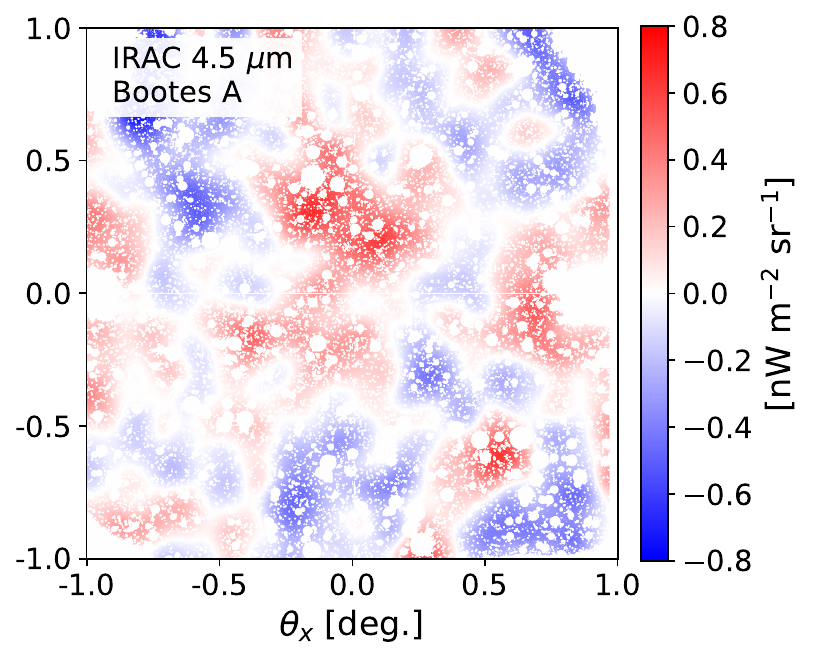}
    \caption{\emph{CIBER} (left column) and \emph{Spitzer} (right column) fluctuation maps for the \bootes\ A field. Each map is low pass-filtered with a Gaussian kernel of width $\sigma=3$\arcmin\ and high pass-filtered with a Gaussian of width $\sigma=15$\arcmin. All maps are regridded to 1.1 $\mu$m detector coordinates. Note that the \emph{CIBER} maps as presented contain errors from the FF correction.}
    \label{fig:ciber_spitzer_bootesA}
\end{figure*}

Having described the measurements from \emph{CIBER} and \emph{Spitzer}, we now use the auto- and cross-power spectra between $1-5$ $\mu$m to probe the correlation structure and spectral characteristics of the measured intensity fluctuations. All results in this section adopt the same masking selection as in \S \ref{sec:ciber_spitzer}, namely masking any source with $J<16.0$, $H<15.5$, or $L<15.0$. Also following \S \ref{sec:ciber_spitzer} we restrict the results in this section to scales $\ell > 875$. In Figure \ref{fig:ciber_spitzer_bootesA} we show the \emph{CIBER} and \emph{Spitzer} maps of \bootes\ A, regridded to 1.1 $\mu$m detector coordinates. The maps are masked and then low- and high-pass filtered to highlight fluctuations on scales $3\arcmin < \theta < 15\arcmin$.

\begin{figure*}
\centering
\includegraphics[width=0.95\linewidth]{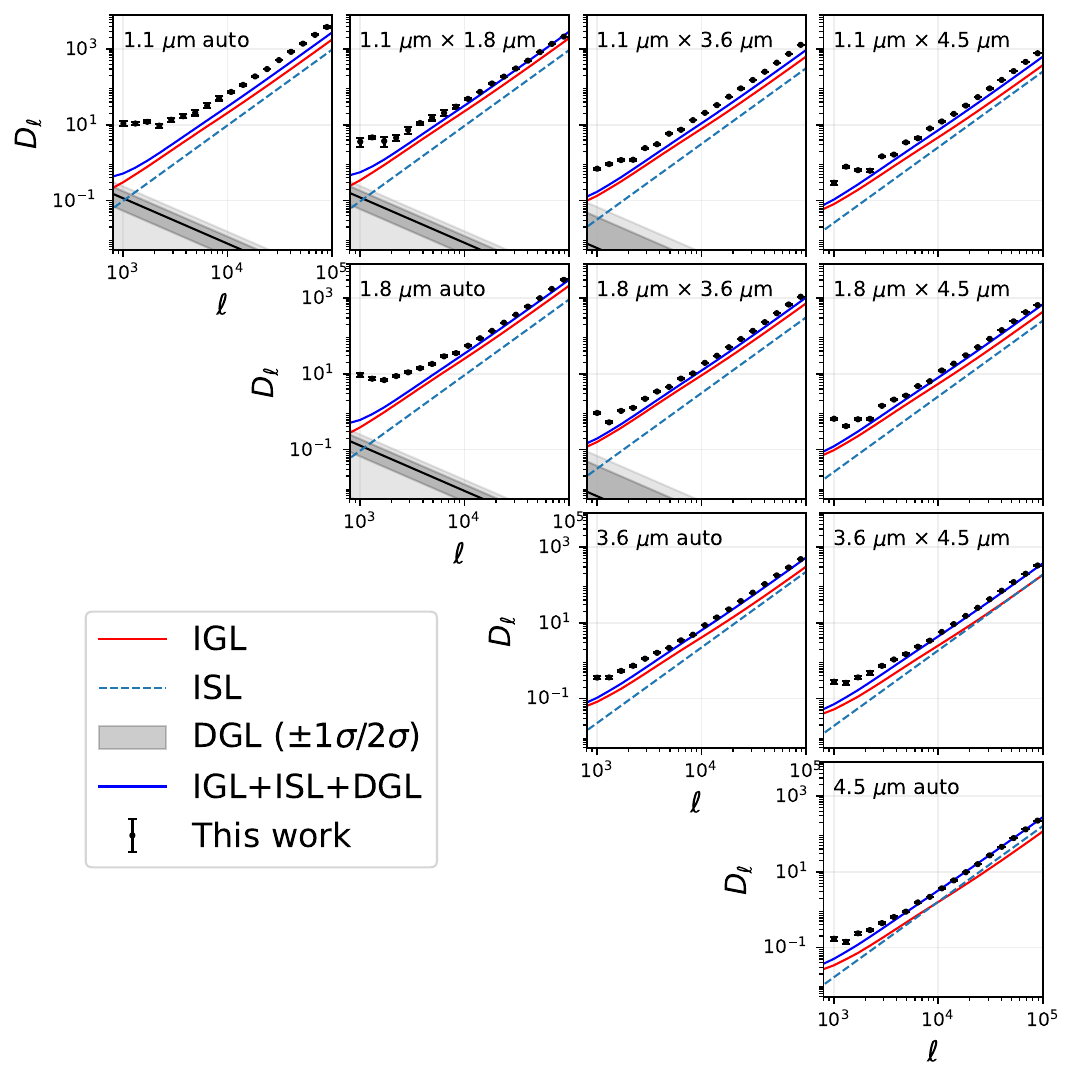}

\caption{\emph{CIBER} and \emph{Spitzer} auto- and cross-power spectra for 1.1 $\mu$m, 1.8 $\mu$m, 3.6 $\mu$m and 4.5 $\mu$m. The \emph{CIBER}-only power spectra are derived from a weighted average of all five \nth{4} flight science fields, while all results with \emph{Spitzer} are derived from the two \bootes\ fields. Following \S \ref{sec:ciber_spitzer} we only present measurements for multipoles above $\ell_{min} = 875$. The different curves indicate the model predictions for ISL (light blue), IGL (red) and DGL (black, with $1\sigma/2\sigma$ uncertainties), and their sum (dark blue). The astronomical mask removes sources with $J<16$, $H<15.5$ or $L<15$.}

\label{fig:ciber_spitzer_supercl}
\end{figure*}

We plot the full set of auto- and cross-power spectra in Fig. \ref{fig:ciber_spitzer_supercl} alongside IGL, ISL and DGL predictions. While the small-scale measurements are largely consistent with expectations from Poisson noise, nearly all power spectra show departures from Poisson noise and model predictions on scales $\ell\leq 5000$ ($\theta > 2\arcmin$), with the exception of the 4.5 $\mu$m auto power spectrum. The departures are largest at \emph{CIBER} wavelengths, where the observed fluctuation power is an order of magnitude higher than IGL at $\ell=1000$. The addition of the $L<15$ masking criterion yields IGL+ISL predictions that slightly underpredict the 1.1 $\mu$m auto- and cross-spectra with 3.6/4.5 $\mu$m.

\begin{figure}
    \centering 
    \includegraphics[width=\linewidth]{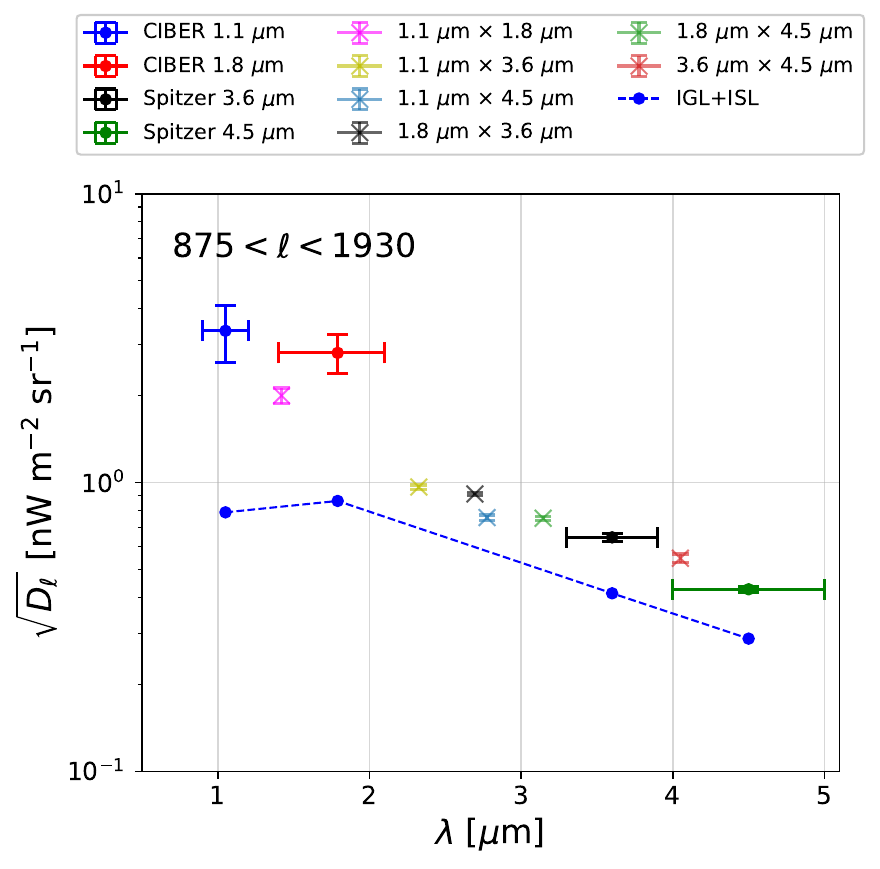}
    \caption{Wavelength dependence of surface brightness fluctuations from \emph{CIBER} and \emph{Spitzer} auto- and cross-spectra for multipoles $875<\ell<1930$. The central wavelengths of the plotted cross-spectrum measurements (crosses) are the means of the pairs of wavelengths from each cross-correlation. We also show the IGL+ISL auto-power spectrum-based predictions (dashed blue), calculated over the same multipole range.}
    \label{fig:ciber_spitzer_excess_fluc}
\end{figure}

In Figure \ref{fig:ciber_spitzer_excess_fluc} we show the wavelength dependence of fluctuations for bandpowers spanning $875<\ell<1930$ ($\theta \approx 5\arcmin-11\arcmin$), measured through auto- and cross-power spectra at 1.1 $\mu$m, 1.8 $\mu$m, 3.6 $\mu$m and 4.5 $\mu$m. Our independent measurements confirm that the amplitude of fluctuations increases sharply toward shorter NIR wavelengths \citep{zemcov14}. The cross-power shows various degrees of coherence between bands and instruments which we expand on shortly. 

\begin{figure}
    \centering 
    \includegraphics[width=\linewidth]{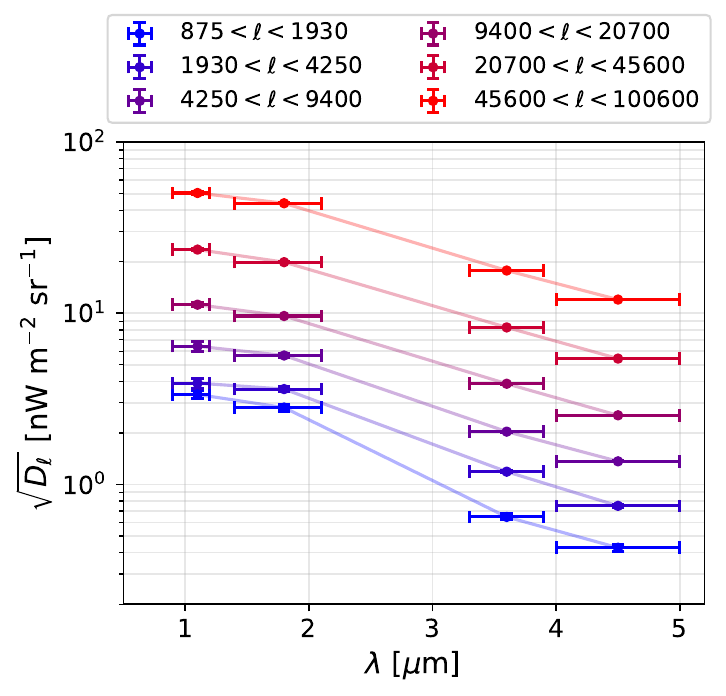}
    \caption{Wavelength dependence of surface brightness fluctuations from \emph{CIBER} and \emph{Spitzer} auto-power spectra, for a set of broad bandpowers spanning $875<\ell<100000$. The horizontal errorbars indicate the effective wavelength range spanned by the \emph{CIBER} and \emph{IRAC} filters. We find that the angular fluctuations on large scales are bluer than the observed and predicted Poisson fluctuations for $\ell < 4000$.}
    \label{fig:ciber_spitzer_fluc_vs_scale}
\end{figure}

In Figure \ref{fig:ciber_spitzer_fluc_vs_scale} we show the same plot but for auto-spectrum measurements in six broad bandpowers between $\ell=875$ and $\ell=100000$. One can see that the color of the fluctuations becomes progressively bluer, going from high to low $\ell$. For our lowest $\ell$ bandpower ($875<\ell<1930$), $\sqrt{D_{\ell}}$ decreases by a factor of eight between 1.1 $\mu$m and 4.5 $\mu$m, compared to a factor of four for the $45600<\ell<100600$ bandpower. Notably, this trend is only across the combined \emph{CIBER} and \emph{Spitzer} wavelengths, i.e., $D_{\ell}^{1.1}/D_{\ell}^{1.8}$ and $D_{\ell}^{3.6}/D_{\ell}^{4.5}$ have much smaller variation across the same bandpowers. A comparison of independent NIR broad-band fluctuation measurements on scales $\theta>100\arcsec$ that includes \emph{AKARI} 2.4 $\mu$m and 3.2 $\mu$m data suggests a Rayleigh-Jeans type spectrum \citep{seo_akari}, however
measurements with higher spectral resolution and coverage in the $2-3$ $\mu$m range are needed for further insight.

\begin{figure}
\centering
    \includegraphics[width=\linewidth]{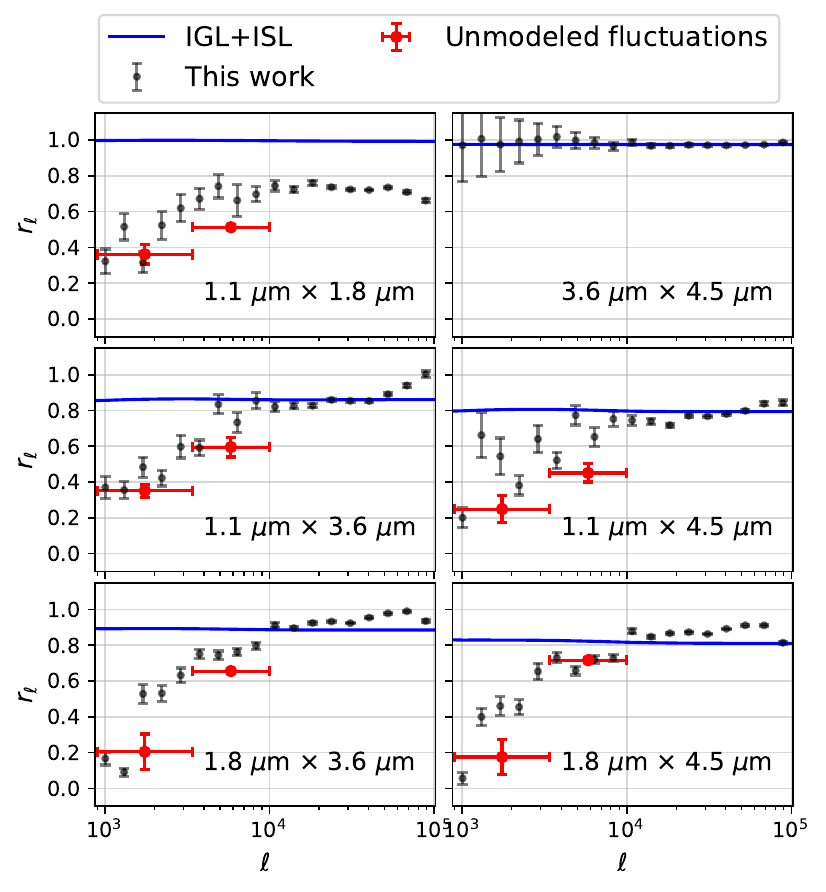}
    \caption{Cross-correlation coefficients of \emph{CIBER} and \emph{Spitzer} band combinations as a function of angular multipole. Black points indicate the field-averaged estimates from the total measured fluctuation power. The solid blue curves indicate predictions from our IGL+ISL model, which have small but non-zero variation in $r_{\ell}$. For all band combinations aside from 3.6 $\mu$m $\times$ 4.5 $\mu$m, we find that fluctuations across wavelengths de-correlate going from $\ell=10000$ to $\ell=1000$. The red points indicate the binned $r_{\ell}$ estimates after subtracting predicted IGL+ISL+DGL contributions from the observed power spectra.}
    \label{fig:ciber_spitzer_corrcoef}
\end{figure}

We examine the coherence of observed fluctuations across wavelengths by calculating the correlation coefficient $r_{\ell}$. In Figure \ref{fig:ciber_spitzer_corrcoef} we plot $r_{\ell}$ for all six band combinations. On small scales ($\ell >10000$), the power spectra are dominated by Poisson fluctuations, and so $r_{\ell}$ is expected to be constant on these scales. We find the observed small-scale cross-correlation coefficients vary by $<10\%$ on these scales, with the exception of a few high-$\ell$ bandpowers. The highest $\ell$ bandpowers are subject to larger systematic errors than the naive reported uncertainties on those scales (see \S \ref{sec:auto_field_consistency}), which may explain some of the $r_{\ell}$ variations for $\ell \gtrsim 50000$. For nearly all band combinations, we find that $r_{\ell}$ departs from the Poisson level, decreasing between $\ell=10000$ and $\ell=1000$. The exception is 3.6 $\mu$m $\times$ 4.5 $\mu$m, for which both the predicted and measured $r_{\ell}$ is near unity. Our IGL+ISL model predicts a small level of variation in $r_{\ell}$, however much smaller than from the observed fluctuations. We note that, while not shown in Fig. \ref{fig:ciber_spitzer_corrcoef}, the 1.1 $\mu$m $\times$ 1.8 $\mu$m correlation coefficient on scales $\ell < 875$ increases, with $\langle r(300<\ell<875)\rangle = 0.61 \pm 0.15$. 

On large scales, the model predictions for IGL, ISL and DGL show a high degree of correlation. We find that, after removing these contributions from the observed auto- and cross-power spectra, the unmodeled fluctuations have slightly lower coherence, shown in coarse bandpowers (red). We find a similar drop in $r_{\ell}$ between high- and low-$\ell$ on intermediate scales, albeit with reduced amplitude.

%% file: sections/modl_interp.tex
\section{Astrophysical modeling and interpretation}

\label{sec:modl_interp}
\subsection{Model predictions}
\label{sec:astro_modl}
\subsubsection{Integrated Galactic Light (IGL)}
\label{sec:igl}

The IGL predictions shown in Figs. \ref{fig:ciber_auto}, \ref{fig:cl_vary_mask}, \ref{fig:ps_result_2MASS_only}, \ref{fig:ciber_cross}, \ref{fig:ciber_cross_vs_mag}, \ref{fig:ciber_spitzer_perfield} and \ref{fig:ciber_spitzer_supercl} will be described in detail in a forthcoming paper (Mirocha et al., in prep.). In brief, these models are semi-empirical: they assume that the properties of dark matter halos are known perfectly (using the \cite{tinker2010} form of the mass function), and parameterize the efficiency of star formation, dust attenuation, and quenching as a function of halo mass and redshift. Then, the free parameters of the model are constrained via MCMC fits to observations of the stellar mass function, the star-forming main sequence, and UV luminosity functions over a broad range of redshifts, $z\sim 0-10$. The SEDs of every galaxy in the model are constructed so as to preserve self-consistency with the calibration datasets, which assume the Bruzual \& Charlot stellar population synthesis models, exponentially-declining star formation histories, fixed solar metallicity, and a Calzetti dust law. With a calibrated galaxy model in hand, EBL fluctuations are generated with a custom halo model framework. The contributions of star-forming and quiescent centrals and satellites are included, under the simplifying assumption that central and satellite galaxies have identical properties at a given (sub)halo mass and redshift.

\subsubsection{Integrated Stellar Light (ISL)}

We model ISL using the TRILEGAL model \citep{trilegal}, which generates synthetic ISL catalogs derived from deep stellar number counts and a model of the galaxy morphological components. TRILEGAL has been shown to reproduce observed counts from 2MASS and deeper stellar catalogs to within 50\% (and typically better) with the exception of fields close to the Galactic plane. On degree angular scales, ISL clustering has been shown to be negligible in the TRILEGAL model, however there are variations in the Poisson level that correlate with Galactic latitude. TRILEGAL captures these variations with catalog realizations unique to each field's sky coordinates. We estimate the ISL Poisson level for each field using ten TRILEGAL realizations per field, predicted for the \emph{CIBER} and \emph{Spitzer} filters. We use the fiducial TRILEGAL model\footnote{\url{http://stev.oapd.inaf.it/cgi-bin/trilegal}}, which assumes a Chabrier log-normal initial mass function, dust extinction following an exponential disc profile with scale height $h_{z,dust}=110$ pc, and a morphology composed of thin disc, thick disc and triaxial bulge components. The TRILEGAL multi-band outputs allow us to generate predictions consistent with the different source masking selections used throughout this work. At our fiducial masking depth, ISL is sub-dominant to IGL but is non-negligible. 

We caution that the IGL and ISL models are derived from separate, model-dependent fits to various photometric datasets and stellar mass functions, i.e., they are not derived self-consistently. This means that there may be some level of systematic error in the combined IGL+ISL predictions, which can be exacerbated by issues such as star-galaxy separation. We do not attempt to quantify such effects, however they may become more relevant in modeling higher-sensitivity measurements.

\subsection{Astrophysical interpretation}

As shown in \S \ref{sec:super_cl} and the preceding sections, our fiducial measurements on small scales are consistent with Poisson predictions from IGL and ISL. However, the IGL+ISL+DGL model is in poor agreement with both the observed \emph{CIBER} and \emph{Spitzer} fluctuations on scales $\theta \gtrsim 2\arcmin$. We consider a number of potential astrophysical explanations for the unmodeled fluctuations from \emph{CIBER} and \emph{Spitzer}. 

\subsubsection{Zodiacal light}
Zodiacal light is expected to be spatially smooth on sub-degree scales \citep{arendt16}. We find no significant correlations between large-angle fluctuation power and the ZL monopole intensities across the five science fields. We do find a correlation coefficient between 1.1 $\mu$m and 1.8 $\mu$m that is less than one, which disfavors a single correlated foreground. ZL is also disfavored due to the coherence of fluctuations between \emph{CIBER} and \emph{Spitzer} cross-correlations, which observe the same sky at different times of year. 


\subsubsection{Diffuse galactic light}
DGL has been invoked in previous studies as an explanation for the low-$\ell$ signal observed by \emph{CIBER} and other optical/NIR measurements \citep{mitchellwynne}. In this work, we detect cross-power between \emph{CIBER} and IRAS-derived CSFD maps below $\ell < 2500$ at 5.3$\sigma$ and 4.8$\sigma$ for 1.1 $\mu$m and 1.8 $\mu$m, respectively, recovering $\nu b_{\nu}$ estimates consistent with existing studies. Our results suggest that scattered DGL contributes $\sim 5\%$ to the observed large-angle \emph{CIBER} fluctuation power and $<10\%$ at 95\% confidence. We also find DGL to be negligible for the \emph{Spitzer} measurements. As with ZL, the \emph{CIBER} per-field bandpowers are uncorrelated with the DGL monopole intensity, disfavoring a potential secondary DGL component that is correlated with ISM dust grains but not seen by \emph{IRAS}. Thermal dust emission is disfavored due to the short wavelengths observed by \emph{CIBER}. For example, blackbodies that peak at 1.1 $\mu$m and 1.8 $\mu$m would have implied dust temperatures of $2000-3000$ K, surpassing the sublimation temperature for many types of common grains.

\subsubsection{Non-linear galaxy clustering}
The measurements of this work are sensitive to scales with contributions from linear and non-linear galaxy clustering. In physical units, the \emph{CIBER} and \emph{Spitzer} measurements probe galaxy clustering on $\sim$ few tens of Mpc scales for $\ell_{min}=300$ and go well into the non-linear regime. The IGL predictions in this work (described in \S \ref{sec:igl}) incorporate non-linear clustering through a halo modeling framework that includes both centrals and satellites but assumes that satellite galaxy properties are identical to centrals at fixed redshift and halo mass. Relaxing these assumptions may affect the predicted one-halo and two-halo predictions, however we do not explore such IGL model variations in this work. Nonetheless, our IGL predictions grossly underestimate the \emph{CIBER} fluctuation power (by 1.5 dex at $\ell=1000$), such that a significant modification to the IGL model would be needed to fully explain the discrepancy. The IGL model is also inconsistent with the \emph{Spitzer} auto- and cross-power spectrum measurements, underpredicting the observed power by a factor of $\gtrsim 2$ at $\ell=1000$.

\subsubsection{IHL}
Intra-halo (and inter-halo) light (IHL) are predicted to contribute power to angular fluctuations on scales relevant to this analysis. A basic IHL interpretation of the \emph{CIBER} and \emph{Spitzer} measurements would imply that the color of IHL fluctuations (integrated over the line of sight) is bluer than that of IGL. However, further interpretation is complicated due to a variety of factors including: the challenges of accurately estimating the accreted stellar component of massive galaxies \citep{sanderson18}; the multiple conventions for what constitutes ``IHL" \citep{cooper10, pillepich14, elias18, proctor23}; the paucity of extended light measurements, and the complexity of astrophysics that potentially generates and distributes IHL. In \cite{cheng22}, the modeled IGL+IHL power spectrum toward large angular scales converges to the two-halo IGL power spectrum, scaled by $\left(\frac{1}{1-f_{IHL}}\right)^2$. However, the IHL mass and redshift dependence, along with the assumed DM halo profile and IGL clustering, need to be modeled consistently in order to robustly infer the IHL abundance, and to probe the mechanisms that produce/distribute IHL on few-arcminute to degree scales.



Recently, the kinetic Sunyaev-Zel'dovich (kSZ) effect was detected at 13$\sigma$ through stacking using ACT and DESI photometric galaxies \citep{boryana_ksz}. This work found evidence for large baryonic feedback on scales $R<4\arcmin$, excluding models where gas traces the dark matter distribution at 40$\sigma$ but in tantalizing agreement with the original ``high-feedback" version of the Illustris hydrodynamical simulations. Further investigation is needed to assess whether the feedback processes driving gas out of galaxies are connected to processes that would source IHL on similar scales.

\subsubsection{Annihilating/decaying dark matter} Annihilating and/or decaying dark matter scenarios, such as from axion-like particles \citep[ALPs;][]{alp} have been proposed to explain absolute and fluctuation-based measurements in the NIR. These models typically assume that the DM is comprised of a mono-energetic species, for which the momentum dispersion is small relative to their mass, resulting in distinct spectral line signatures. The measurement of common large-angle fluctuations in \emph{CIBER} and \emph{Spitzer} cross-power spectra disfavors such models as an explanation for the observed fluctuations. At optical wavelengths, \cite{wang_23}
searched for DM line signals in the outskirts of \emph{DESI}-observed galaxies and obtained upper limits constraining the putative line surface brightness to be two orders of magnitude lower than that of the measured EBL. ALP models that predict a much broader energy spectrum would in principle alleviate these discrepancies, however the viability of such models is unclear.

\subsubsection{Direct collapse black holes (DCBHs)} DCBHs formed at $z\geq 13$ have been hypothesized to produce emission that can redshift into the NIR, depending on their redshift distribution. The cumulative emission from Compton-thick BHs is predicted to sharply peak near observed wavelengths $\lambda\sim 2$ $\mu$m \citep{yue13}. While the partial coherence between \emph{Spitzer} and X-ray observations admits a DCBH interpretation \citep{cappelluti_17}, the model is inconsistent with the observed amplitude and correlation of the \emph{CIBER} fluctuations, without modification to the photoevaporation conditions that suppress DCBH formation after $z\sim 13$. The coherence of the \emph{CIBER} and \emph{Spitzer} fluctuations measured in this work further limits the portion of \emph{Spitzer} fluctuations that can be explained by DCBHs. 



%% file: sections/conclusion.tex
\section{Conclusion}
\label{sec:conclusion}

In this work we present new fluctuation measurements in the NIR using imaging data from \emph{CIBER}'s fourth and final flight. Building on the analysis of \cite{zemcov14}, we address two important systematics necessary for unbiased recovery of sky fluctuations, namely errors in source masking and correction of FF errors. By using estimates of the FF from stacking the science fields, we bypass the use of field differences used in previous analysis and determine that after FF bias correction, the remaining FF errors increase our recovered power spectrum uncertainties on scales $500<\ell<2000$ by less than 20\% (presented in Paper I). 

With these improvements in combination with the higher quality fourth flight data, we arrive at our main conclusions:
\begin{itemize}
    \item We detect large-angle ($\theta > 5\arcmin$) NIR fluctuations at 1.1 $\mu$m and 1.8 $\mu$m auto-power spectra at 14.2$\sigma$ and 18.1$\sigma$, respectively, and in 1.1 $\mu$m $\times$ 1.8 $\mu$m cross-power spectra at 9.9$\sigma$, representing a five- to ten-fold increase sensitivity at these wavelengths on several arcminute to degree scales.
    \item We confirm at high significance that scattered DGL, as traced by the CSFD extinction maps from \cite{csfd}, explains $\lesssim 5\%$ of the observed \emph{CIBER} fluctuations on large scales. 
    \item We measure departures from Poisson fluctuations in \emph{Spitzer} auto-power and \emph{CIBER} $\times$ \emph{Spitzer} cross-power spectra for multipoles $875 < \ell < 5000$. A characterization of the SDWFS mosaicing process is needed to improve measurements on large angular scales, through estimation of the mosaic transfer function and mitigation of associated systematics.
    \item The auto-power across the \emph{CIBER} and \emph{Spitzer} measurements on large scales is bluer than the Poisson noise from stars and galaxies, however the observed de-correlation at low-$\ell$ disfavors explanations that invoke a single correlated sky component.
\end{itemize}

In this work we probe mask-signal correlations empirically by varying the size of masks around point sources. We find perturbing the source masking radius function by $\pm 25\%$ about our fiducial case has a negligible impact on the large-scale clustering, however reverting to the masking function used in \zem\ has a more significant impact on scales $1000 < \ell < 10000$. In this limit, interpretation of models that describe the extended light distribution of galaxies will require a careful treatment of such correlations, along with precise mitigation of extended PSF and instrumental effects. 

Beyond the auto- and cross-power spectra presented in this work, another tool for deciphering EBL fluctuations is cross-correlation with tracers of the matter distribution. Cross-correlations of \emph{CIBER} maps with photometric galaxy samples offer a more direct probe of EBL fluctuations correlated with linear and non-linear galaxy clustering, and can disentangle the evolution of fluctuations with redshift information, i.e., through EBL redshift tomography \citep{ykc_uv, ykc_tsz}. Furthermore, targeted cross-correlations that select on galaxy properties, such as shape and color, offer a path to constraining the environmental dependencies of IHL as predicted by hydrodynamical simulations \citep[e.g.,][]{proctor23}. With a larger effective area, cross-correlations with CMB lensing will be feasible and would provide a strong test on the nature of the large-angle fluctuations measured in this work.

Existing and near-future experiments will map the NIR EBL over the full sky with significantly improved sensitivity, spectral coverage and resolving power. In this analysis we use data from five science fields of varying galactic and ecliptic latitude covering an area of 20 deg$^2$, with an effective total exposure time of 220 seconds. \emph{CIBER}-2, the second generation of \emph{CIBER}, has three H2RG detectors and six windowpane filters for imaging at $0.5-2.5$ $\mu$m \citep{shirahata_16, nguyen_16, takimoto2020, matsuura24}. \emph{CIBER}-2 has flown three times from White Sands, New Mexico in June 2021, June 2023 and May 2024. Along with increased sensitivity and spectral coverage extending to optical wavelengths, \emph{CIBER}-2 maps enable multi-wavelength cross-correlations that can help further disentangle the colors of local and extragalactic astrophysical components. Furthermore, \emph{CIBER}-2's coverage of the COSMOS field in the third flight will enable a host of galaxy cross-correlations with deep NIR catalogs.

While we focus in this work on fluctuation measurements in the NIR, measurements at optical wavelengths provide complementary information that may inform the interpretation of NIR fluctuation anisotropies. The \emph{VERTECS} experiment (Visible Extragalactic background RadiaTion Exploration by CubeSat) is set to launch in the summer of 2025 and is designed to pursue optical EBL fluctuation measurements over a $6^{\circ} \times 6^{\circ}$ instantaneous FOV \citep{sano24_vertecs, takimoto24_vertecs}.

The Spectro-Photometer for the History of the Universe, Epoch of Reionization, and Ices Explorer (\emph{SPHEREx}) is the next NASA Medium Class Explorer mission which is planned for launch in early 2025 \citep{dore14, korngut_spherex}. \emph{SPHEREx} will conduct a two-year, all-sky survey in 102 bands spanning 0.75-5 $\mu$m, dramatically increasing the volume of data available for intensity mapping \citep{crill2020_spie}. The regions with highest surface brightness sensitivity will be two 100 deg$^2$ regions covering the North and South Ecliptic Poles. Deep field mosaics with \emph{SPHEREx} will be orders of magnitude more sensitive on several arcminute angular scales due to significantly lower $1/f$ noise \citep{Heaton_2023, Nguyen_2025}, lower ZL levels and a daily cadence over the poles throughout the nominal two-year survey. These maps are expected to be largely photon noise-dominated, which may ameliorate some of the challenges faced in this analysis. Both external and \emph{SPHEREx}-internal galaxy redshift catalogs \citep{feder23} will enable tomographic analyses of NIR EBL anisotropies \citep{cheng22_cnib_tomography}. Furthermore, \emph{SPHEREx} intensity maps covering the full sky will have the sensitivity to characterize the dependence of fluctuations on foregrounds such as those correlated with the galactic plane and the solar system. 

If the excess large-angle intensity fluctuations detected in this work are indeed astrophysical in origin, upcoming datasets will measure the same signal with exquisite sensitivity. In this limit, stringent control of survey systematics and robust astrophysical predictions will drive progress towards a consensus view in the NIR of our local neighborhood and the extragalactic background light.   

%% file: sections/acknowledgements.tex
\acknowledgements
We thank the CIBER-1 instrument team for collecting the CIBER imager data used in this work. We also thank Yi-Kuan Chiang for help integrating the CSFD extinction maps into this analysis, as well as Tzu-Ching Chang, Olivier Doré, Ari Cukierman and Brandon Hensley for helpful discussions. 

This material is based upon work supported by
the National Aeronautics and Space Administra-
tion under APRA research grants NNX10AE12G,
NNX16AJ69G, 80NSSC20K0595, 80NSSC22K0355, and
80NSSC22K1512. J.M. was supported by an appointment to the NASA Postdoctoral Program at the Jet Propulsion Laboratory/California Institute of Technology, administered by Oak Ridge Associated Universities under contract with NASA.

This publication makes use of data products from the Two Micron All Sky Survey, which is a joint project of the University of Massachusetts and the Infrared Processing and Analysis Center/California Institute of Technology, funded by the National Aeronautics and Space Administration and the National Science Foundation.

The Pan-STARRS1 Surveys (PS1) and the PS1 public science archive have been made possible through contributions by the Institute for Astronomy, the University of Hawaii, the Pan-STARRS Project Office, the Max-Planck Society and its participating institutes, the Max Planck Institute for Astronomy, Heidelberg and the Max Planck Institute for Extraterrestrial Physics, Garching, The Johns Hopkins University, Durham University, the University of Edinburgh, the Queen's University Belfast, the Harvard-Smithsonian Center for Astrophysics, the Las Cumbres Observatory Global Telescope Network Incorporated, the National Central University of Taiwan, the Space Telescope Science Institute, the National Aeronautics and Space Administration under Grant No. NNX08AR22G issued through the Planetary Science Division of the NASA Science Mission Directorate, the National Science Foundation grant No. AST-1238877, the University of Maryland, Eotvos Lorand University (ELTE), the Los Alamos National Laboratory, and the Gordon and Betty Moore Foundation.

\software{\texttt{matplotlib}, \texttt{numpy}, \texttt{scipy}, \texttt{sklearn}, \texttt{pyfftw}}

%% file: sections/field_diff.tex
\section{CIBER field difference auto spectra}
\label{sec:field_diff}

To assess the level of residual FF error in our fiducial auto-power spectra, we calculate an alternative estimate from field differences (the fiducial method from \zem). As FF errors primarily couple to the mean intensity of each map, such errors cancel out in a field difference if the fields have similar sky brightnesses. We perform field differences on the two \bootes\ fields from the fourth flight as these have mean sky brightnesses with $\Delta I/I \sim 5\%$, corresponding to a relative power spectrum bias $\propto (\Delta I/I)^2$. We construct a union mask from the individual fields, leading to lower unmasked fractions of $\sim 50\%$ and $\sim 40\%$ for 1.1 $\mu$m and 1.8 $\mu$m, respectively. We bypass the need for a FF noise bias correction and apply mode coupling corrections that only depend on the mask and map filtering.

\begin{figure}
    \centering
    \includegraphics[width=0.48\linewidth]{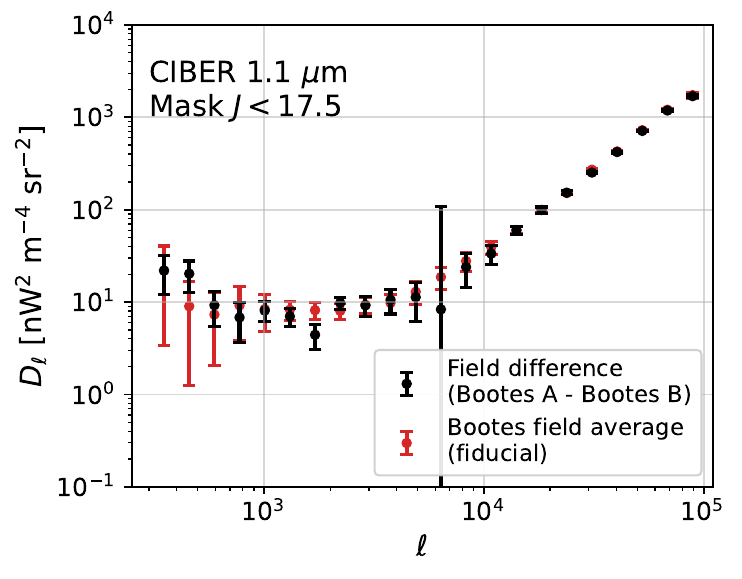}
    \includegraphics[width=0.48\linewidth]{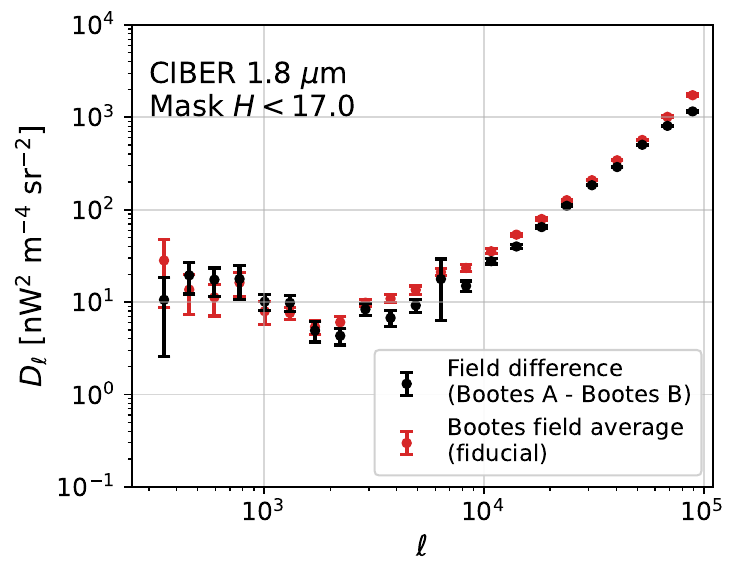}
    \caption{Comparison of auto-power spectrum measurements for the two \bootes\ fields using our fiducial method (red) and the method of field differences (black), for which FF errors are mitigated. We find broad consistency between methods on large and small angular scales, validating our treatment of FF errors.}
    \label{fig:field_differences}
\end{figure}

In Figure \ref{fig:field_differences} we show the results of this comparison. We find strong agreement between methods on large scales for both bands. On small scales, the 1.8 $\mu$m difference spectra are slightly lower than the fiducial results, which may indicate a small error in the noise bias subtraction. Due to the more aggressive union mask, mode coupling is more severe for the field differences, leading to degraded sensitivity on intermediate scales. Nonetheless, this comparison validates that residual FF errors do not have a significant impact on the measured large-angle fluctuation power. 

%% file: sections/halfexp_crosscorr.tex
\section{Half-exposure cross-correlations}
\label{sec:halfexp_cross}
As a complementary consistency check on the CIBER auto-power spectra, we compute the cross-spectra of half-exposure pairs used in \S \ref{sec:noisemodl_validation} with the exception of elat30, for which read noise is prohibitively large at half-exposure length. This cross-correlation test can help identify the contribution of transient phenomena, such as cosmic ray events, that inject flux in the full slope fits but are uncorrelated in time. The noise in each half-exposure is uncorrelated, however instrument noise in $\hat{FF}$ contributes coherently to each FF-corrected half-exposure. We estimate and correct for this FF noise bias in a similar fashion to the fiducial measurements. 

\begin{figure*}
    \centering 
    \includegraphics[width=0.48\linewidth]{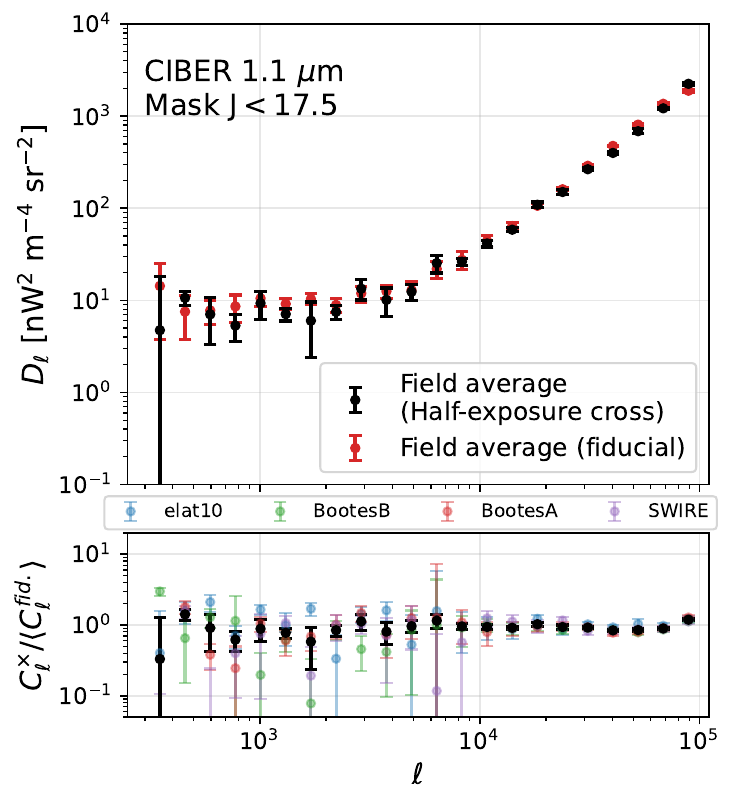}
    \includegraphics[width=0.48\linewidth]{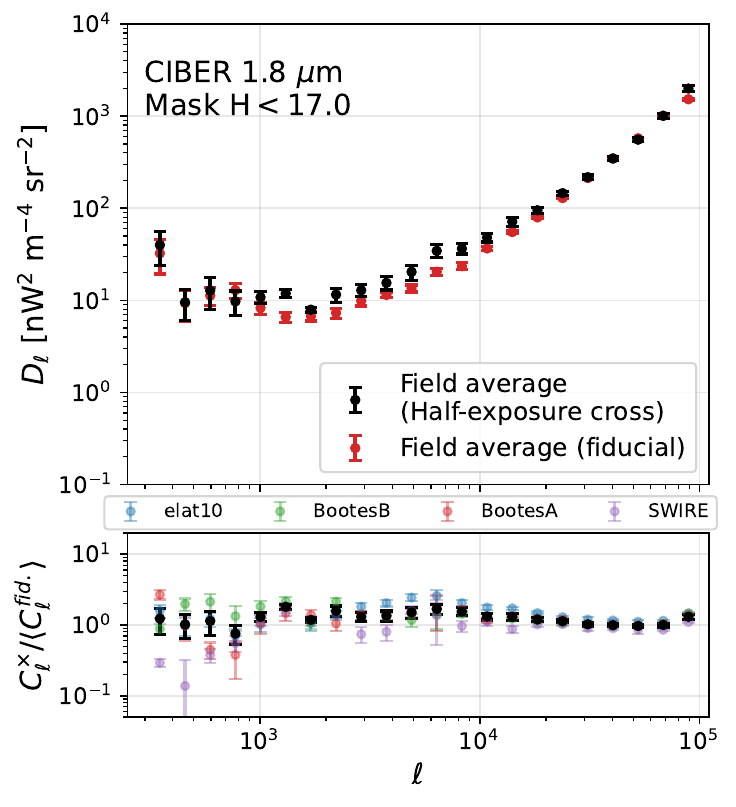}
    \caption{Comparison of auto-power spectrum measurements using the fiducial method (red) and the method of half-exposure cross-correlations (black). We use the same field weights for both sets of power spectra when calculating the field average. In the bottom panels we plot the ratio of half-exposure cross-power spectra and the fiducial, field-averaged power spectrum.}
    \label{fig:halfexp_crosscorr}
\end{figure*}

In Figure \ref{fig:halfexp_crosscorr} we show the resulting cross-spectra compared with our fiducial auto spectrum measurements. The cross-spectra across the four fields are internally consistent with the exception of a few bandpowers. Furthermore, for both 1.1 and 1.8 $\mu$m, the weighted averages from the four fields are consistent with our fiducial auto spectra. The agreement on scales $\ell > 10000$ is a strong indication that transient phenomena do not contribute significantly to our measurements and do not explain the discrepancy between our measurements and model predictions. The discrepancy between methods on scales $2000<\ell<10000$ for 1.8 $\mu$m may be driven by effects in the read noise-dominated regime that are not captured by our model at half-exposure length, such as non-stationarity and larger FF errors. 

\subsection{``Shift" cross-correlations}
We further utilize the half-exposure pairs by computing shifted cross-correlations, in which we shift one half-exposure from the other by integer pixel offsets ($\Delta x, \Delta y$) and re-calculate the cross-spectra. On small scales, we define the power spectrum de-correlation response $\mathcal{R}_{\ell}$ as 
\begin{equation}
    \mathcal{R}_{\ell}(\Delta x, \Delta y) = \frac{C_{\ell}^{\times}(\Delta x, \Delta y)}{C_{\ell}^{\times}(\Delta x=0, \Delta y=0)}.
\end{equation}
The shape of $\mathcal{R}_{\ell}$ is sensitive to both Poisson signal and instrument noise. By estimating the same shift cross-correlations from mock data, we can assess whether the observed response is consistent with unmasked point sources or another noise component. 

\begin{figure}
    \centering
    \includegraphics[width=0.85\linewidth]{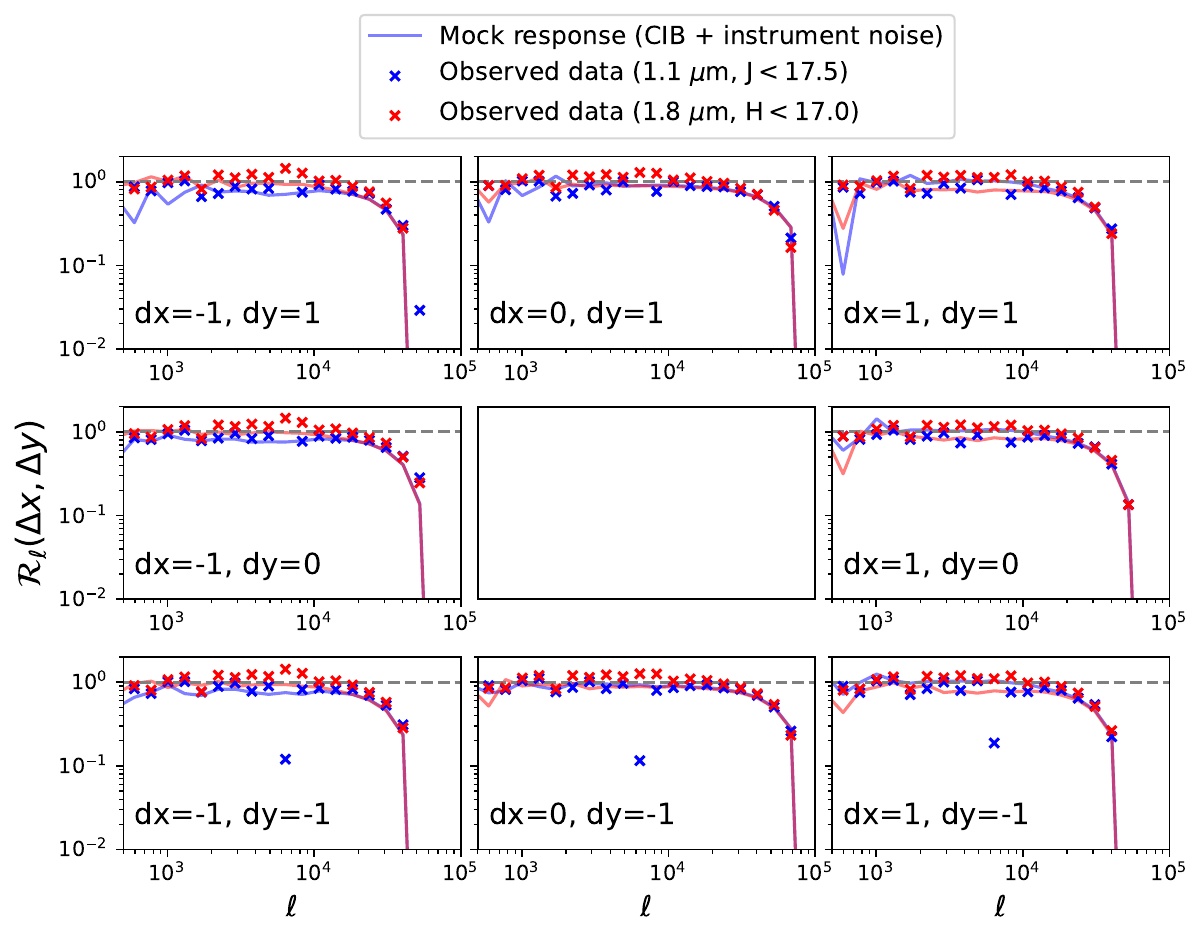}

    \caption{Fractional response of half-exposure cross-power spectra to integer pixel shifts between maps, evaluated on mocks with point sources and noise (solid curves) and on observed data (crosses). The agreement between observed and mock responses support the conclusion that the observed small-scale power is consistent with unmasked sources.}
    \label{fig:shift_cl}
\end{figure}

In Figure \ref{fig:shift_cl} we show $\mathcal{R}_{\ell}$ for pixel shifts $(\Delta x, \Delta y) \in [-1, 0, 1]$. In all cases we recompute masks as the union of each shifted and unshifted mask. We estimate $\mathcal{R}_{\ell}$ from a set of fifty point source mocks that are combined with 500 noise realizations. The response of noiseless point sources rolls off according to the shape of the beam, and has the largest impact on the mock response. While photon noise is uncorrelated, we find that the read noise introduces slightly correlated behavior. In particular, for both bands we find that shifts in the $\pm y$ direction (i.e. the read direction in each array) lead to slightly higher $\mathcal{R}_{\ell}$ than those in the $x$ direction.

We find that the observed response agrees closely with the mock predictions and exhibits the same asymmetric response. Because the 1.1 $\mu$m and 1.8 $\mu$m band detectors are clocked by 90 degrees with respect to each other, pointing variation between half-integrations would lead to orthogonal responses in detector coordinates across the two bands, a scenario inconsistent with the observed response. While in \S \ref{sec:ciber_results} we consider several potential explanations for the discrepancy between the auto-power spectrum measurements and IGL+ISL predictions, these results provide a strong indication that the observed small-scale PS signal is dominated by unmasked point sources. 

%% file: sections/image_filter.tex
\section{Choice of image filtering}
\label{sec:filterchoice}

\begin{figure}
    \centering
    \includegraphics[width=0.49\linewidth]{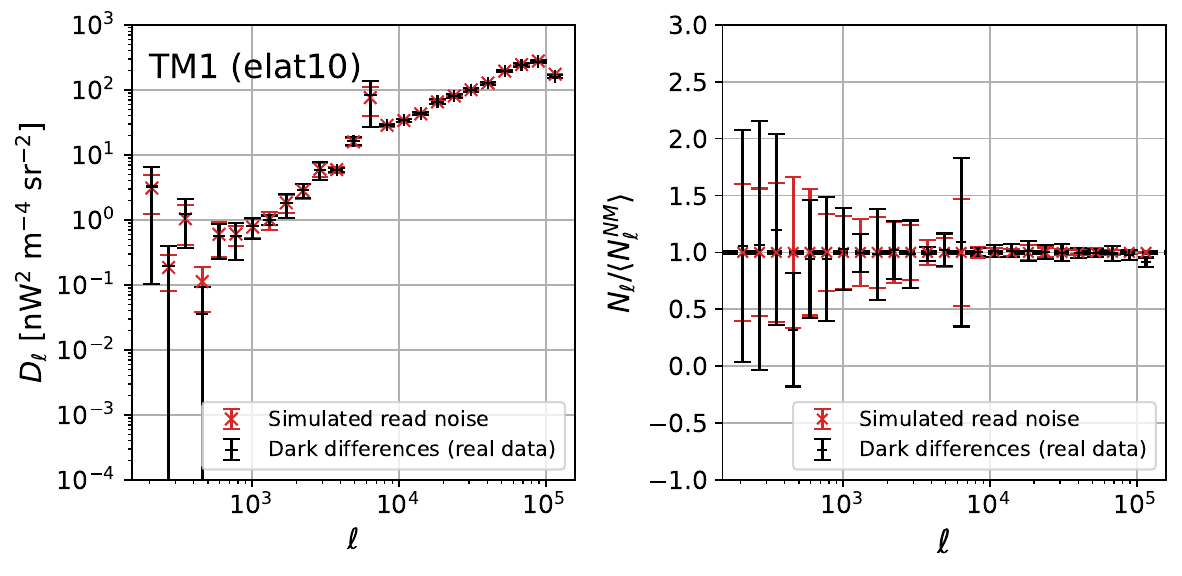}
    \includegraphics[width=0.49\linewidth]{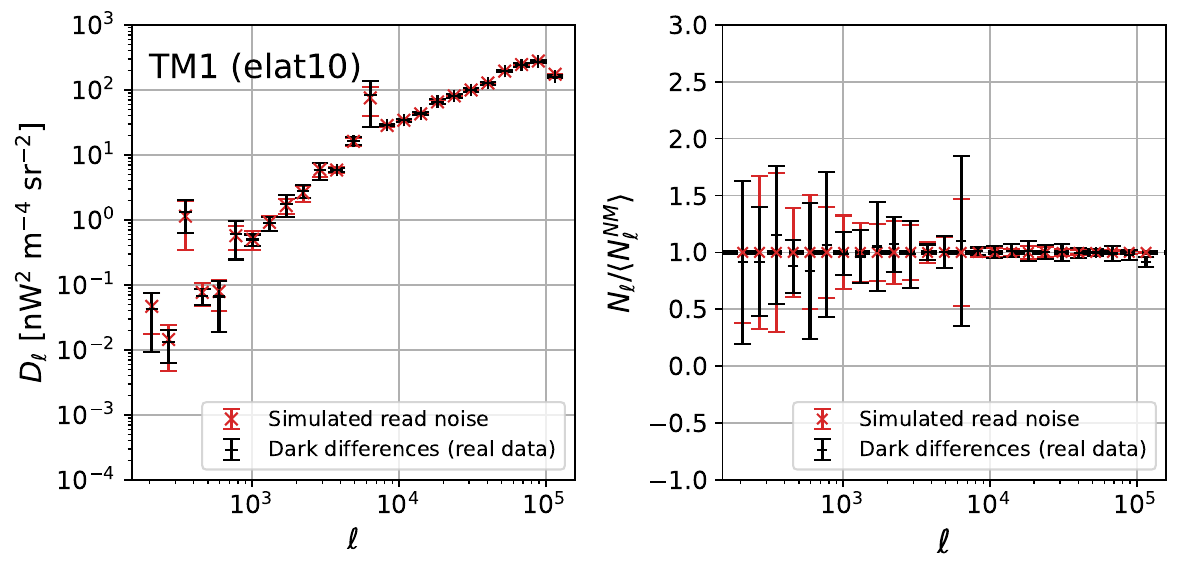}
    \includegraphics[width=0.49\linewidth]{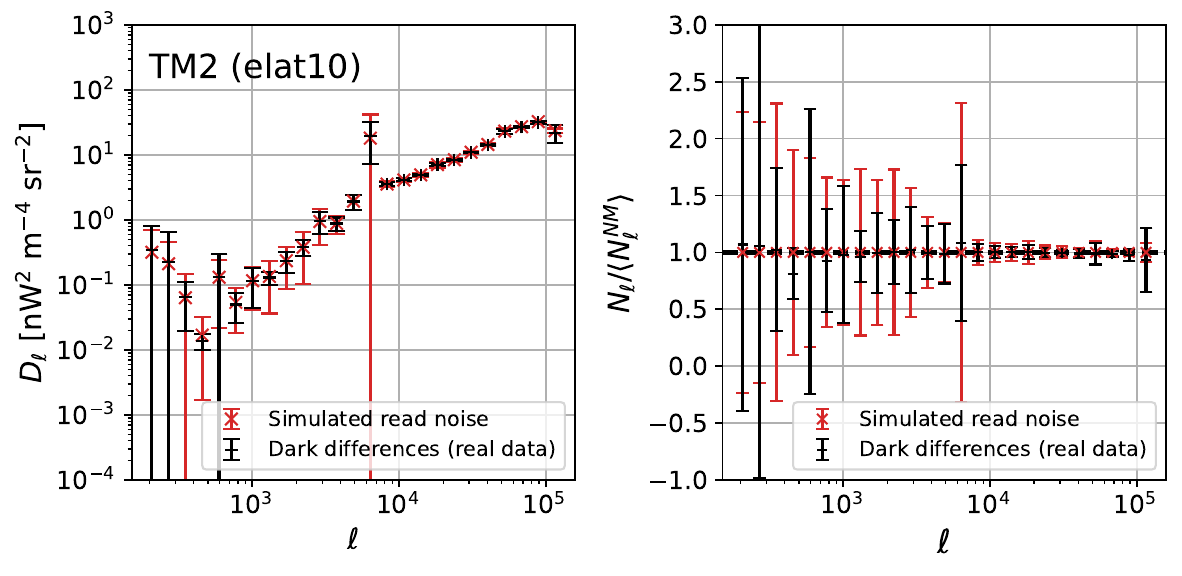}
    \includegraphics[width=0.49\linewidth]{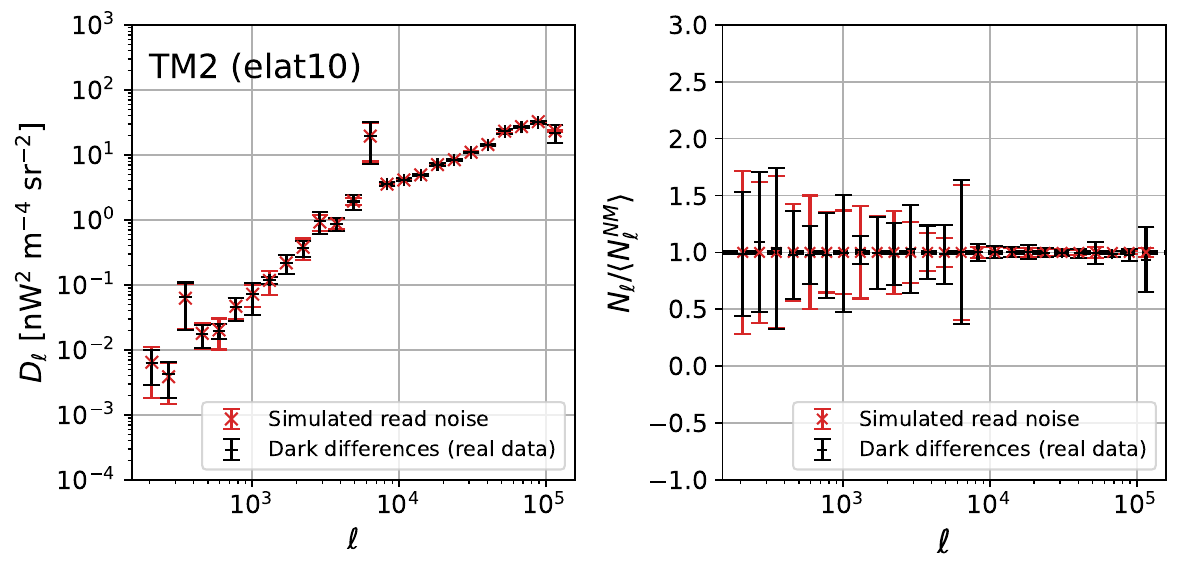}
    \caption{Validation of read noise models for \emph{CIBER} imagers with full array (left) and with per-quadrant (right) mean subtraction. Each set of results employ gradient filtering over the full arrays. The left panels of each set of plots show the mean unweighted power spectrum of exposure differences taken before flight (black) and Gaussian read noise model realizations (red). Each right-hand panel shows the dispersion of power spectra, normalized to $\langle N_{\ell}^{model}\rangle$. We plot the noise power spectra in sky units, however without correction for the beam transfer function $B_{\ell}$.}
    \label{fig:darkdiff_read_TM1_diff_filter}
\end{figure}

In this section we compare the derived read noise models against dark data with full array and per-quadrant offset fitting applied. The results from these two filtering configurations are shown in the left (full array) and right (per-quadrant) sets of panels in Fig. \ref{fig:darkdiff_read_TM1_diff_filter}. Due to the quadrant-specific electrical effects discussed in \S \ref{Sec:ps_estimation}, the read noise model derived from exposure differences with full array offset fitting struggle to reproduce the mean and variance of the differences. By applying per-quadrant offset fitting to the dark exposures, we reduce variance across realizations considerably for $\ell<1000$ and obtain a read noise model that more accurately captures the noise statistics of the dark differences, with the exception of one bandpower centered at $\ell=1300$.




%% file: sections/dgl_zl_mono_bpcorr.tex
\section{Correlation of large-angle CIBER fluctuations with ZL and DGL intensity monopoles}
\label{sec:dgl_zl_monopole_bpcorr}

\begin{figure*}
    \centering
    \includegraphics[width=0.7\linewidth]{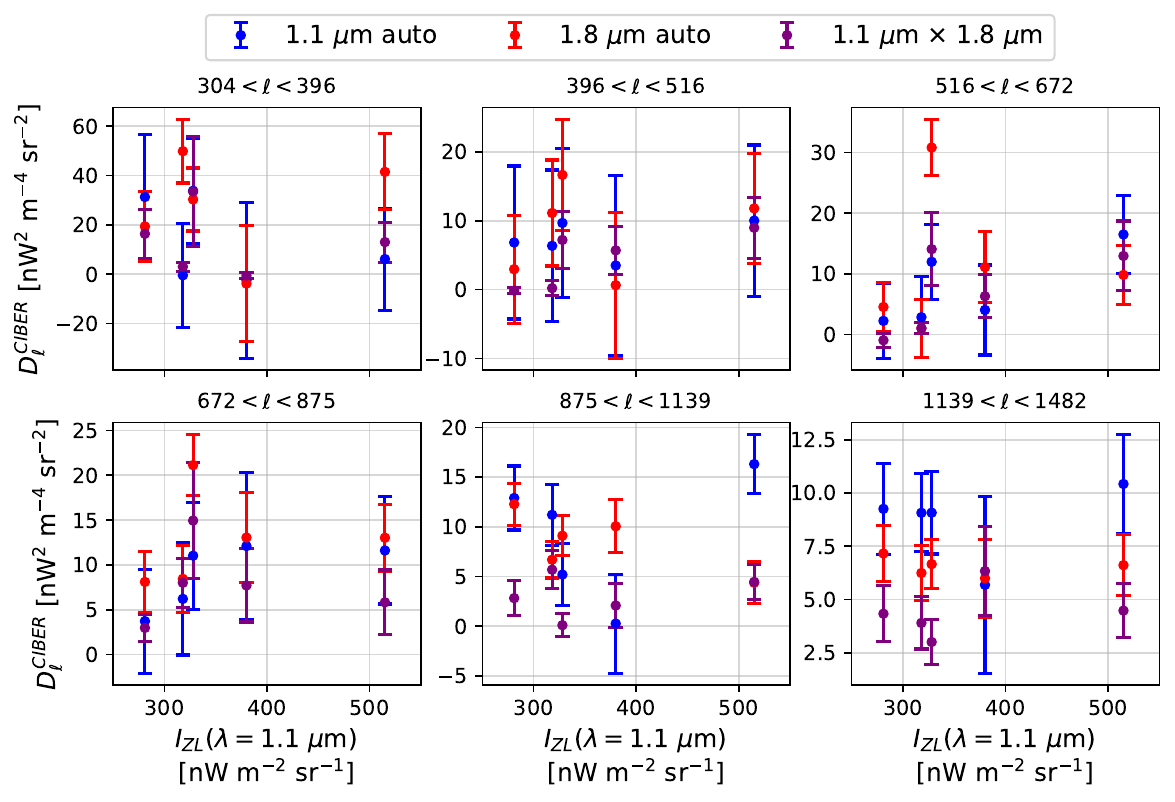}
    \includegraphics[width=0.7\linewidth]{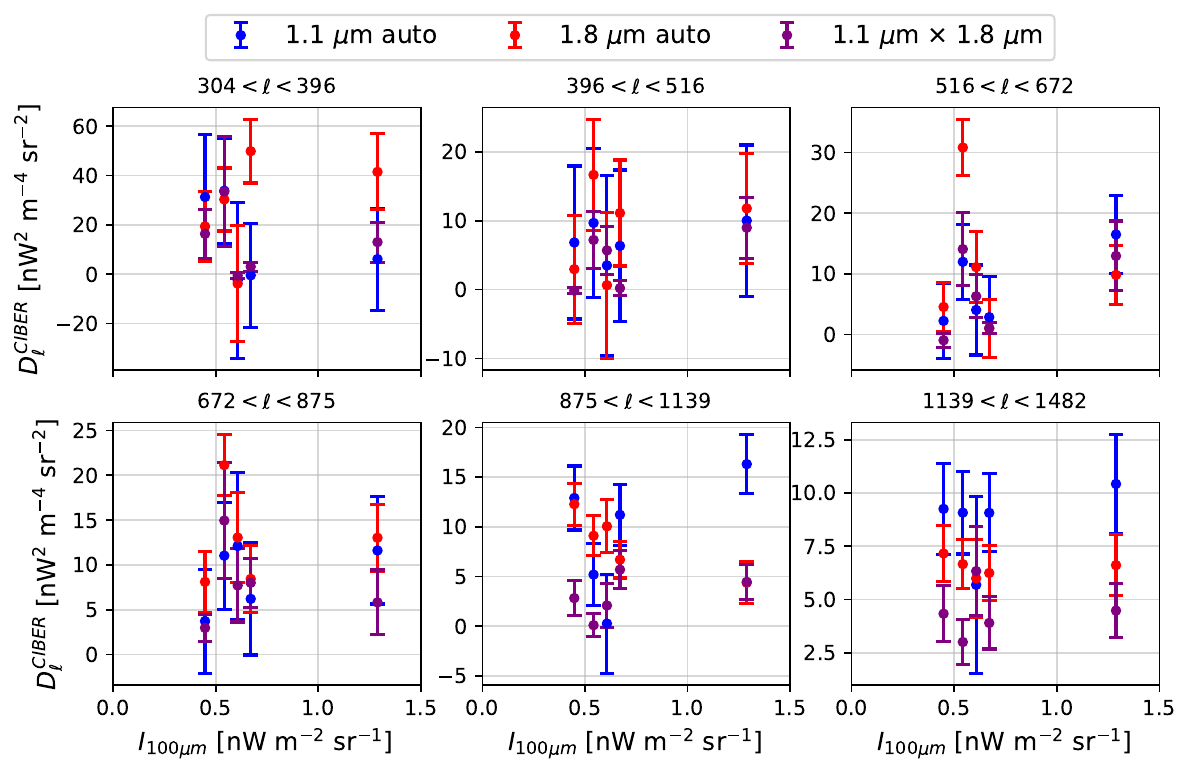}
    \caption{Correlation of \emph{CIBER} 1.1 $\mu$m (blue), 1.8 $\mu$m (red) and 1.1 $\mu$m $\times$ 1.8 $\mu$m (purple) power spectrum bandpowers with respect to ZL intensity from \cite{kelsall} (top panel) and DGL intensity (bottom). The lack of correlation rules out the presence of additional foreground fluctuations that are correlated with ZL and DGL but are not captured by existing tracer maps/models.}
    \label{fig:dgl_zl_monopole_bpcorr}
\end{figure*}

While the ZL fluctuation power on large scales is expected to be small and scattered DGL is measured to be small from cross-correlation, it is possible that additional components of either foreground contribute power to the \emph{CIBER} auto- and cross-spectrum measurements. Assuming that additional foregrounds may correlate with the overall dust content on solar system and Galactic scales, we check for correlations between our measured large-angle bandpowers ($\ell < 1500$) and the ZL and DGL intensity monopoles. In Figure \ref{fig:dgl_zl_monopole_bpcorr} we show the results of this test by ordering the five \emph{CIBER} science fields in $I_{ZL}(\lambda=1.1$ $\mu$m$)$ and $I(\lambda=100$ $\mu$m$)$, for both auto- and cross-spectra. Across the lowest six bandpowers there is no clear evidence for correlation with fluctuation power, reinforcing the finding that ZL and DGL are unlikely to source the observed \emph{CIBER} large-angle fluctuations.

%% file: sections/spitzer_mopex_selfcal.tex
\section{Comparison of SDWFS mosaics}
\label{sec:sdwfs_mosaic_app}
\begin{figure}
    \centering
\includegraphics[width=0.9\linewidth]{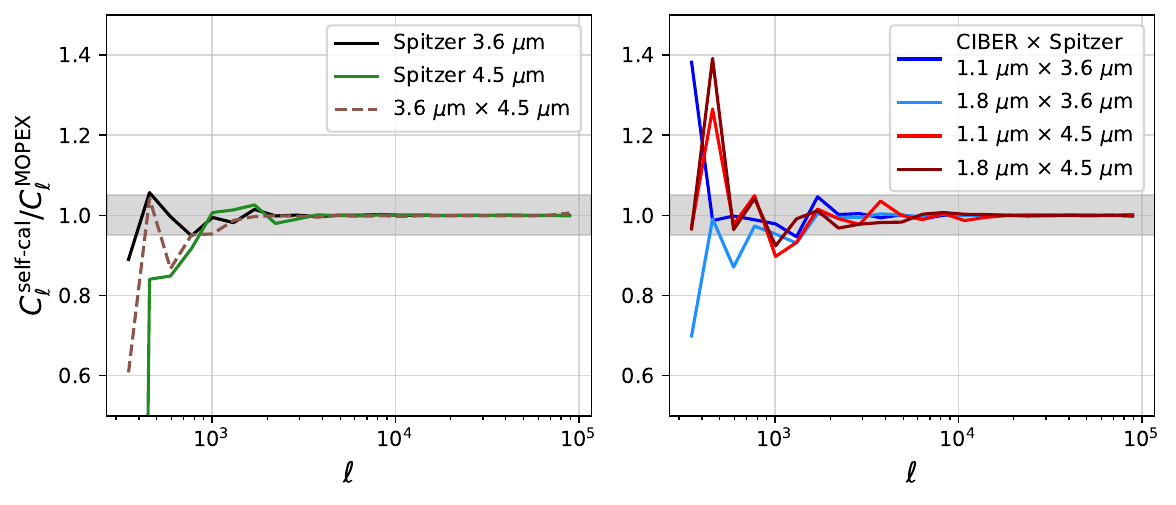}

    \caption{Comparison of auto- and cross-power spectra using different \emph{Spitzer} mosaicing algorithms. In the left panel we show \emph{Spitzer} internal auto- and cross-power spectrum ratios, while in the right panel we show the same but for \emph{CIBER} $\times$ \emph{Spitzer} cross-power spectra. The grey bands indicate $\pm 5\%$ deviations in recovered power.}
    \label{fig:sdwfs_mosaic_compare_ps}
\end{figure}

To assess the sensitivity of our \emph{Spitzer}-based measurements to choice of mosaicing algorithm, we compute SDWFS auto- and cross-power spectra in the \bootes\ fields from both the SDWFS fiducial mosaics (generated with the MOPEX algorithm) and those from \cite{cooray12} that were generated using a self-calibration method (used in our fiducial results). We plot the ratio of power spectra for different band combinations in Fig. \ref{fig:sdwfs_mosaic_compare_ps}, finding considerable differences on scales $\ell < 1000$. In particular, the \emph{Spitzer} 3.6 $\mu$m and 4.5 $\mu$m auto- and cross-power spectra show consistently lower power when using the self-cal mosaics, suggesting the differences are not driven by noise. The \emph{CIBER} $\times$ \emph{Spitzer} cross-power spectra show more complicated behavior among the lowest $\ell$ bandpowers. These results highlight the importance of mosaicing choices when reconstructing modes much larger than the IRAC field of view. However, without a more detailed characterization of mosaicing transfer functions, it is difficult to interpret inconsistencies between codes and whether biases exist in both sets of fluctuation measurements.

